\documentclass[a4paper,11pt]{article}
\usepackage{jcappub} 
\usepackage{lineno}
\usepackage[dvipsnames]{xcolor}
\usepackage{xfrac}
\usepackage[T1]{fontenc} 
\usepackage{aas_macros}
\usepackage{natbib}
\usepackage{bm}
\usepackage{hyperref}
\usepackage{subcaption}
\usepackage{mathrsfs}

\usepackage{xcolor}
\usepackage{soul}

\arxivnumber{1234.56789} 
\title{Baryonification V: Thermal Sunyaev-Zel'dovich maps from dark-matter-only lightcones}

\author[a]{Michael Kova\v{c},}
\author[a]{Andrina Nicola,}
\author[b]{Jozef Bucko,}
\author[c]{Siyu Bian,}
\author[c]{Aurel Schneider,}
\author[d,e]{Robert Reischke,}
\author[a]{Jessica A. Cowell,}
\author[b]{Alexandre Refregier,}
\author[f, g]{Sambit K. Giri,}
\author[h,i]{Matthieu Schaller,}
\author[i]{Joop Schaye}

\affiliation[a]{Jodrell Bank Centre for Astrophysics, Department of Physics and Astronomy, The University of Manchester, Manchester M13 9PL, UK}
\affiliation[b]{Institute for Particle Physics and Astrophysics, ETH Zurich, Wolfgang-Pauli-Strasse 27, 8093 Zurich, Switzerland}
\affiliation[c]{Department of Astrophysics, University of Zurich, Winterthurerstrasse 190, 8057 Zurich, Switzerland}
\affiliation[d]{Ruhr University Bochum, Faculty of Physics and Astronomy, Astronomical Institute (AIRUB), German Centre for Cosmological Lensing, 44780 Bochum, Germany}
\affiliation[e]{Argelander-Institut für Astronomie, Universität Bonn, Auf dem Hügel 71, D-53121 Bonn, Germany}
\affiliation[f]{Van Swinderen Institute for Particle Physics and Gravity, University of Groningen, Nijenborgh 4, 9747 AG Groningen, The Netherlands}
\affiliation[g]{Department of Astronomy and Oskar Klein Centre, AlbaNova, Stockholm University, SE-10691 Stockholm, Sweden}
\affiliation[h]{Leiden Observatory, Leiden University, PO Box 9513, 2300 RA Leiden, the Netherlands}
\affiliation[i]{Lorentz Institute for Theoretical Physics, Leiden University, PO box 9506, 2300 RA Leiden, the Netherlands}

\emailAdd{michael.kovac@postgrad.manchester.ac.uk}

\abstract{Baryonic feedback is a major theoretical uncertainty for Stage-IV weak lensing (WL) surveys. The thermal Sunyaev-Zel'dovich (tSZ) effect traces the hot gas that feedback redistributes, so combining it with weak lensing can calibrate feedback and improve cosmological constraints. Using this in a simulation-based inference framework requires correlated lensing and Compton-$y$ maps across many cosmologies and feedback scenarios. In this paper we present two methods to generate Compton-$y$ maps from dark-matter-only $N$-body light cones using the component-wise baryonification (BFC) model, which produces separate gas and stellar fields and computes the gas pressure from hydrostatic equilibrium. We first show that the two methods, painting the spherically symmetric pressure profiles onto halos or redistributing the signal along the baryonified gas, agree to within a few per cent. We then calibrate the model to the fiducial and strongest AGN feedback FLAMINGO simulations by jointly fitting the gas mass fraction and thermal pressure profiles, measuring the non-thermal pressure from the simulated gas velocities rather than leaving it as a free parameter. The predicted Compton-$y$ profiles match the profiles from FLAMINGO within halo-to-halo scatter, and when painted onto the FLAMINGO halos we reproduce the tSZ power spectrum of both feedback models to $\sim5\%$ for $500\lesssim\ell\lesssim3000$. Applied to a \emph{CosmoGridV1-like} $N$-body simulation, our maps agree with FLAMINGO within $\sim10\%$ for $300\lesssim\ell\lesssim8000$. The remaining differences come from the few hundred most massive clusters, which dominate the tSZ signal: removing them from both halo catalogs brings the two paintings into agreement. Our method therefore allows for accurate and efficient forward models of the tSZ signal starting from $N$-body simulations and can thus be applied directly to cosmological simulation suites such as CosmoGridV1. This provides an important step towards joint simulation-based inference analyses of weak lensing and tSZ for surveys such as Euclid, LSST, and the Simons Observatory.}

\begin{document}
\maketitle
\flushbottom

\section{Introduction}\label{sec: Introduction}

Energetic feedback from active galactic nuclei (AGN) and supernovae redistributes gas within and around halos and suppresses the matter power spectrum on non-linear scales~\citep{van_Daalen_2011, Chisari_2019, van_Daalen_2020, Schaye_J_2023}. Since these scales carry much of the lensing signal, the strength of baryonic feedback is one of the leading theoretical uncertainties~\citep{Semboloni_2011, Schneider_2020} for Stage-IV weak-lensing (WL) surveys such as Euclid\footnote{\url{https://www.esa.int/Science_Exploration/Space_Science/Euclid}}, the Rubin Observatory Legacy Survey of Space and Time (LSST)\footnote{\url{https://www.lsst.org/}}, and the Roman Space Telescope\footnote{\url{https://roman.gsfc.nasa.gov/}.}. Observables that trace the baryons directly, in particular the hot, ionized gas, probe this redistribution and can therefore constrain the feedback that WL analyses would otherwise have to marginalize over. 

The thermal Sunyaev-Zel'dovich (tSZ) effect is especially well suited to constrain feedback. It arises from inverse Compton scattering of Cosmic Microwave Background photons off hot electrons, and its amplitude, the Compton-$y$ parameter, is the line-of-sight integral of the thermal electron pressure $P_{\rm e}\propto n_{\rm e}T_{\rm e}$. The signal therefore depends both on where the gas is located and on how much thermal energy has been injected by feedback. Compton-$y$ maps have been produced by \textit{Planck}, the Atacama Cosmology Telescope (ACT), and the South Pole Telescope (SPT), with deeper maps expected from the Simons Observatory (SO), and the angular power spectrum $C_\ell^{yy}$ has been measured over a wide range of scales~\citep{Planck_tsz_2016, Choi_act_tsz_2020, Reichardt_SPT_2021, Efstathiou_McCarthy_25}. Its amplitude scales steeply with the matter fluctuations, $C_\ell^{yy}\propto\sigma_8^{7}$~\citep{Komatsu_2002}, while its shape at $\ell\gtrsim1000$ traces the same feedback that suppresses the matter power spectrum~\citep{Battaglia_2012}. Hydrodynamical simulations generally overpredict the measured $C_\ell^{yy}$ unless a low-$S_8$ cosmology is adopted~\citep{McCarthy_2014, Mccarthy_2018, McCarthy_2025}, illustrating how closely cosmology and feedback are intertwined in this observable. Moreover, since the integrated Compton-$y$ signal of a halo scales roughly as $M^{5/3}$, $C_\ell^{yy}$ at $\ell\lesssim1000$ is dominated by a few massive, low-redshift clusters, making the field strongly non-Gaussian and subject to large sample variance. Some analyses therefore treat the brightest clusters separately~\citep{Rotti_Bolliet_Chluba_2021}.

Simulation-based inference (SBI), which learns the likelihood from forward-simulated maps~\citep{Fluri_J_2022, Jeffrey_2024, Thomsen_2025}, instead accounts for this non-Gaussianity directly and can exploit map-level and higher-order statistics. SBI also provides a natural framework for combining probes. Combining the tSZ effect with WL can calibrate feedback while tightening cosmological constraints~\citep{Hojjati_2017, Osato_2020, Troester_2022, Gatti_2022, Pandey_2024, McCarthy_2024}: the two probes respond differently to $\sigma_8$ and feedback, and overlap only partially in the halos they are most sensitive to~\citep{Lucie-Smith_2025}. A joint SBI analysis of tSZ, WL, and their cross-correlations with upcoming data from SO and Stage-IV lensing surveys therefore requires thousands of mocks of all probes, generated consistently from the same underlying large-scale structure while varying both cosmology and feedback.

Cosmological hydrodynamical simulations~\citep{Dubois_HorizonAGN_2014, leBrun2014owls, Mccarthy_2016bahamas, Springel_2017, Dolag2016sz, Schaye_J_2023} provide such correlated predictions, but their cost restricts them to one or a few cosmologies and limited volumes, far from the $\sim20{,}000$ simulations of suites such as CosmoGridV1~\citep{Kacprzak_T_2023}. Moreover, their subgrid models, although calibrated to similar data, still disagree on the matter power spectrum suppression at the tens-of-percent level~\citep[e.g.][]{van_Daalen_2020, Schaller_2025}, so no single simulation can serve as ground truth, and cosmology and feedback must be varied simultaneously.

Baryonification (BFC), first introduced in Ref.~\citep{Schneider2015_code}, provides an efficient alternative to hydrodynamical simulations. BFC models displace particles in a dark-matter-only (DMO) $N$-body simulation according to physically motivated parametric gas, stellar, and dark matter density profiles, with several variants that share this principle but differ in their profiles and parametrisation~\citep{Schneider2019_code, Arico_2020_BCM, Arico_2020_BACCO, Anbajagane_D_2024, Schneider_A_2025}. Because BFC only post-processes a DMO simulation, it adds little to the cost of the $N$-body run and can be used with existing simulation suites. The model parameters control the radial shapes of the gas and matter profiles, which in turn set physically meaningful quantities such as the halo gas fraction. This allows for retuning the model to different feedback scenarios and to sample its parameters alongside the cosmological parameters, rather than fixing them to a single subgrid model. BFC reproduces the suppression of the matter power spectrum found in a wide variety of hydrodynamical simulations to within a 1--2\%~\citep{Giri2021_BCemu, Schneider_A_2025, Zhou_A_2025} and has been applied in various combinations to cosmological and astrophysical inference using WL, X-ray gas fractions, and kinetic SZ data~\citep{Schneider2022_WL_xray_kSZ, Grandis2023, Bigwood2024, Kovac_2025, Siegel_2026, Bucko_2026}. 

While standard BFC relies on snapshots, large simulation suites usually keep only projected lightcone shells. BFC has therefore been modified to act directly on these shells~\citep{Fluri_J_2022, Anbajagane_D_2024}, and in this form it has been used to baryonify entire simulation suites designed for SBI, such as CosmoGridV1~\citep{Kacprzak_T_2023, Fluri_J_2022}. However, these shell-based implementations rely on earlier or different BFC models that displace the total matter and do not provide separate gas and stellar fields. In these implementations, the gas enters only through the analytic profiles used to compute the displacement.

Assuming hydrostatic equilibrium (HSE), the gas pressure and temperature follow directly from BFC, because the gas profile of each halo is linked to its total mass profile. Ref.~\citep{Arico_G_2024} combined this approach with the baryonic correction model (BCM) of Ref.~\citep{Arico_2020_BCM} and fitted it to the pressure power spectra of the BAHAMAS simulations~\citep{Mccarthy_2016bahamas, Mccarthy_2018}, recovering their tSZ and WL power spectra to within $\sim10\%$ and $\sim1\%$, respectively. Ref.~\citep{Anbajagane_D_2024} applied BFC at the map level using \textsc{BaryonForge}, which baryonifies lightcone shells and paints pressure profiles onto halos, enabling efficient modeling of higher-order correlations in the WL and tSZ fields. Both \textsc{BaryonForge} and the BCM of Ref.~\citep{Arico_2020_BCM} build on earlier versions of the BFC model~\citep{Schneider2015_code, Schneider2019_code}. A related approach, baryon pasting, assigns gas directly to the halos or particles of an $N$-body simulation using analytic HSE models, where the gas density and pressure follow a polytropic equation of state~\citep{Osato_K_2023}. Without a separate gas field, however, shell-based methods can assign the pressure only as spherically symmetric profiles at halo positions, rather than following the displaced gas.

Using the tSZ effect in SBI requires Compton-$y$ maps that are fast to generate for entire simulation suites and validated against hydrodynamical simulations. Existing BFC-based tSZ maps have not been validated against a modern hydrodynamical suite at the level of both Compton-$y$ profiles and $C_\ell^{yy}$ for different feedback strengths, the BFC parameters have been calibrated to mass fractions, mass-ratio profiles, or power spectra rather than to pressure profiles, and the non-thermal pressure is at least partly fixed to generic prescriptions. In this work, we address these points using the component-wise BFC model of Ref.~\citep{Schneider_A_2025} (hereafter S25), which splits each $N$-body particle into a dark-matter and a baryon component and displaces them separately, yielding individual dark-matter, gas, and stellar density fields together with a pressure field from HSE. To obtain accurate tSZ predictions, we calibrate the model to the fiducial (L1\_m9) and strongest AGN feedback (f$_{\rm gas}-8\sigma$) FLAMINGO simulations~\citep{Schaye_J_2023, Kugel_2023} by jointly fitting the halo gas mass-fraction profiles and the pressure profiles. Additionally, we measure the non-thermal pressure from gas velocities in the FLAMINGO simulations rather than treating it as a free parameter. We then adapt the existing shell-BFC method~\citep{Fluri_J_2022, Anbajagane_D_2024} by implementing the new S25 model, with a full description and validation presented in a companion paper (Schneider, Bucko, Kova\v{c} et al. in prep.). We build Compton-$y$ maps with two methods: painting spherically symmetric pressure profiles at halo positions, and a gas-weighted method that weights the pressure profile using the baryonified gas maps from shell-BFC. The two methods yield consistent power spectra. We compare our maps to FLAMINGO at the level of both Compton-$y$ profiles and power spectra, and show that we reproduce their results.

This paper is structured as follows. Sec.~\ref{sec: theory modeling tsz} summarizes the tSZ effect, its halo-model power spectrum, and the BFC model; Sec.~\ref{sec: hydrosims} describes the FLAMINGO simulations used for comparison and how the BFC model is calibrated to them; Sec.~\ref{sec: bfc sims} presents the framework used to generate our simulations and how they are baryonified; Sec.~\ref{sec: maps} presents the two tSZ map-making methods; Sec.~\ref{sec: pipeline validation} describes our pipeline validation; Sec.~\ref{sec: Comparison to hydro} compares our generated tSZ maps against the FLAMINGO simulations; and Sec.~\ref{sec: conclusion} discusses the application to the CosmoGridV1 suite and concludes.

\section{Modeling the tSZ signal}\label{sec: theory modeling tsz}

In this section, we will first review the thermal Sunyaev-Zel'dovich (tSZ) effect~\citep{SZ_effect_1970}, in particular the modeling of the Compton-$y$ signal and the halo-model calculation of its angular power spectrum in Sec.~\ref{subsec: tsz effect} and Sec.~\ref{subsec: tsz power spectrum}. Both take the thermal electron pressure profile of a halo, as a function of halo mass and redshift, as their central input. We then summarize the BFC model in Sec.~\ref{subsec: baryonification}, using the thermal electron pressure profile from the BFC model of S25, and we summarize the components relevant to the tSZ signal, especially the electron pressure profile. 

\subsection{Thermal Sunyaev-Zel'dovich effect}\label{subsec: tsz effect}

The thermal Sunyaev-Zel'dovich (tSZ) effect is a secondary cosmic microwave background (CMB) anisotropy caused by the inverse Compton scattering of CMB photons off hot, free electrons primarily located in the intracluster medium (ICM) of galaxy clusters. Because the tSZ spectral modification follows a known, universal frequency dependence, multi-frequency observations allow us to isolate this component and extract the line-of-sight integrated thermal electron pressure, parameterized directly by the dimensionless Compton-$y$ signal as a function of angular position from the center of the cluster $\bm{\theta}$:

\begin{equation}\label{eq: compton-y}
    y(\bm{\theta}) = \frac{\sigma_\mathrm{T}}{m_\mathrm{e} c^2}\int_{\rm los} \mathrm{d}l\; \mathscr{P}_{\rm e}\!\left(\sqrt{l^2 + d_\mathrm{A}(z)^2\theta^2}\right) \;,
\end{equation}
where $\sigma_\mathrm{T}$ is the Thomson cross-section, $m_\mathrm{e} c^2$ the electron rest-mass energy, $\mathscr{P}_{\rm e}$ the thermal electron pressure, and $d_\mathrm{A}(z)$ the angular diameter distance at redshift $z$. Since $y$ traces the integrated thermal electron pressure of the gas, which baryonic feedback processes such as AGN heating and gas expulsion strongly shape, the tSZ effect is a sensitive probe of both the distribution of hot gas around halos and feedback strength.

In this work, we obtain $\mathscr{P}_{\rm e}$ directly from the BFC model described in Sec.~\ref{subsec: baryonification}, which provides the thermal electron pressure profile of each halo as a function of mass and redshift. To avoid double-counting gas between adjacent shells, we set the line-of-sight integration range in Eq.~\eqref{eq: compton-y} to the shell thickness $L$.

\subsection{tSZ power spectrum}\label{subsec: tsz power spectrum}

While the spatial Compton-$y$ map traces the line-of-sight integrated thermal pressure, the angular auto-power spectrum efficiently captures its statistical properties, $C_\ell^{yy}$. Measuring $C_\ell^{yy}$ provides a dual probe of cosmological parameters and baryonic astrophysics. Since the Compton-$y$ signal is dominated by massive clusters, its overall normalization is strongly sensitive to $\sigma_8$~\citep{Komatsu_2002}, while its small-scale shape ($\ell \gtrsim 1000$) traces the gas pressure profile $\mathscr{P}_\mathrm{e}(r)$ and hence baryonic feedback. Compton-$y$ maps are built from multi-frequency CMB data (e.g., \textit{Planck}, ACT, SPT) using component separation methods~\citep{Delabrouille_2009}, which make use of the distinctive tSZ spectrum: a decrement below $\approx 217\,\mathrm{GHz}$ and an increment above it. $C_\ell^{yy}$ is then measured from these maps with pseudo-$C_\ell$ estimators that correct for masking, beam smoothing, and partial sky coverage.

Since the tSZ signal is contained almost completely by gas bound in dark matter halos, $C_\ell^{yy}$ can be predicted well through the halo model~\citep{Seljak_2000, Peacock_2000, Cooray_2002}. Within the halo model, the power spectrum of two fields $U(\bm{x})$ and $V(\bm{x})$ can be described as the sum of a one-halo term, from the correlations within the halo, and a two-halo term, from correlations between surrounding halos:

\begin{equation}
    P_{UV}(k,z) = P^{1\mathrm{h}}_{UV}(k,z) + P^{2\mathrm{h}}_{UV}(k,z),
\end{equation}
with
\begin{align}
    P^{1\mathrm{h}}_{UV}(k,z) &= \int \mathrm{d}M\, \frac{\mathrm{d}n}{\mathrm{d}M}(M,z)\, u(k,z|M)\, v(k,z|M) \,, \\
    P^{2\mathrm{h}}_{UV}(k,z) &= \langle b\,u \rangle(k,z)\, \langle b\,v \rangle(k,z)\, P_\mathrm{lin}(k,z) \,,
\end{align}
where we have defined the bias-weighted profile integral
\begin{equation}
    \langle b\,u \rangle(k,z) \equiv \int \mathrm{d}M\, \frac{\mathrm{d}n}{\mathrm{d}M}(M,z)\, b(M,z)\, u(k,z|M) \,,
\end{equation}
and analogously for $v$. $\mathrm{d}n/\mathrm{d}M$ is the halo mass function, $b(M,z)$ the linear halo bias, $P_{\rm lin}(k,z)$ the linear matter power spectrum, and $u$ and $v$ the Fourier-space profiles of $U$ and $V$ for a halo with mass $M$.
For the tSZ auto-power spectrum, $u=v=\tilde{\mathscr{P}}_{\rm e}$, the Fourier transform of the isotropic electron pressure profile,
\begin{equation}\label{eq: uk}
    \tilde{\mathscr{P}}_{\rm e}(k,z|M) = 4\pi \int_0^\infty \mathrm{d}r\, r^2\, \frac{\sin(kr)}{kr}\, \mathscr{P}_{\rm e}(r|M,z) \;,
\end{equation}
The angular power spectrum $C_\ell^{yy}$ is obtained from the standard power spectrum $P_{\rm ee}(k,a)$ via the Limber-approximated projection
\begin{equation}\label{eq: Cl Limber}
    C_\ell^{yy} = \int \mathrm{d}\chi\, \frac{W_y(\chi)^2}{\chi^2}\, {P}_{\rm ee}\!\left(k = \frac{\ell+1/2}{\chi},\, z(\chi)\right) \;,
\end{equation}
where $\chi$ is the comoving distance and the Compton-$y$ radial kernel is
\begin{equation}\label{eq: tsz kernel}
    W_y(\chi) = \frac{\sigma_\mathrm{T}}{m_\mathrm{e} c^2}\,\frac{1}{1+z(\chi)} \;. 
\end{equation}

We estimate the theoretical uncertainty of the halo-model prediction from the covariance of $C_\ell^{yy}$, the sum of a Gaussian and a connected non-Gaussian term~\citep{Komatsu_2002, Bolliet_2018},
\begin{equation}\label{eq: cov Cl}
    \mathrm{Cov}(\ell,\ell') = \frac{2\,\big(C_\ell^{yy}\big)^2}{(2\ell+1)\,f_{\rm sky}}\,\delta_{\ell\ell'}
    + \frac{1}{4\pi f_{\rm sky}} \int \mathrm{d}\chi\, \frac{W_y(\chi)^4}{\chi^6}\,
      T^{1\mathrm{h}}_{\rm ee}\!\left(\frac{\ell+1/2}{\chi}, \frac{\ell'+1/2}{\chi}, z(\chi)\right) \;,
\end{equation}
with $f_{\rm sky}$ being the sky fraction and $T^{1\mathrm{h}}_{\rm ee}$ the one-halo term of the trispectrum. Since $C_\ell^{yy}$ is dominated by rare, massive clusters~\citep{Komatsu_2002} we keep only the one-halo term of the trispectrum,
\begin{equation}\label{eq: T1h}
    T^{1\mathrm{h}}_{\rm ee}(k,k',z) = \int \mathrm{d}M\, \frac{\mathrm{d}n}{\mathrm{d}M}(M,z)\,
    \tilde{\mathscr{P}}_{\rm e}^{\,2}(k,z|M)\, \tilde{\mathscr{P}}_{\rm e}^{\,2}(k',z|M) \;.
\end{equation}
The theoretical uncertainty on $C_\ell^{yy}$ is the square root of the diagonal of Eq.~\eqref{eq: cov Cl},
\begin{equation}\label{eq: sigma Cl}
    \sigma^2\big(C_\ell^{yy}\big) = \frac{2\,\big(C_\ell^{yy}\big)^2}{(2\ell+1)\,f_{\rm sky}} + \mathrm{Cov}^{\rm cNG}(\ell,\ell) \;,
\end{equation}
where $\mathrm{Cov}^{\rm cNG}$ is the second term of Eq.~\eqref{eq: cov Cl} and $f_{\rm sky}=1$ as we assume a full-sky map.

\subsection{Baryonification}\label{subsec: baryonification}

From a dark-matter-only input, the BFC model of S25 predicts the dark-matter, gas, and stellar density fields of each halo, together with its thermal pressure profile. In the following, we outline the method (Sec.~\ref{subsubsec: bfc-overview}) and then give the gas density and pressure profiles that source the tSZ signal (Sec.~\ref{subsubsec: bfc-profiles}). For the dark-matter and stellar profiles and the complete parameter set, we refer the reader to S25.

\subsubsection{Overview of the Baryonification method}\label{subsubsec: bfc-overview}

The BFC model modifies gravity-only $N$-body simulations so that the matter distribution in and around haloes resembles that of a hydrodynamical simulation, without explicitly following gas physics. Each halo is described by spherically symmetric density profiles for the initial gravity-only and final baryonified state. The latter accounts for gas expelled by AGN feedback, gas that cooled to form galaxies, and the dark-matter back-reaction due to changes in the gravitational potential. The BFC parameters $\pmb{\theta}_\mathrm{bfc}$ control all final-state profiles. Around each halo, we then move particles radially so their distribution changes from the former gravity-only profile to the baryonified profile described by $\pmb{\theta}_\mathrm{bfc}$. Throughout this paper, we define the halo mass, $M_{200}$, as the mass enclosed within a radius $r_{200}$ centered on the group or cluster where the mean density equals 200 times the critical density of the Universe, unless stated otherwise.

Following the component-wise displacement from S25, each particle in the gravity-only output is first replaced by a dark matter and baryon particle, which share the same initial position but will be displaced differently. The masses are renormalized to match the cosmic dark matter and baryon fractions. The radial displacement for each host halo center is then defined by

\begin{equation}\label{eq: displacement}
\begin{aligned}
d_{\mathrm{dm}}(\pmb{r}_{\mathrm{i,dm}}|M_{200},c_{200},\pmb{\theta}_{\mathrm{bfc}}) &= \pmb{r}_{\mathrm{f,dm}}(M_{\mathrm{dm}}) - \pmb{r}_{\mathrm{i,dm}}(M_{\mathrm{dm}}), \\
d_{\mathrm{bar}}(\pmb{r}_{\mathrm{i,bar}}|M_{200},c_{200},\pmb{\theta}_{\mathrm{bfc}}) &= \pmb{r}_{\mathrm{f,bar}}(M_{\mathrm{bar}}) - \pmb{r}_{\mathrm{i,bar}}(M_{\mathrm{bar}}),
\end{aligned}
\end{equation}
and thus depends on the initial radius $r_{\rm i}$ of the particle, the halo mass $M_{200}$, the halo concentration $c_{200}$, and the BFC parameters $\pmb{\theta}_\mathrm{bfc}$. The functions $\pmb{r}_{\rm i,f}(M)$ are the inverse cumulative mass profiles of the initial gravity-only and final baryonified states,

\begin{equation}
M_{\mathrm{i,f}}(r) = 4\pi\int_0^r \mathrm{d}s\, s^2 \rho_{\mathrm{i,f}}(s)\, ,
\label{eq:cumulative_mass}
\end{equation}
where $r$ is the distance to the halo center, and $\rho_{\mathrm{i,f}}$ are the individual initial and final density profiles described in the next Sec.~\ref{subsubsec: bfc-profiles}.

\subsubsection{Density Profiles}\label{subsubsec: bfc-profiles}

The initial density profiles of dark matter and baryons share the same functional form and differ only by their cosmic fraction. The profile combines a one-halo term, described by a truncated NFW profile~\citep{Navarro_1996, Navarro_1997, Baltz_2009}, and a two-halo contribution representing the large-scale environment. 

\begin{equation}\label{eq: rho1h+rho2h}
\begin{split}
    \rho_\mathrm{dm,i}(r) &= f_{\rm dm}[\rho_{\rm nfw}(r) + \rho_{\rm 2h}(r)], \quad \rho_\mathrm{dm,f}(r) = \rho_{\rm dm}(r) + f_{\rm dm}\rho_{\rm 2h}(r)\;, \\
    \rho_\mathrm{bar,i}(r) &= f_{\rm bar}[\rho_{\rm nfw}(r) + \rho_{\rm 2h}(r)], \quad \rho_\mathrm{bar,f}(r) = \rho_{\rm gas}(r) + \rho_{\rm stars}(r) + f_{\rm bar}\rho_{\rm 2h}(r)\,,
\end{split}
\end{equation}
where the cosmic dark matter and baryon fractions are defined by $f_{\rm dm} = \Omega_{\rm dm}/\Omega_\mathrm{m}$ and $f_{\rm bar} = \Omega_{\rm b}/\Omega_\mathrm{m}$. The final state profiles of the gas and stellar components are further divided into $\rho_{\rm gas}=\rho_{\rm hga}+\rho_{\rm iga}$ and $\rho_{\rm stars}=\rho_{\rm cga}+\rho_{\rm sga}$, where hga, iga, cga, and sga represent the diluted hot gas, the central cold gas (i.e., ISM), central galaxies, and satellite galaxies, respectively. 

Since the tSZ signal is primarily sensitive to the hot ionized gas, we focus on this component below. In the BFC model, the hot gas is modeled as
\begin{equation}\label{eq: rhoHGA}
\rho_{\rm hga}(r) = \rho_{\rm hga,0}
                    \left[1+x_{\rm c}^{\alpha}\right]^{-\beta(M_{200})/\alpha}
                    \left[1+x_{\rm t}^{\gamma}\right]^{-\delta/\gamma}\,,
\end{equation}
where $x_{\rm c}=r/r_{\rm c}$ and $x_{\rm t}=r/r_{\rm t}$. The core radius is defined as $r_{\rm c}=\theta_{\rm co}r_{200}$, the truncation radius as $r_{\mathrm{t}} = \varepsilon r_{200}$ with $\varepsilon = \varepsilon_0 - \varepsilon_1 \nu$, with $\varepsilon_0 = 4.0$, $\varepsilon_1=0.5$, and $\nu = \delta_\mathrm{c}/\sigma(M)$ being the peak height of the halo with $\delta_\mathrm{c}=1.686$. $\theta_{\rm co}$, and $\delta$ regulate the inner and outer parts of the hot gas profile, while the mass-dependent slope $\beta$ of the profile is defined through
\begin{equation}\label{eq: rhoHGA beta}
\beta(M_{200}) = \frac{3(M_{200}/M_{\rm c})^{\mu}}
                    {1+(M_{200}/M_{\rm c})^{\mu}}\,.
\end{equation}
The parameters $M_{\rm c}$ and $\mu$ control the pivot mass and mass dependence of the gas profile. Lower-mass halos generally have flatter gas density profiles (lower $\beta$) because AGN-driven gas expulsion is stronger.

The hot gas fraction is determined from the cosmic baryon fraction after subtracting the stellar and cold gas contributions,
\begin{equation}
f_{\rm gas} = f_{\rm bar} - f_{\rm star} - f_{\rm iga}\,.
\end{equation}
For a detailed description of the dark matter, stellar, and cold gas parametrization, we refer the reader to Ref.~\cite{Schneider_A_2025}.

\subsubsection{Pressure Profile}\label{subsubsec: pressure-profile}
Assuming that galaxy clusters and groups are in HSE, we can obtain the total pressure profile through the thermal (hot) gas density profile and the enclosed total mass profile
\begin{equation}
\frac{\mathrm{d}\mathscr{P}_{\rm tot}}{\mathrm{d}r} = -\rho_{\rm hga}(r)\,\frac{GM_{\rm tot}(r)}{r^2}\,,
\label{eq: Ptot}
\end{equation}
HSE holds only for spherically symmetric, virialized, stable halos that are not undergoing mergers or residing in turbulent regions, and it is less reliable for low-mass halos (see, e.g., Ref.~\cite{Oppenheimer:2018, Braspenning_2025}).
Furthermore, non-thermal pressure from effects such as bulk motions and turbulence due to gas accretion or mergers on the outskirts contributes to the total pressure. This becomes particularly pronounced at large halo radii~\citep{Shaw_L_2010, Nelson_2014, Angelinelli_2020}. We therefore define the total pressure as the sum of thermal and non-thermal components: $\mathscr{P}_\mathrm{tot}=\mathscr{P}_\mathrm{th}+\mathscr{P}_\mathrm{nth}$. Since the tSZ effect arises exclusively from thermal electron pressure, we must subtract this non-thermal contribution from Eq.~\eqref{eq: Ptot} to isolate the thermal contribution. Following the work of Ref.~\cite{Shaw_L_2010}, the fraction of non-thermal to total pressure can be modeled as

\begin{equation}\label{eq: Pnth}
    \frac{\mathscr{P}_\mathrm{nth}}{\mathscr{P}_\mathrm{tot}}(z)=\alpha_{\rm nth}(z)\left(\frac{r}{R_{200}}\right)^{n_\mathrm{nth}}\;,
\end{equation}
with $n_{\rm nth}$ defining the slope of the non-thermal pressure and $\alpha_{\rm nth}(z)=\alpha_{\rm nth,0}f(z)$ being the redshift dependent amplitude, with $\alpha_{\rm nth,0}$ being the amplitude at redshift $z=0$. $f(z)$ is an increasing or decreasing function of redshift given by:

\begin{equation}\label{eq: pressure profile f(z)}
    f(z)=\mathrm{min}\left[(1+z)^{\beta_{\rm nth}},\;(f_\mathrm{max}-1)\;\mathrm{tanh}(\beta_{\rm nth} z)+1\right]\;,
\end{equation}
with $f_{\rm max} = 4^{-n_{\rm nth}}/\alpha_{\rm nth,0}$ and $\beta_{\rm nth}$ a free parameter regulating the evolution rate. While Ref.~\cite{Shaw_L_2010} originally calibrated $\alpha_{\rm nth,0}=0.18$, $\beta_{\rm nth}=0.5$, and $n_{\rm nth}=0.8$ against a fixed set of hydrodynamical simulations (values that were also adopted in S25), the amount of non-thermal pressure support in halos is expected to depend on the underlying feedback physics, since stronger (or more recent) energy injection alters the dynamical state of the gas. We therefore do not adopt these values as universal, but instead re-calibrate $\alpha_{\rm nth,0}$, $\beta_{\rm nth}$, and $n_{\rm nth}$ separately for each feedback model considered in this work, such that the resulting thermal pressure profiles agree with the non-thermal pressure fraction measured directly in the corresponding FLAMINGO simulations~\cite{Schaye_J_2023} (see Sec. \ref{ssec: flamingo_sims}); the resulting values of $\alpha_{\rm nth,0}$, $\beta_{\rm nth}$, and $n_{\rm nth}$ are given in Table~\ref{tab: parameters}. Appendix~\ref{app:pnt} describes this calibration procedure in detail. We note that Eq.~\eqref{eq: Pnth} can become larger than one, i.e., the thermal pressure negative, in the far outskirts of the halo, well beyond $r_{200}$, where the non-thermal correction formally exceeds the total pressure; in this regime we set $\mathscr{P}_{\rm th}(r)=0$, consistent with the expectation that thermal pressure support vanishes before non-thermal support does.

Finally, by assuming a fully ionized medium with primordial abundances, the electron pressure can then be estimated with

\begin{equation}\label{eq: P_e}
    \mathscr{P}_{\rm e}=\left(\frac{2+2X_{\rm H}}{3+5X_{\rm H}}\right)\mathscr{P}_{\rm th},
\end{equation}
where $X_{\rm H}=0.76$ is the hydrogen mass fraction from primordial abundances (see e.g. \cite{Dodelson_2003}). All BFC model parameters, including their values, are shown in Tab.~\ref{tab: parameters}.

For all halo model calculations, we use \textsc{PyCCL}~\citep{Chisari_E_2019} together with our implementation of the BFC profiles.

\begin{table}[t!]
\renewcommand{\arraystretch}{1.5} 
\setlength{\tabcolsep}{8pt} 
\begin{center}
\begin{tabular}{cp{8cm}cc} \hline\hline
{Parameter} & Description & {L1\_m9} & {$f_\mathrm{gas}-8\sigma$} \\ \hline
\multicolumn{4}{l}{\textit{Fitted to the FLAMINGO gas mass-fraction and pressure profiles}} \\
\hline
$\theta_{\mathrm{co}}$ & Core radius of the gas profile relative to $r_{200}$ (Eq.~\ref{eq: rhoHGA}) & 0.276 &  0.351\\
$\log_{10}\: M_\mathrm{c}[\mathrm{M_{\odot}}/h]$ & Pivot mass for the mass-dependent slope of the gas profile (Eq.~\ref{eq: rhoHGA} \& Eq.~\ref{eq: rhoHGA beta}) & 13.153 & 13.660 \\
$\mu$ & Transition steepness of mass-dependent gas parameter (Eq.~\ref{eq: rhoHGA} \& Eq.~\ref{eq: rhoHGA beta}) & 0.899 & 0.922 \\
$\delta$ & Characterizes the transition from the inner to the outer part of the gas profile (Eq.~\ref{eq: rhoHGA}) & 4.574 & 4.830 \\
$\eta$ & Total stellar fraction within a halo (Eq. 2.21 in~\cite{Schneider_A_2025}) & 0.066 & 0.073 \\     
$d\eta$ & Fraction of central galaxies with respect to the total stellar fraction (Eq. 2.21 in~\cite{Schneider_A_2025}) & 0.215 & 0.226 \\
$N_{\mathrm{star}}$ & Stellar fraction amplitude (Eq. 2.21 in~\cite{Schneider_A_2025}) & 0.026 & 0.028 \\
$c_\mathrm{iga}$ & Cold gas fraction (Eq. 2.23 in~\cite{Schneider_A_2025}) & 0.122 & 0.107 \\
\hline
\multicolumn{4}{l}{\textit{Fixed: measured from the FLAMINGO gas velocities}} \\
\hline
$\alpha_{\mathrm{nth, 0}}$ & Non-thermal pressure amplitude at $z=0$ (Eq.~\ref{eq: Pnth}) & 0.135 & 0.107 \\
$n_{\mathrm{nth}}$ & Slope of the non-thermal pressure (see Eq.~\ref{eq: Pnth}) & 0.80 & 0.54 \\
$\beta_{\mathrm{nth}}$ & Redshift evolution for the non-thermal pressure amplitude (see Eq.~\ref{eq: pressure profile f(z)}) & 1.575 & 0.938\\\hline\hline
\end{tabular}
\end{center}
\caption{Summary of the BFC model parameters, with a brief description, and the best-fit parameters for the fiducial L1\_m9 and strongest feedback $f_{\rm gas}-8\sigma$ FLAMINGO simulations (Sec.~\ref{subsec: bfc fitting}) at redshift $z=0$.}
\label{tab: parameters}
\end{table}

\section{Hydrodynamical Simulations}\label{sec: hydrosims}

Our comparison is based on the FLAMINGO simulation suite, which we briefly describe below, with emphasis on the lightcone shell output that we use for our comparisons. \\
\\
\subsection{The FLAMINGO simulations}
\label{ssec: flamingo_sims}
The FLAMINGO suite (Ref.~\cite{Schaye_J_2023, Kugel_2023, Helly_2026}) consists of more than 20 cosmological hydrodynamical simulations run with the \textsc{SWIFT} code~\cite{Schaller_2024}, covering two main box sizes (1 Gpc and 2.8 Gpc) and three mass resolutions (gas particle masses of $1.34 \times 10^{8}$, $1.07 \times 10^{9}$, and $8.56 \times 10^{9}$ $\mathrm{M}_\odot$). The simulations explore multiple astrophysical feedback models and different cosmologies to robustly test baryonic effects on large-scale structure. In this work, we use four simulations from the FLAMINGO suite: the fiducial DMO run (L1$\_$m9 DMO), the fiducial (L1$\_$m9) run, the strongest-feedback ($f_\mathrm{gas}-8\sigma$) hydrodynamical run, and the fiducial run in the larger $2.8$ Gpc box (L2p8\_m9), which we use to estimate the sample variance of the tSZ power spectrum in a 1\,Gpc volume. We adopt the L1\_m9 resolution for the fiducial case because all of the FLAMINGO feedback-strength variations, including $f_\mathrm{gas}-8\sigma$, were run only at this resolution. The halos were identified using HBT-HERONS~\citep{Forouhar_2025}, which builds on friends-of-friends (FoF) groups~\citep{Davis_1985} and tracks their bound substructures through time. The resulting catalogues were analyzed with \textsc{SOAP}~\citep{McGibbon_2025}.\\
The hydrodynamical simulations feature $N=1800^3$ dark matter and baryonic particles and $1000^3$ neutrino particles, with $m_\mathrm{gas}=1.07\times 10^9\,\mathrm{M}_\odot$, $m_\mathrm{CDM}=5.65\times 10^9\,\mathrm{M}_\odot$, a comoving gravitational softening length of $\varepsilon_{\mathrm{com}}=22.3\,\mathrm{ckpc}$, and a physical softening length of $\varepsilon_{\mathrm{max}}=5.7\,\mathrm{kpc}$ with a comoving box of side length $L=1\,\rm Gpc$. The DMO simulation uses $N=1800^3$ dark matter particles with mass $m_\mathrm{CDM}=6.72\times 10^9\,\mathrm{M}_\odot$, while all other parameters remain the same. L2p8\_m9 has the same resolution but $N=5040^3$ dark matter and baryon particles and $2800^3$ neutrino particles in a box of side length $L=2.8\,\rm Gpc$. All simulations are run at the D3A fiducial cosmology (Table~\ref{tab: D3A_cosmo}), which is the same cosmology adopted for our \emph{CosmoGridV1-like} dark-matter-only simulation (Sec.~\ref{subsec: cosmogird-simulation}).\\
All simulations incorporate advanced subgrid models for baryonic physics, including stellar and AGN feedback. The L1\_m9 model is calibrated to match observed gas mass fractions in low-redshift clusters and the galaxy stellar mass function at $z=0.0$ (see Ref.~\cite{Kugel_2023}), while the simulation with the strongest feedback is calibrated to the same low-redshift galaxy stellar mass function but with gas fractions lower by $8\sigma$ as compared to the fiducial run.\\

For our comparisons, we primarily use the publicly available FLAMINGO lightcone shell maps rather than the full particle snapshots, since these directly match the shell-based output of our pipeline (Sec.~\ref{sec: maps}). FLAMINGO provides these shells as \textsc{HEALPix} maps at two resolutions, $N_{\rm side}=4096$ and $N_{\rm side}=4\times4096=16384$; throughout this work we use the $N_{\rm side}=4096$ maps for consistency with the resolution of our own simulated maps.

\subsection{Calibrating BFC to FLAMINGO}\label{subsec: bfc fitting} 
To test the model's self-consistency, we fit the 3D gas mass-fraction and thermal pressure profiles and then use them to predict the Compton-$y$ profiles and the tSZ power spectrum. Since the non-thermal pressure and the hot gas density profile affect the thermal pressure similarly, we measure the non-thermal pressure through the velocity dispersion (Ref.~\citep{Nelson_2014}) from the hydrodynamical simulations directly instead of letting it free. Because a simultaneous fit across mass bins with the parametric form of Ref.~\citep {Shaw_L_2010} is not possible, we fix the parameters to the most relevant mass bin (highest mass with enough halos across multiple redshifts, $10^{14.0}$--$10^{14.5}\mathrm{M}_\odot$). After fixing the non-thermal pressure parameters, we fit the gas mass-fraction profiles together with the 3D thermal pressure profiles of each feedback model with the BFC model\footnote{In the profile fits of S25 only the mass-fraction profiles were used, and the pressure was predicted from these fits} to get the best-fit parameters for the BFC model. We find that the resulting parameter constraints are considerably tighter than those of S25, owing to the additional constraining power of the pressure profiles around and beyond $r_{200}$, while remaining fully consistent with the S25 posteriors. Likewise, the corresponding matter power spectrum suppression lies within the uncertainty band of S25. This step is necessary as the tSZ power spectrum is very sensitive to all scales of the electron pressure profile. In contrast, the mass-fraction profiles are mainly sensitive to the central part, and the error bars increase strongly at larger radii (see Fig. 6 in S25), while the error bars for the pressure profile remain tighter at these larger radii (see Fig. 12 in S25). All parameters are listed in Table ~\ref{tab: parameters}. For the stellar parameters, which have very little effect on the pressure profile, we adopt the best-fit parameters from S25.

\section{Baryonified $N$-body simulations}\label{sec: bfc sims}
Validating our pipeline and testing it against the FLAMINGO simulations requires a DMO $N$-body simulation with the same underlying cosmology, which serves as the input for baryonification and for constructing Compton-$y$ maps. We use a dedicated \textsc{PKDGRAV3} run, built with the CosmoGridV1 pipeline using the same cosmology as the fiducial FLAMINGO simulations (D3A; see Tab.~\ref{tab: D3A_cosmo}), and HBT-HERONS~\citep{Forouhar_2025} to identify halos in our simulation. Since we use this simulation also for internal consistency tests of the pipeline, comparing snapshot- and shell-based BFC (Sec.~\ref{sec: pipeline validation}), we need it both as three-dimensional snapshots and as projected lightcone shells. 

\subsection{CosmoGridV1-like simulation}\label{subsec: cosmogird-simulation}
The CosmoGridV1 suite~\cite{Kacprzak_T_2023} is a large set of DMO $N$-body simulations (with $L_{\rm box}=900\,h^{-1}\mathrm{Mpc}$ and $N_\mathrm{part} = 832^3$) designed for map-level cosmological inference. It combines a grid of $\sim 2500$ cosmologies sampled across the $w$CDM parameter space ($\Omega_\mathrm{m}$, $\sigma_8$, $H_0$,  $w_0$, $n_s$, $\Omega_\mathrm{b}$ sampled on a Sobol sequence) with multiple random phase realizations at fixed cosmology, each evolved with \textsc{PKDGRAV3}~\cite{Potter_D_2017}, which also identifies halos on the fly with a FoF algorithm~\cite{Davis_1985}, and post-processed into past-lightcone shells suitable for producing full-sky or partial-sky observable maps (e.g.\ weak-lensing convergence). This combination of broad cosmology coverage and realization statistics, with a total of $\sim 20,000$ simulations, makes CosmoGridV1 well suited for SBI, as shown for WL and galaxy clustering in Ref.~\citep{Fluri_J_2022, Thomsen_2025}. In future work, we plan to use the full suite to generate large numbers of mock tSZ maps spanning the CosmoGridV1 cosmology grid.

Before that production stage, however, we first need to validate our baryonification and map-making pipeline against a hydrodynamical simulation. None of the CosmoGridV1 grid points coincide with the FLAMINGO fiducial cosmology, so for this comparison we instead generate a single dedicated DMO simulation, run with the identical CosmoGridV1 pipeline and numerical setup, but at the FLAMINGO fiducial cosmology. We refer to this as our \emph{CosmoGridV1-like} simulation; it uses the same machinery we will later apply across the full CosmoGridV1 grid, so the validation presented here also serves as a check of that pipeline ahead of the production run. Note that, unlike FLAMINGO, whose initial conditions fix the amplitudes of the largest-scale modes to the mean of the power spectrum~\cite{Schaye_J_2023}, our \emph{CosmoGridV1-like} simulation uses standard Gaussian random initial conditions, so its large-scale modes are subject to the full cosmic variance.\\

We run the simulation at the D3A cosmology (see Table~\ref{tab: D3A_cosmo}), matching the cosmological parameters of the FLAMINGO hydrodynamical simulations described in Sec.~\ref{sec: hydrosims}. The simulation follows $N_\mathrm{p} = 832^3 \approx 5.76\times10^8$ particles in a periodic box of comoving side length $L_{\rm box}=900\,h^{-1}\mathrm{Mpc}$. As in CosmoGridV1, the massive neutrinos (a single species with $m_\nu=0.06$\,eV, $\Omega_\nu=0.0014$) are not followed as particles but are included as a linear species in \textsc{PKDGRAV3}, following Ref.~\citep{Tram_2019}. This differs from FLAMINGO, which evolves the neutrinos as particles (Sec.~\ref{ssec: flamingo_sims}). The particle mass is
\begin{equation}
    m_\mathrm{p} = \rho_{\rm crit,0}\,\Omega_{\rm cb} \frac{L_{\rm box}^3}{N_p} \approx 1.07\times10^{11}\,h^{-1}M_\odot\,,
\end{equation}
where $\rho_{\rm crit,0}=2.775\times10^{11}\,h^2 M_\odot\,\mathrm{Mpc}^{-3}$. To identify halos in our \emph{CosmoGridV1-like} simulation, we used HBT-HERONS with a minimal particle Number of $N_{\rm min}=100$, which leads to a minimal halo mass of $M_{\rm h,min}\approx10^{13}\, \mathrm{M_\odot}/h$.

\begin{table}
\centering
\begin{tabular}{ccccccccc}
\hline
Name & $h$ & $\Omega_\mathrm{m}$ & $\Omega_\mathrm{b}$ & $\sum m_\nu c^2$ & $n_{\rm s}$ & $A_{\rm s}$ & $\sigma_8$ & $S_8$ \\
\hline
D3A & 0.681 & 0.306 & 0.0486 & 0.06 eV & 0.967 & $2.099\times10^{-9}$ & 0.807 & 0.815 \\
\hline
\end{tabular}
\caption{Cosmological parameters of the D3A cosmology used for the \emph{CosmoGridV1-like} validation simulation, matched to the fiducial FLAMINGO cosmology.}
\label{tab: D3A_cosmo}
\end{table}

\subsection{Lightcone construction}\label{subsec: lightcone-construction}
We build the past lightcone from our snapshot sequence by using periodic copies of the simulation box. We place the observer at the origin of the box at (0,0,0), redshift $z=0$, and place identical copies of the box side by side along the three-dimensional grid. Each snapshot provides a spherical shell of particles between an inner and outer comoving distance, $[\chi_{\rm in}, \chi_{\rm out}]$, corresponding to the redshift interval. Because distant shells grow larger than a single box, multiple replicated boxes are needed to fill a single shell. To do this, we process each shell in two steps. First, we select candidate boxes by checking the observer's distance to their eight corners. If a box lies entirely inside the inner radius or entirely outside the outer radius, we reject it. We keep any other box that intersects the shell or is completely enclosed by it. Second, we select the particles within the kept boxes that fall into the shell by calculating their exact distance to the observer. Combining the selected particles from all retained boxes fills the spherical shell. Replication starts at $\chi=L_{\rm box}$, i.e.\ at $z\approx0.33$ for the \emph{CosmoGridV1-like} box ($L_{\rm box}=900\,h^{-1}\mathrm{Mpc}$) and $z\approx0.24$ for FLAMINGO L1 ($L_{\rm box}=681\,h^{-1}\mathrm{Mpc}$). Up to $z=2$ ($\chi\approx3.6\,h^{-1}\mathrm{Gpc}$), the box is replicated about four and five times along the line of sight, respectively. We then project these particles onto a two-dimensional HEALPix map along the line of sight. This pipeline generates all lightcone products, i.e., the particle and halo lightcones. This lightcone generation method differs from the CosmoGridV1 pipeline but matches the map products used in the FLAMINGO simulation suite.

\subsection{Snapshot-BFC}\label{subsec: snapshot-bfc}
In the standard snapshot-based approach (Snapshot-BFC), the BFC model (Sec.~\ref{subsec: baryonification}) is applied directly in three dimensions. For each snapshot of our \emph{CosmoGridV1-like} simulation, the particles around each halo are displaced according to the analytic profiles described in Sec.~\ref{subsec: baryonification}, producing new snapshots containing dark-matter, gas, and star particles. We then construct the past lightcone from these baryonified snapshots using the lightcone-construction pipeline (Sec.~\ref{subsec: lightcone-construction}). 

\subsection{Shell-BFC}\label{subsec: shell-bfc}
For a single simulation, we can apply snapshot-BFC and construct a past lightcone from it. But for a large simulation suite, such as CosmoGridV1, storing all redshift snapshots is not feasible; therefore, we need a method that can be applied directly to shells, which are usually stored. The shell-BFC method uses DMO maps and the halo lightcone directly as input. Therefore, we first build the particles' and halos' past lightcone before applying the shell-BFC method at the 2D level. Conceptually, shell-BFC applies the same displacement as the snapshot-based method, but in projection. It maps gravity-only halo profiles to their baryonified states through displacements along the shell surface. The approach was introduced for application on the CosmoGridV1 suite in~\citep{Fluri_J_2022} and extended in the \textsc{BaryonForge} package by Ref.~\citep{Anbajagane_D_2024}. We build on these approaches but implement the new component-wise S25 model that displaces baryons and DM individually. With this, we can generate correlated shells of DM, gas, and stars from a DMO $N$-body simulation. Here we summarize the method, while a detailed description and validation of Shell-BFC will be presented in a forthcoming paper (Schneider, Bucko, Kova\v{c} et al. in prep.). To validate it for the purpose of this work, we will apply both shell-BFC and snapshot-BFC to the same DMO simulation and compare the resulting $C_\ell^{yy}$ (Sec.~\ref{sec: pipeline validation}). 

In shell-BFC, each halo is first moved to the center of the shell, and the mass profile is integrated along the line of sight over a depth set by the shell thickness $L$,

\begin{equation}\label{eq: projected-mass}
    M^{\rm p}_{A}(R) = 2\int_0^{R}\!\!2\pi R'\,{\rm d}R'
    \int_0^{L/2}\!\!{\rm d}s\;\rho_{ A}\!\left(\sqrt{s^2+R'^2}\right)\,,
    \qquad { A}=\{{\rm dm},\,{\rm bar}\}\,.
\end{equation}

Unlike methods that use a fixed multiple of the halo size, as in Ref.~\citep{Fluri_J_2022}, we tie the projection depth to the shell to prevent double counting between adjacent slices. 

Since a displacement using pixels is not possible on fixed HEALPix grids where pixels are larger than the scales over which feedback redistributes matter, we represent every non-empty pixel with tracer particles. To restore the necessary sub-pixel resolution, we apply adaptive sampling according to how close the pixels are to the centers of haloes, with the pixels that contain a halo center being divided into 16 tracers, the neighboring pixels into 4, and all the others into 1. The mass is distributed among these tracers in accordance with the projected NFW profile of the nearby halo so that the discrete field matches the continuous profile before the displacement takes place. This method directs the majority of the computational work to the pixels closest to the halo center that are modified the most while keeping the larger, low-density regions at their native resolution.

The use of a thin, curved shell leads to edge effects which are missing in conventional 3D methods. Since the halo profiles are projected assuming the halo to be located on the mid-plane of the shell, a halo whose actual position is close to the edge of the shell has only a portion of its profile within the shell. In order to make up for this, the displacement function is multiplied by a correction factor, $I_{\rm A}$.

\begin{equation}\label{eq: impact-factor}
    I_{A} = \frac{\int {\rm d}r\,r^2\,\rho_{A}(r)\,f_{\rm shell}(r)}
                     {\int {\rm d}r\,r^2\,\rho_{A}(r)}\,,
\end{equation}

\noindent where $\rho_{\rm A}$ refers to the one-halo profile for dark matter or baryons, and $f_{\rm shell}(r)$ is the fraction of a sphere of radius $r$ (with the center at the halo's true comoving position) that falls within the range of the shell. As a result, $I_{\rm A}$ tends towards 1 when the haloes are far from boundaries and decreases to about 1/2 in the case of haloes which are situated directly on an edge. Since each halo is only treated within its own shell, any part of an edge halo that ends up in a neighboring shell remains in place, an effect that affects only a small percentage of the haloes located within roughly a virial radius of an edge.

We then collect the displaced tracers onto separate output maps. To preserve the sub-pixel resolution of the displacements, which a nearest-pixel assignment would discard, we distribute each tracer’s mass over the four neighbouring pixels using bilinear weights.

\section{Constructing Compton-$y$ maps}\label{sec: maps}
In the present section we explain how the theoretical BFC pressure profiles from Sec.~\ref{sec: theory modeling tsz} are combined with the output of the simulations in Sec.~\ref{sec: bfc sims} in order to produce full-sky Compton-$y$ maps. We implement two map-generating methods of increasing complexity: \emph{painting}, in which the spherically symmetric theoretical profiles are placed directly at the positions of the halos using only the halo catalog, and the second is referred to as \emph{gas-weighted}, in which the same theoretical profiles are locally rescaled using the gas distribution of the baryonified shells. We will now describe each of these in turn.

\subsection{Painted Compton-$y$ maps}\label{subsec: painting}
In the painting method, which has also been used to construct tSZ maps in~\citep{Osato_K_2023, Anbajagane_D_2024}, we assign to each halo in the lightcone halo catalog (Sec.~\ref{subsec: cosmogird-simulation}) a spherically symmetric electron pressure profile $\mathscr{P}_{\rm e}(r)$, which has been calculated using the BFC model (Sec.~\ref{subsubsec: pressure-profile}) at the halo's given $M_{200}$, concentration $c_{200}$, and redshift $z$.
Each profile is then projected along the line of sight,

\begin{equation}
    \mathscr{P}_\mathrm{e}^{\rm 2D}(R) = 2\int_0^{L/2} \mathscr{P}_\mathrm{e}\!\left(\sqrt{R^2+x^2}\right)\,\mathrm{d}x\,.
\end{equation}
$R$ is the projected comoving distance from the halo center and $L$ is the comoving shell thickness of the shell the halo sits in. This effectively moves the halo to the center of the shell and ensures that the projection does not double-count contributions from adjacent shells. The projected profile is converted to a dimensionless Compton-$y$ profile through

\begin{equation}
    y(\theta) = \frac{\sigma_\mathrm{T}}{m_\mathrm{e} c^2}\,\mathscr{P}_\mathrm{e}^{\rm 2D}(R = \chi\,\theta)\,,
\end{equation}

We paint $y(\theta)$ onto every \textsc{HEALPix} pixel within an angular radius $\theta_{\rm max}=f_{200}\,r_{200}/\chi$ of the halo, with $\chi$ the comoving distance and $f_{200}=5$, beyond which we find the tSZ power spectrum to be converged. A pixel at angular separation $\theta_p$ from the halo center samples the profile at $y(\theta_p)$. To obtain the pixel-averaged signal, we subdivide each pixel into $16$ sub-pixels, evaluate $y$ at each sub-pixel center, and assign the mean value to the parent pixel. This is especially necessary for low-mass and/or high-redshift halos, where a halo only subtends a few pixels.

After painting all halos into their respective shells, we co-add the shells to form the final Compton-$y$ map.  

\subsection{Gas-weighted Compton-$y$ maps}\label{subsec: gas-weighted}

The painting method described in the previous section assigns each halo a spherically symmetric Compton-$y$ profile based on halo mass and redshift. Real halos are not spherical: their gas is stretched along the local filament, disturbed by recent mergers, and clumped around substructure. The baryonified gas shell (Sec.~\ref{sec: bfc sims}), produced by shell-BFC, carries this anisotropic gas distribution. With the gas-weighted method, we include the effect of asphericity of gas shells on the Compton-$y$ profiles.

In angular annuli, we weight the painted profile in each pixel  $y^{(p)}$ by the gas profile at that pixel  $g^{(p)}$, i.e.

\begin{equation}\label{eq: gas-weight}
    y^{(p)} \;\longrightarrow\; y^{(p)}\,\frac{g^{(p)}}{\bar g_b}\,,
\end{equation}
where $\bar g_b$ is the mean gas profile across all pixels in a given annulus. Since the weight $g^{(p)}/\bar g_b$ has a mean of unity in each annulus, Equation~\eqref{eq: gas-weight} redistributes the total Compton-$y$ budget fixed by the painted maps among the pixels according to the simulated gas.  The rescaled map thus preserves the mean BFC radial Compton-$y$ profile while reproducing the halo's asphericity and the effects of substructure within its area of coverage. As in the case of the painting method, the map is constructed shell by shell, and the shells are co-added to produce the final Compton-$y$ map.

\section{Pipeline validation}\label{sec: pipeline validation}

Before comparing our Compton-$y$ maps to FLAMINGO, we test the two map-level steps of our pipeline on the \emph{CosmoGridV1-like} simulation: the shell-BFC gas shells that enter the gas-weighted method, and the gas-weighted method itself against the simpler painting. All maps in this section use the HBT-HERONS halo lightcone and the L1\_m9 best-fit parameters of Table~\ref{tab: parameters}. They are made at $N_{\rm side}=2048$, the resolution of CosmoGridV1.

\begin{figure}
    \centering
    \includegraphics[width=0.5\linewidth]{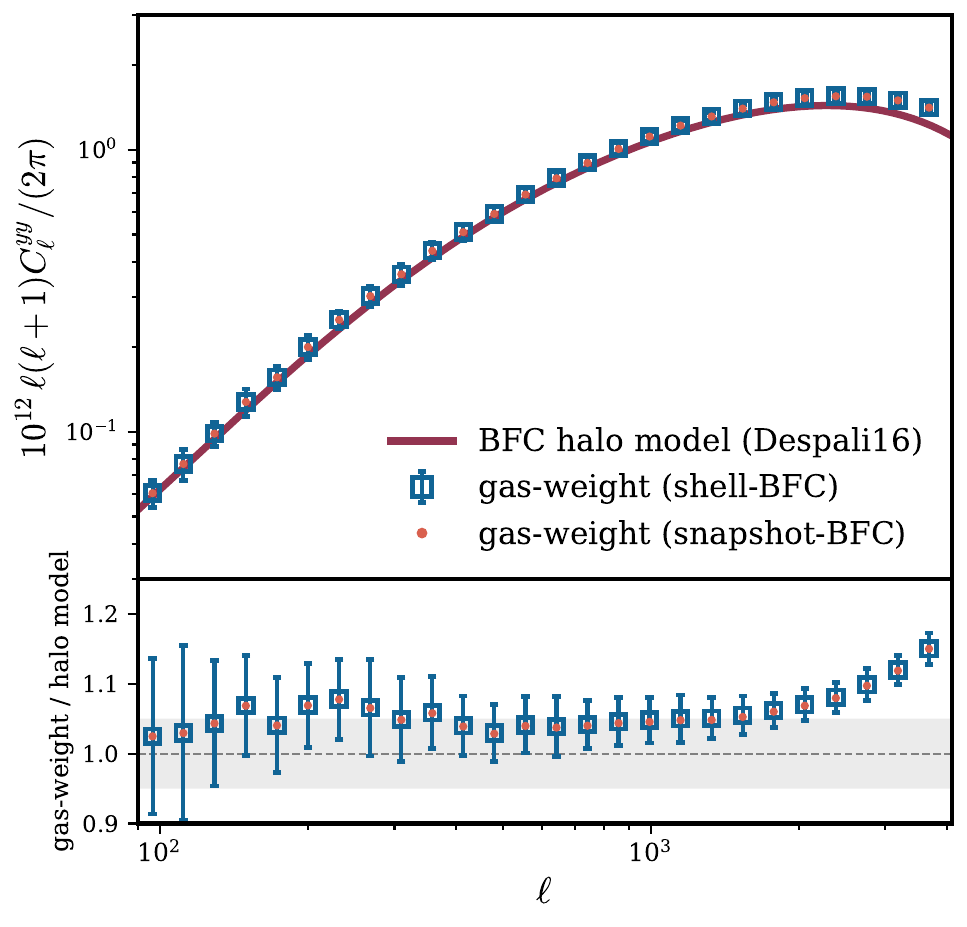}
    \caption{Compton-$y$ power spectrum of two gas-weighted maps from the \emph{CosmoGridV1-like} simulation ($N_{\rm side}=2048$, L1\_m9 best-fit parameters), using gas shells from shell-BFC (blue squares) and snapshot-BFC (orange dots). The solid line shows the BFC halo-model prediction using the Despali16~\citep{Despali_2016} HMF. The lower panel shows the ratio of the maps to the halo model, with the grey band marking $\pm5\%$. Error bars show the standard deviation of the individual $C_\ell$ within each $\ell$ bin. The two gas-weighted spectra agree to better than $0.2\%$ over $90\leq\ell\leq4096$.}
    \label{fig: snapshot vs shell gas-weight}
\end{figure}

The shell-BFC in this work is validated on the Compton-$y$ power spectrum. We generate two gas-weighted maps (Sec.~\ref{subsec: gas-weighted}) that differ only in the gas shells used for the weighting. One uses the shell-BFC shells (Sec.~\ref{subsec: shell-bfc}), where we construct the past lightcone from the DMO snapshots and then baryonify the shells; the other uses snapshot-BFC, where we baryonify the DMO snapshots first and then build the past lightcone and project them onto shells (Secs.~\ref{subsec: snapshot-bfc} and~\ref{subsec: lightcone-construction}). This comparison tests if the two BFC methods distribute the gas around halos in the same way, which is exactly the information used in the gas-weighted method. We show the results of the first test in Fig.~\ref{fig: snapshot vs shell gas-weight} and find that the two spectra agree to better than 0.2\% over $90\leq\ell\leq4096$, so for constructing Compton-$y$ maps, shell-BFC gas shells (blue squares) are equivalent to those from snapshot-BFC (orange dots). A more detailed validation of shell-BFC, including the suppression of the matter power spectrum, halo density profiles, and weak-lensing statistics, will be presented in a companion paper (Schneider, Bucko, Kova\v{c} et al., in prep.).

\begin{figure}
    \centering
    \begin{minipage}[c]{0.6\linewidth}
        \centering
        \includegraphics[width=\linewidth]{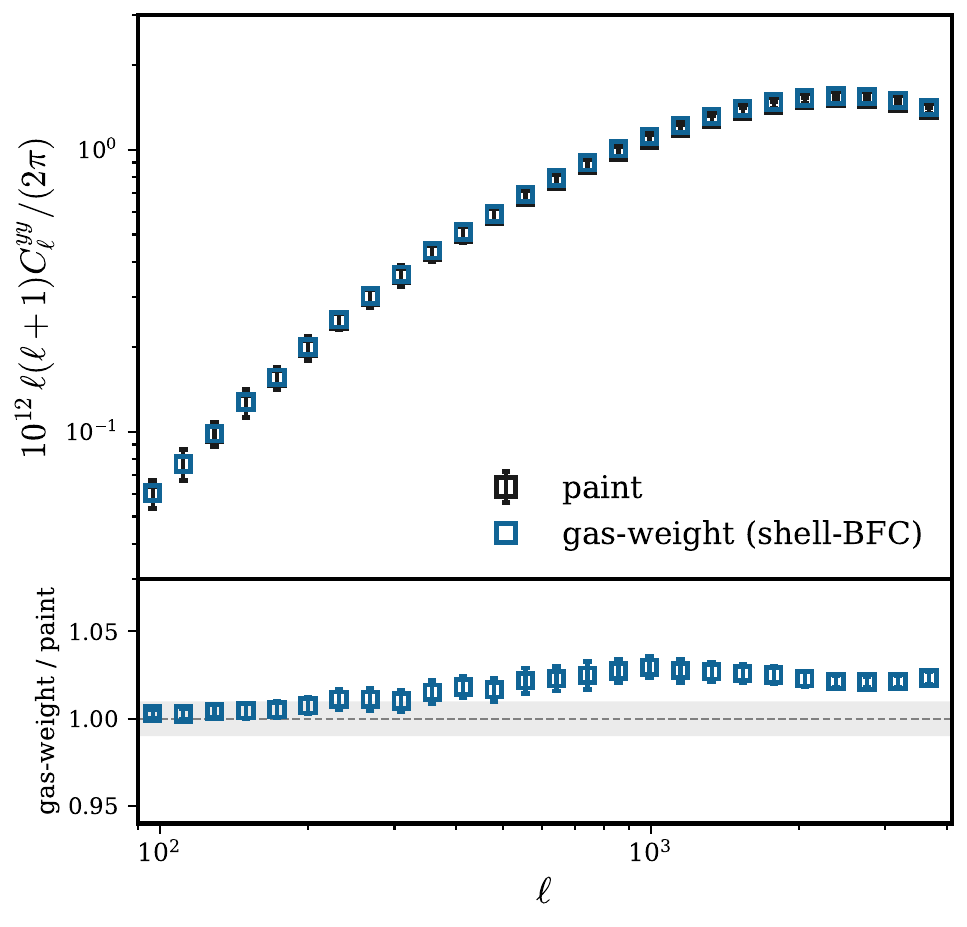}
    \end{minipage}\hfill
    \begin{minipage}[c]{0.3\linewidth}
        \centering
        \includegraphics[width=\linewidth]{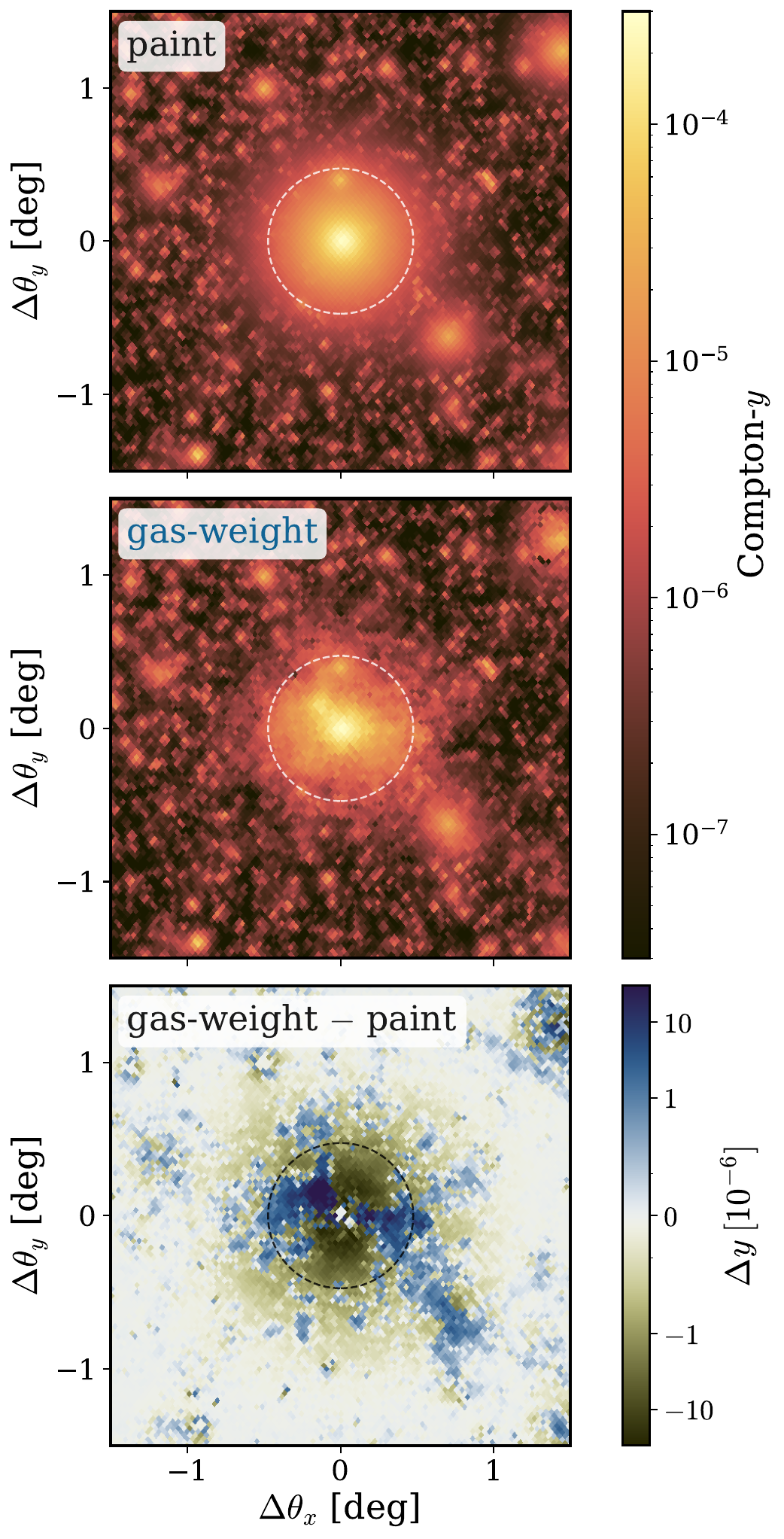}
    \end{minipage}

    \caption{Comparison of the two map-making methods using the HBT-HERONS halo lightcone of our \emph{CosmoGridV1-like} simulation with the fiducial L1\_m9 parameters at $N_{\rm side}=2048$. \textit{Left}: gas-weighted spectra using the shell-BFC (blue squares) and painting (black) with their ratio (grey band: $\pm1\%$). \textit{Right}: $3^\circ\times3^\circ$ gnomonic cut-outs centred on an HBT-HERONS cluster: the painted map (top), the gas-weighted map using the shell-BFC gas shells (middle) and their difference (bottom). The dashed circle marks $r_{200}$. Gas weighting moves signal along the anisotropic gas distribution, here towards a filament to the lower right.}
    \label{fig: paint vs gas-weight}
\end{figure}

We next compare the gas-weighted method with the shell-BFC gas shells against the painting method (Sec.~\ref{subsec: painting}) to generate Compton-$y$ maps. The top row of Fig.~\ref{fig: paint vs gas-weight} shows both maps in a $3^\circ\times3^\circ$ field around a massive cluster ($M_{200c}=1.8\times10^{15}\,h^{-1}{\rm M_\odot}$, $z=0.086$). The painted cluster is spherically symmetric by construction, while the gas-weighted map follows the anisotropic gas of the baryonified shell. As the difference map shows, signal moves out of regions where the gas density is below its annulus average and into a filament to the lower right, as well as into substructure within $r_{200}$.

The bottom of Fig.~\ref{fig: paint vs gas-weight} compares the painting (black) and gas-weighted (blue) methods at the power spectrum level. The two approaches agree within $2\%$, with differences being smallest at large scales and flattening around $\ell\sim 1000$ at the $2\%$ level. Furthermore, the gas-weighted method yields a systematically higher tSZ power spectrum compared to the painting method. This systematic offset arises because the gas-weighted method scales the smooth pressure profile with the local density in each annulus. While this preserves the average pressure profile per halo, local density fluctuations add spatial variance, which can boost power across all resolved scales compared to purely smooth painted profiles.  

Neither approach can be seen as the ground truth, and the comparison above does not by itself establish which is more accurate. The painting method assumes smooth, spherically symmetric profiles and therefore neglects halo asphericity and substructure. The gas-weighted method captures these effects to some extent, since it modulates the pressure profile by the local gas density of the baryonified shells, and is therefore expected to provide a more realistic description of the small-scale gas distribution. For the purposes of this paper, the $\sim2\%$ difference between the two methods is well below the precision of current tSZ power spectrum measurements~\citep{Efstathiou_McCarthy_25}. We therefore adopt the painting method for the remainder of this work, as it is computationally cheaper, easily adapted to different resolutions, and does not require full baryonified lightcones.

\section{Comparison to Hydrodynamical Simulations}\label{sec: Comparison to hydro}
We will now test the ability of our model to reproduce the Compton-$y$ profiles and the tSZ power spectrum from two different FLAMINGO simulations, corresponding to the fiducial feedback and strong feedback regimes (L1\_m9 and f$_{\rm gas}-8\sigma$). First, in Sec.~\ref{subsec: comptony_profiles} we check the agreement of Compton-$y$ profiles across halo mass and redshift bins. Second, in Sec.~\ref{subsec: tsz pk results} we compare the tSZ power spectra and show that our model reproduces the FLAMINGO results for both feedback models.

\subsection{Compton-$y$ profiles}\label{subsec: comptony_profiles}

\begin{figure}
    \centering
    \includegraphics[width=\textwidth]{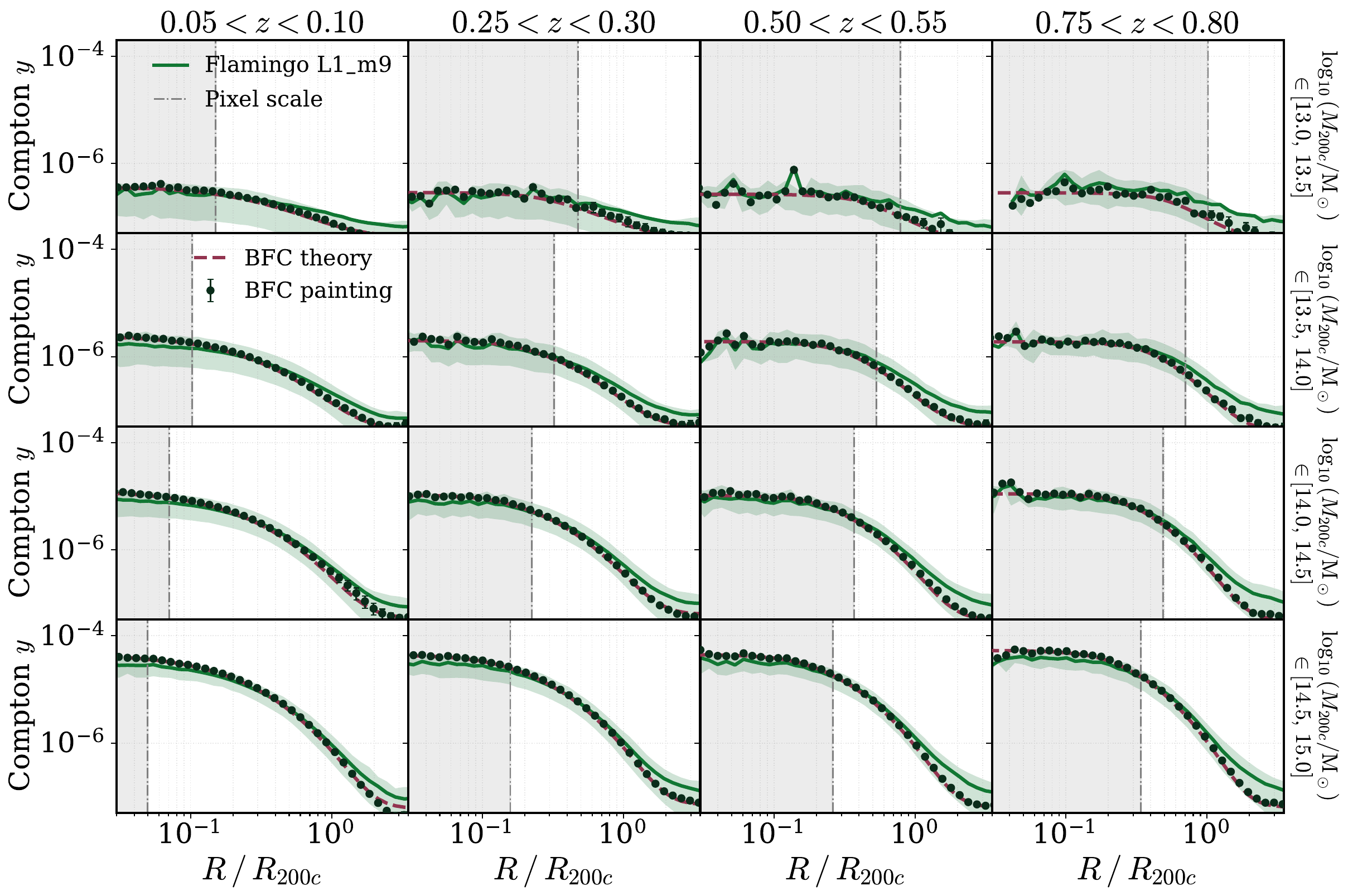}
    \caption{Stacked Compton-$y$ profiles as a function of projected radius $R/R_{200c}$ for the fiducial L1\_m9 simulation (green) and for our painted map of the \emph{CosmoGridV1-like} simulation (black points). Columns show four redshift slices between $z=0.05$ and $0.80$ and rows show four bins of $\log_{10}(M_{200c}/{\rm M_\odot})$ between 13 and 15. The green line is the mean profile of up to 1000 randomly selected FLAMINGO halos per mass bin and lightcone shell, and the shaded band is their 16th-84th percentile halo-to-halo scatter. The painted halos are weighted so that their mean mass matches the FLAMINGO sample in each bin (Sec.~\ref{subsec: comptony_profiles}); error bars show the standard error of the weighted mean. The red dashed line is the BFC prediction at the FLAMINGO mean mass, convolved with the pixel window of the map. All curves use the $z=0$ best-fit parameters of Table~\ref{tab: parameters}. The gray region marks radii smaller than one $N_{\rm side}=4096$ pixel.}

    \label{fig: m9 compton-y profiles}
\end{figure}

To validate our pressure modeling, we compare our painted Compton-$y$ maps to the FLAMINGO Compton-$y$ maps with the same pixel resolution of $N_{\rm side}=4096$. We use the best-fit parameters for each feedback model from the fitted 3D gas mass-fraction and thermal pressure profiles, as described in Sec.~\ref{subsec: bfc fitting} and shown in Table~\ref{tab: parameters}, to generate our Compton-$y$ maps using the painting method described in Sec.~\ref{subsec: painting}. We then use the resulting Compton-$y$ maps to measure halo Compton-$y$ profiles in different mass and redshift bins. In the FLAMINGO simulations, we randomly select $N=1000$ halos for each mass bin and redshift shell. If there are fewer than $N$ halos, we use all halos in that bin. We then stack the halos in each redshift and mass bin and calculate the mean Compton-$y$ profile and its 68$\%$ uncertainty. Since the Compton-$y$ scales strongly with mass, $\propto M^{5/3}$, it is essential that the measured halos in our generated maps have the same mean mass as the measured FLAMINGO sample. To achieve this, we assign a weight $w_i = M_i^\theta$ to each halo $i$. For every $(z, M)$ bin, we solve for $\theta$ via a 1D root-finding algorithm such that the weighted mean mass matches the FLAMINGO mean mass. We then apply these weights consistently across all subsequent stacking calculations, error estimates, and effective sample sizes. 

Fig.~\ref{fig: m9 compton-y profiles} shows the resulting comparison of Compton-$y$ profiles for the fiducial L1\_m9 feedback scenario, on a grid of four redshift bins (columns) in the range $z=[0.05,0.80]$ and four halo mass bins (rows) spanning $M_{\rm 200c}=[10^{13}, 10^{15}]\,\mathrm{M_\odot}$. The solid green line shows the mean Compton-$y$ profile measured in the L1\_m9 simulation, with the shaded region indicating the 68\% scatter across the sample. Black points show the corresponding measurement from our painted maps. The red dashed line shows the theoretical BFC prediction evaluated at the mean mass of the measured L1\_m9 halos and convolved with the pixel window function. The Compton-$y$ profiles of our generated maps lie well within the scatter of the measured L1\_m9 simulation but tend to underpredict the hydrodynamical profile around $r_{200}$ for all mass bins and redshifts. The number of high-mass halos decreases at higher redshift, and we use the best-fit parameters from $z=0$ for all redshifts, which drive the slightly larger disagreement in the high-mass, high-redshift bins. The gray shaded region marks the map pixel scale, below which resolution effects dominate, and the recovered profiles can no longer be trusted.

\begin{figure}
    \centering
    \includegraphics[width=\textwidth]{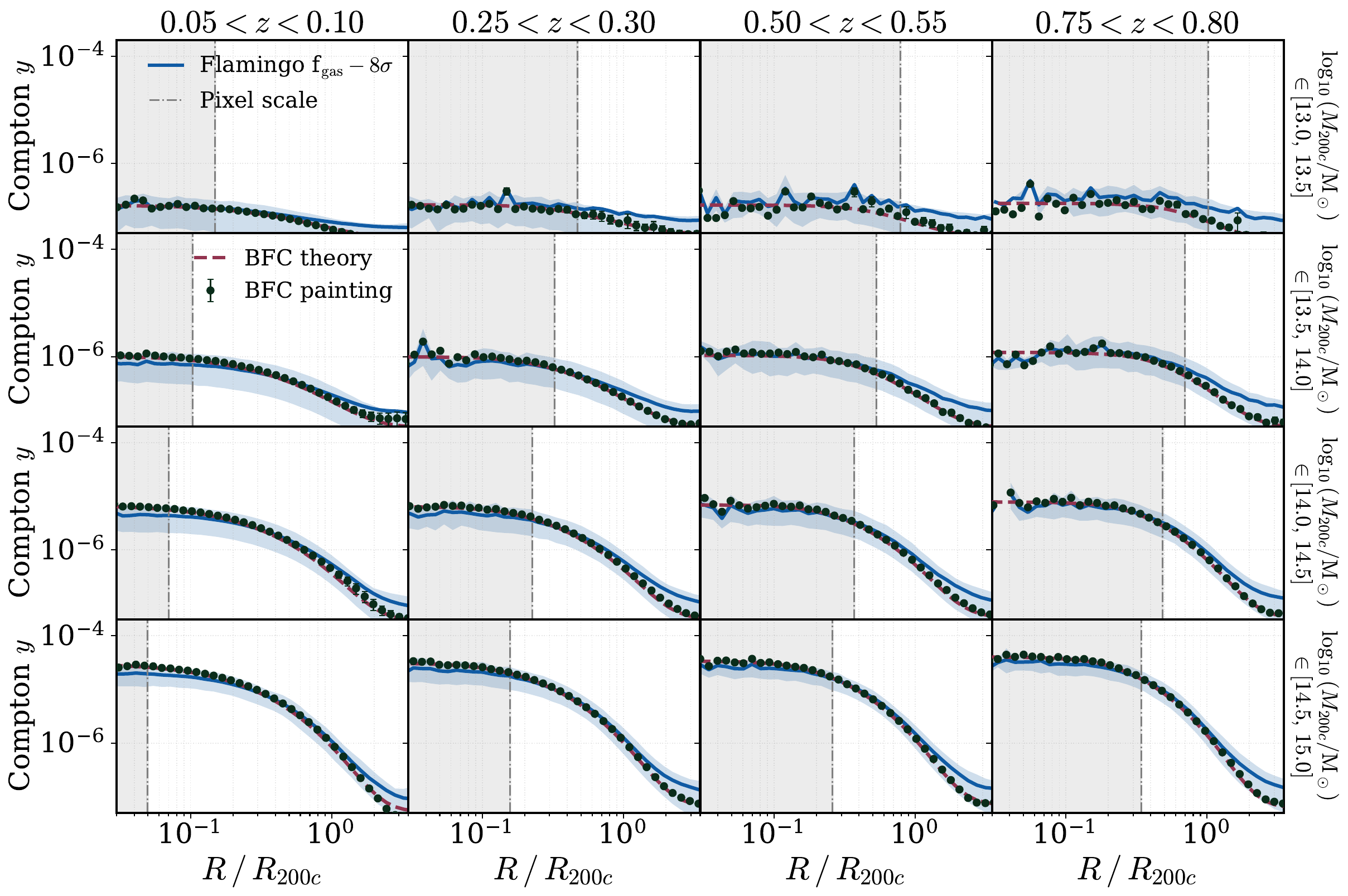}
    \caption{Same as Fig.~\ref{fig: m9 compton-y profiles}, but for the $f_{\rm gas}-8\sigma$ simulation (blue), using the $f_{\rm gas}-8\sigma$ best-fit parameters of Table~\ref{tab: parameters}.}

    \label{fig: fs8 compton-y profiles}
\end{figure}

Figure~\ref{fig: fs8 compton-y profiles} shows the same comparison for the $f_{\rm gas}-8\sigma$ simulation. The Compton-$y$ profiles in this stronger-feedback scenario are considerably more extended and have a lower central amplitude than their fiducial counterparts, consistent with AGN feedback pushing gas out to larger radii and suppressing the central pressure. As in the fiducial case, the painted measurements lie within the scatter of the measured $f_{\rm gas}-8\sigma$ Compton-$y$ profiles but are also low around $r_{200}$. This confirms that our pressure modeling approach captures both qualitatively different feedback regimes just as well. As in the L1\_m9 simulation, we find that agreement between our generated maps and the $f_{\rm gas}-8\sigma$ simulation is best in the high-mass, low-redshift regime. This is also the regime most relevant for the tSZ power spectrum (see App.~\ref{app: tsz pk mass and z convergence}).

\subsection{tSZ power spectrum}\label{subsec: tsz pk results}

Having validated the painted Compton-$y$ profiles against the L1\_m9 and $f_{\rm gas}-8\sigma$ measured Compton-$y$ profiles, we now turn to the angular power spectrum $C_\ell^{yy}$, which is one of the statistics employed for cosmological inference with observational data. As before, we test our pipeline for both the L1\_m9 and $f_{\rm gas}-8\sigma$ scenarios. All measured and theoretically calculated tSZ power spectra span redshifts from $z_{\rm min}=0.05$ to $z_{\rm max}=2.00$ and are calculated using $N_{\rm side}=4096$ maps. We chose the lower redshift cut because the closest shells inherit very few halos, which appear large on the sky due to the close distance and vary strongly between simulations.

\begin{figure}
    \centering
    \includegraphics[width=\textwidth]{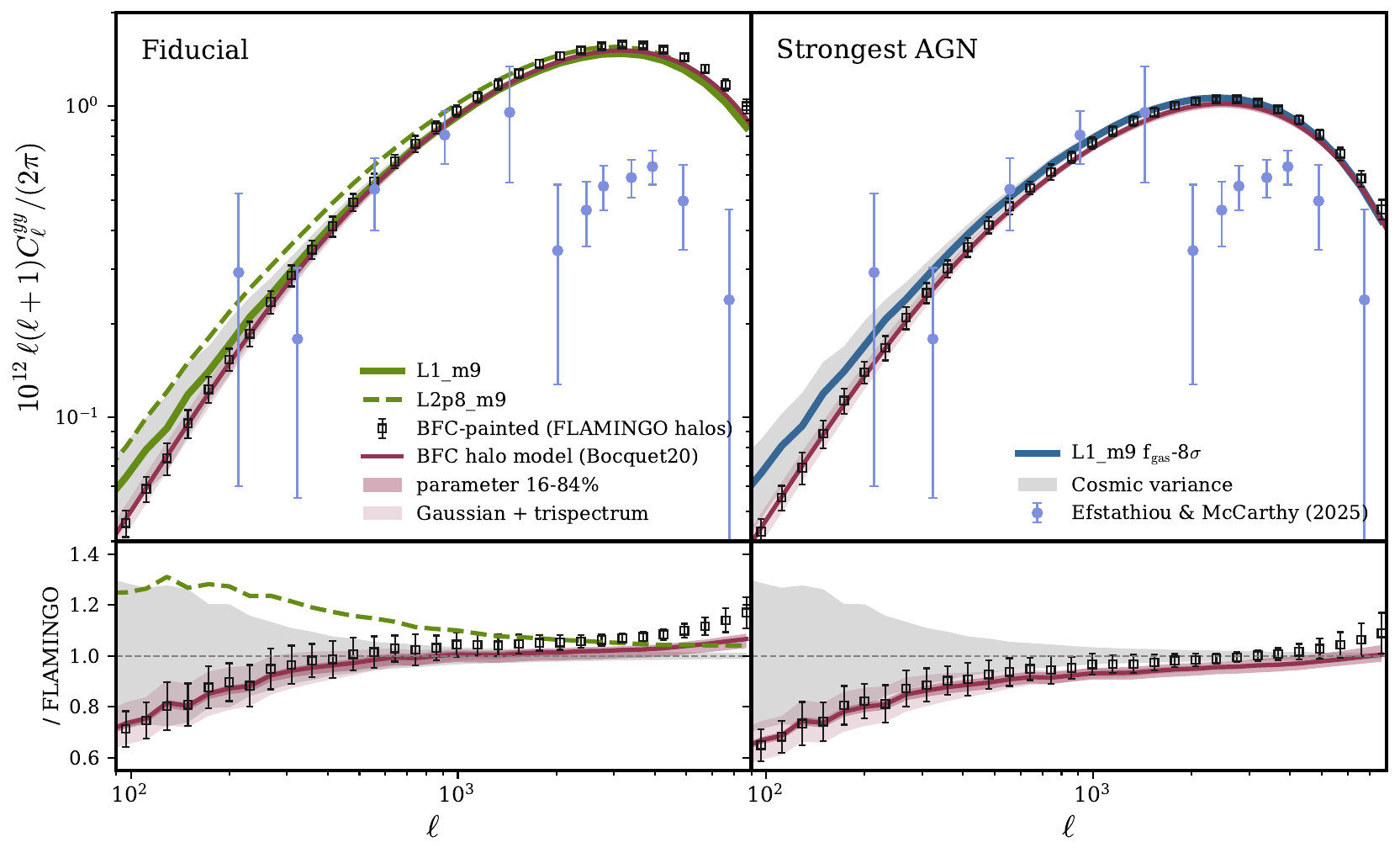}
    \caption{tSZ angular power spectrum $D_\ell^{yy}=\ell(\ell+1)C_\ell^{yy}/(2\pi)$ over $0.05<z<2$ from FLAMINGO and from BFC profiles painted onto the HBT-HERONS halos of the FLAMINGO L1\_m9 DMO simulation, for the L1\_m9 (left) and $f_{\rm gas}-8\sigma$ (right) models. \textit{Top}: FLAMINGO lightcone spectrum (green/blue thick line); painted map (open squares); BFC halo model with the Bocquet20 HMF~\citep{Bocquet_2020} (solid red line), with the 16-84\% uncertainty from BFC parameters (dark band) and the Gaussian plus one-halo trispectrum uncertainty of a full-sky map (light band, Eq.~\eqref{eq: sigma Cl}); and the measurements by Ref.~\citep{Efstathiou_McCarthy_25} (filled circles). The dashed green line in the left panels is the mean of the eight L2p8\_m9 lightcones, which use the same model and resolution as L1\_m9 in a box with 2.8 times the side length; its offset from L1\_m9 reflects sample variance of the most massive clusters in the 1\,Gpc volume (Sec.~\ref{subsec: tsz pk results}). The gray band shows the fractional scatter between these eight lightcones, applied to both models because no strong-feedback run exists in that volume; since all eight are drawn from the same box, it does not include the variance between independent boxes. \textit{Bottom}: ratio to FLAMINGO. All model curves use the painting parameters (Table~\ref{tab: parameters}) and include the $N_{\rm side}=4096$ pixel window.}

    \label{fig: tsz_pk hbt}
\end{figure}

Since $C_\ell^{yy}$ depends on both the pressure model and the halo population, we separate the two effects in three steps. We first paint the halos of FLAMINGO itself, which tests the pressure model at a fixed halo population. We then paint the halos of our \emph{CosmoGridV1-like} simulation. Finally, we remove the most massive halos from both lightcones to isolate the difference between the two halo populations for the largest scales.

\begin{figure}
    \centering
    \includegraphics[width=\textwidth]{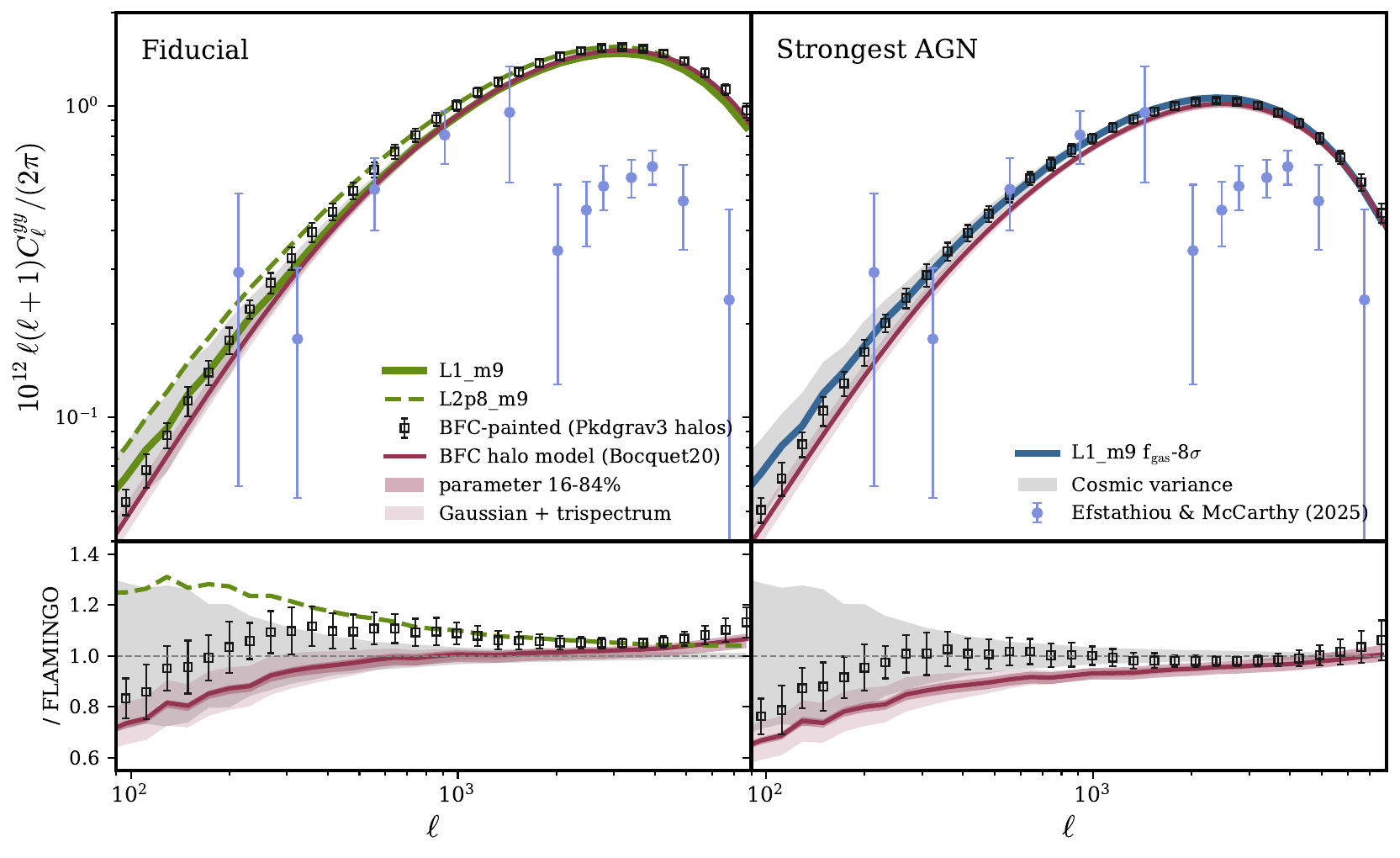}
    \caption{Same as Fig.~\ref{fig: tsz_pk hbt}, but painted onto the HBT-HERONS halos of our \emph{CosmoGridV1-like} simulation.}

    \label{fig: tsz_pk cosmogrid halos}
\end{figure}

We construct a halo lightcone from the HBT-HERONS halos of FLAMINGO L1\_m9 DMO using the procedure described in Sec.~\ref{subsec: lightcone-construction} and paint the BFC pressure profiles of each feedback model onto it. The painted maps thus share their halo population and their realization of the large-scale structure with the FLAMINGO maps, so their differences test the pressure model and the painting rather than the halo population. The left panel of Fig.~\ref{fig: tsz_pk hbt} shows the comparison to the L1\_m9 simulation and the right panel the comparison to $f_{\rm gas}-8\sigma$. The difference to the measured angular tSZ power spectrum falls below $10\%$ for $250\lesssim\ell\lesssim5000$ (and below $5\%$ for $300\lesssim\ell\lesssim3000$) for the L1\_m9 simulation and below $10\%$ for $360\lesssim\ell\lesssim7600$ (below $5\%$ for $500<\ell<5700$) for the $f_{\rm gas}-8\sigma$ simulation. Over these ranges, the differences are smaller than the uncertainties of the tSZ power spectrum reconstructions from \textit{Planck}, ACT, and SPT~\citep{Efstathiou_McCarthy_25}, shown as purple data points. This comparison tests how well BFC reproduces the simulations; it is not a comparison to the data. The measurements are shown for reference only: for $2500\lesssim\ell\lesssim5000$, FLAMINGO itself lies a factor of $\sim2.5$ (L1\_m9) and $\sim1.8$ ($f_{\rm gas}-8\sigma$) above them, and the painted maps inherit this offset because BFC is calibrated on the FLAMINGO profiles. We discuss the comparison with observations in App.~\ref{app: tsz data}. Since BFC is fitted only to the 3D gas mass-fraction and pressure profiles, and neither the Compton-$y$ profiles nor $C_\ell^{yy}$ enter the fit, the agreement with FLAMINGO is a genuine prediction. 

Fig.~\ref{fig: tsz_pk cosmogrid halos} shows the same comparison for the HBT-HERONS halos of our \emph{CosmoGridV1-like} simulation, run with the same cosmology as the FLAMINGO simulations (see Sec.~\ref{subsec: cosmogird-simulation}). The painted maps lie within $10\%$ of FLAMINGO for $150<\ell<7000$ (within $5\%$ for $1500<\ell<5000$) for L1\_m9 and for $200<\ell<7600$ (within $5\%$ for $250<\ell<6000$) for $f_{\rm gas}-8\sigma$. Since the pressure model is the same as in Fig.~\ref{fig: tsz_pk hbt}, any additional differences come from the halo population. We use the same halo finder as FLAMINGO to remove the finder as a source of these differences; a different halo finder can raise the power by up to $\sim10\%$ on all scales (App.~\ref{app: halo finders}).

\begin{figure}
    \centering
    \includegraphics[width=\textwidth]{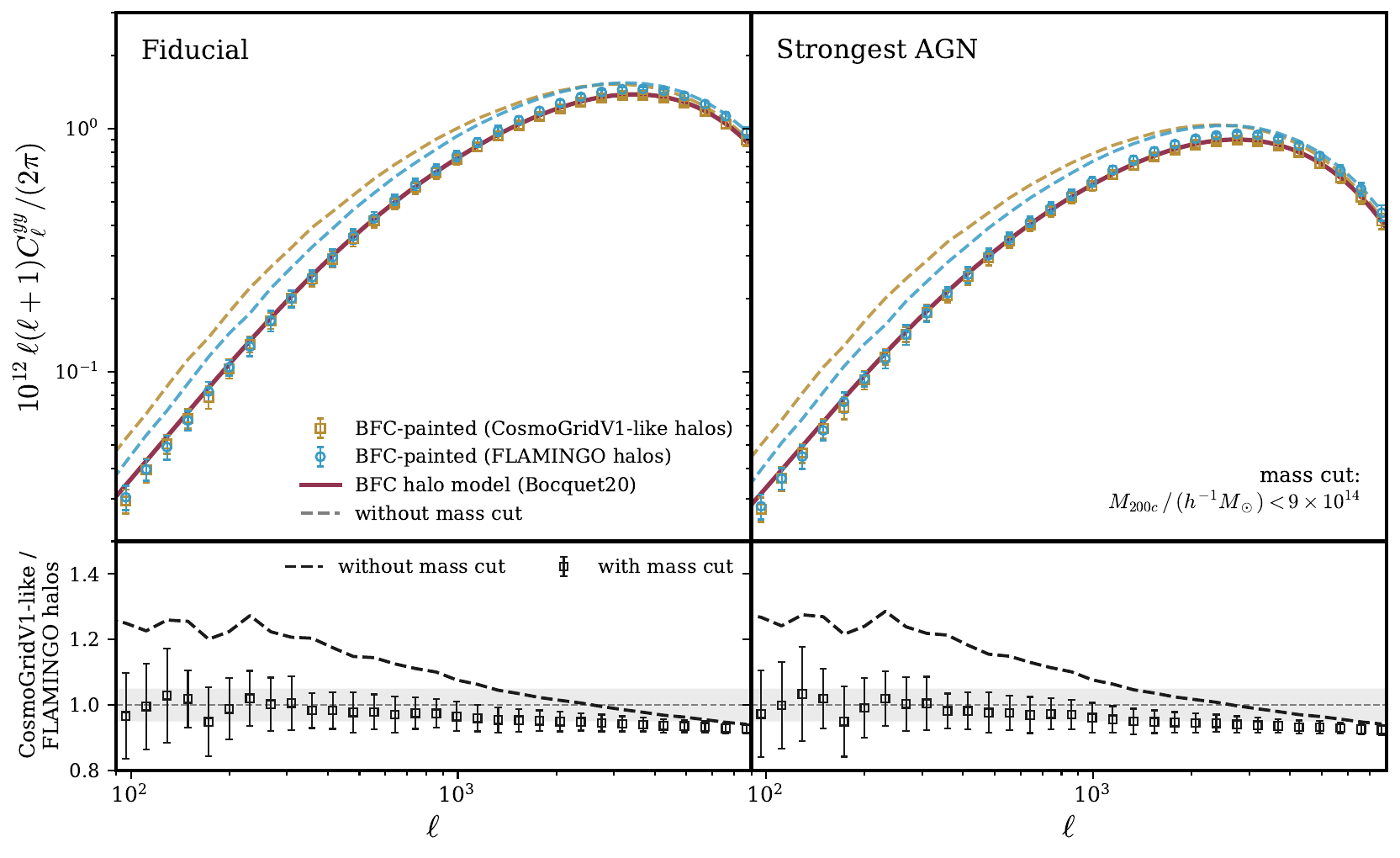}
    \caption{Effect of removing the most massive halos. Both lightcones are painted with the L1\_m9 (left) and $f_{\rm gas}-8\sigma$ (right) parameters after removing all halos with $M_{200c}\geq9\times10^{14}\,h^{-1}{\rm M_\odot}$ (320 halos in the \emph{CosmoGridV1-like} and 277 in the FLAMINGO lightcone). \textit{Top}: painted spectra with the cut for the \emph{CosmoGridV1-like} halos (yellow squares) and the FLAMINGO HBT-HERONS halos (blue circles), the same paintings without the cut (dashed), and the Bocquet20 halo model with the same upper mass limit (solid). \textit{Bottom}: ratio of the \emph{CosmoGridV1-like} to the FLAMINGO painting with (squares) and without (dashed) the cut; the gray band marks $\pm5\%$. With the cut, the two halo populations agree within a few per cent up to $\ell\approx2000$. The remaining difference at higher $\ell$ reflects the lower abundance of \emph{CosmoGridV1-like} halos at higher redshift (App.~\ref{app: hmf difference in simims}).}

    \label{fig: tsz_pk masscut}
\end{figure}

The two simulations differ in their $N$-body code, initial conditions, neutrino treatment, and box size, so their halo populations are not identical, especially at the rare high-mass end. Even with identical physics, a single 1\,Gpc lightcone is subject to large sample variance from these rare clusters. The mean of the eight L2p8\_m9 lightcones lies $\sim10\%$ above L1\_m9 at $\ell\approx1000$ (Fig.~\ref{fig: tsz_pk hbt}, see also Ref.~\citep{McCarthy_2023}). Because the tSZ signal scales steeply with halo mass ($C_\ell^{yy}\propto M^{10/3}$, Ref.~\citep{Komatsu_2002}), a handful of the most massive clusters can dominate the power on large and intermediate scales. Similar to Ref.~\citep{Rotti_Bolliet_Chluba_2021}, we therefore remove all halos with $M_{200c}\geq9\times10^{14}\,h^{-1}{\rm M_\odot}$ from both lightcones, i.e.\ 320 halos from the \emph{CosmoGridV1-like} and 277 from the FLAMINGO lightcone ($\sim10^{-5}$ of all painted halos). As the FLAMINGO maps contain these clusters, we compare the two painted lightcones with each other instead (Fig.~\ref{fig: tsz_pk masscut}). With the cut, their difference drops from 25\% at $\ell\sim100$ to a few per cent up to $\ell\approx2000$, showing that most of the remaining difference between the two simulations is driven by the most massive halos. At higher $\ell$, where these halos contribute little, the cut has no effect and the residual difference of up to 6\% reflects the abundance of lower-mass halos at higher redshift. The same sensitivity to the high-mass tail enters the halo-model prediction: the choice of HMF alone changes $C_\ell^{yy}$ by up to $\sim25\%$ (App.~\ref{app: hmf difference in simims}), comparable to the differences discussed here. This does not preclude precision cosmology with $C_\ell^{yy}$, but it requires the HMF to be calibrated on the same halo finder and mass definition as the pressure model, as done here with Bocquet20 and HBT-HERONS, and any residual HMF uncertainty to be marginalised over jointly with the gas parameters, with which it is degenerate. Since $C_\ell^{yy}\propto\sigma_8^{7}$, a $20\%$ shift corresponds to only $\sim2$--$3\%$ in $\sigma_8$, which sets the required calibration accuracy for future emulators.

For the largest scales $\ell\lesssim200-300$, we see for all cases (Fig.~\ref{fig: tsz_pk cosmogrid halos} and Fig.~\ref{fig: tsz_pk hbt}) that the halo model and also the painted results systematically underpredict the power compared to the results coming from the hydrodynamical simulations, while still being inside the cosmic-variance error band. We suspect that this difference comes from the halo model itself and our pressure modeling, especially on large scales. At $\ell\lesssim200$, the two-halo term starts to contribute non-negligibly to the tSZ power spectrum, and around $\ell\approx 100$ we enter the transition region between the one- and two-halo regimes. In the halo model, this regime is the hardest to model because of halo exclusion and non-linear bias~\citep{Mead_2021, Asgari_2023}. On the other hand, pressure modeling itself introduces several potential sources of discrepancy, most prominent on large scales. Specifically, the assumption of HSE likely breaks down toward halo outskirts where turbulent gas motions dominate, while our use of smooth, spherical profiles neglects halo-to-halo scatter and triaxiality. Furthermore, the hydrodynamical maps capture unmodeled pressure from diffuse gas in filaments and warm-hot intergalactic gas expelled beyond halo boundaries, which our painted maps miss by construction. In summary, the large-scale deficit lies within cosmic variance and most likely reflects the limits of a spherical, halo-bound pressure model in the one-to-two-halo transition rather than the BFC calibration or the painting itself; a dedicated investigation is left for future work.

\section{Conclusions}\label{sec: conclusion}

The joint analysis of Stage-IV surveys like Euclid, LSST, Roman and SO requires analysis methods such as SBI in order to account for systematic uncertainties, which limit these experiments. One of the main challenges with SBI analyses for multi-probe cosmology is that these methods require large simulation suites providing correlated realizations of different cosmological probes across cosmologies and baryonic feedback models. In this paper, we address this challenge by developing an efficient and accurate method to generate tSZ maps from $N$-body simulations based on the component-wise BFC model from S25.

Specifically, we use a \emph{CosmoGridV1-like} $N$-body simulation, run with the FLAMINGO cosmological parameters, and generate tSZ maps in two different ways: first, by painting spherically symmetric BFC pressure profiles at the positions of halos, which requires only the halo catalog; and second, with a gas-weighted method that redistributes the painted signal according to the baryonified gas density, so that the maps follow halo asphericity and substructure. The required gas shells can be obtained either from baryonified snapshots or directly with shell-BFC, which agree to within $0.2\%$.

Moving on to two-point statistics, we first compare the tSZ power spectrum from the painting and gas-weighted methods. The two map-making techniques agree within $2-3\%$, with the gas-weighted maps showing slightly higher power due to the increased variance of the gas field. These differences are below current observational uncertainties~\citep{Efstathiou_McCarthy_25}, and thus we conclude that the painting method is sufficient for generating accurate tSZ simulations. We then compare the BFC-painted tSZ maps to FLAMINGO. Painted onto the halos of the FLAMINGO dark-matter-only run, so that both share the same halo population, our maps reproduce the FLAMINGO tSZ power spectrum to within $5\%$ for $500\lesssim\ell\lesssim3000$ in both feedback models. Painted onto the halos of our \emph{CosmoGridV1-like} simulation, they agree with FLAMINGO within $\sim10\%$ for $\ell\gtrsim300$. At larger scales they give up to $\sim25\%$ more power than the painting of the FLAMINGO halos. This excess comes from the few most massive clusters, which dominate the tSZ signal: removing all halos with $M_{200c}\geq9\times10^{14}\,h^{-1}{\rm M_\odot}$, only $\sim10^{-5}$ of all painted halos, brings the two paintings into agreement to within $5\%$ up to $\ell\sim2000$.

These results show that our tSZ simulations based on $N$-body simulations and BFC are able to self-consistently match hydrodynamical simulations from FLAMINGO across different feedback models. Calibrated only at the level of 3D profiles, we are able to match matter power spectrum suppression, 2D Compton-$y$ profiles, and tSZ power spectra, with maximal differences to within $5\%$ for $500\lesssim\ell\lesssim3000$ once the halo populations are matched. Since the painting requires only a halo catalog, a full-sky $N_{\rm side}=4096$ Compton-$y$ map of the $\sim3.5\times10^7$ halos of the FLAMINGO halo lightcone takes only $\sim190$ CPU hours on top of the $N$-body run. This therefore lays the groundwork for efficient generation of multi-wavelength simulation suites starting from $N$-body simulations, and our pipeline can readily be applied to CosmoGridV1, thus allowing for joint SBI analyses of WL and tSZ. Looking ahead, this paves the way towards multi-probe SBI analyses for Stage-IV surveys such as Euclid, LSST, Roman, and SO.

\begin{acknowledgments}

We would like to thank Jens Chluba, Richard Battye, and David Alonso for numerous insightful conversations about this project and the tSZ in general. This work made use of several open-source tools, including \textsc{NumPy}~\cite{Harris_numpy_2020}, \textsc{SciPy}~\cite{Virtanen_scipy_2020}, \textsc{matplotlib}~\cite{Hunter_matplotlib_2007}, \textsc{HEALPix}~\citep{Gorski_Healpix_2005}. JC and AN acknowledge support from the European Research Council (ERC) under the European Union’s Horizon 2020 research and innovation program with Grant agreement No. 101163128. JB acknowledge support by a grant from the Swiss National Supercomputing Centre (CSCS) under project ID 990 on Alps.
\end{acknowledgments}

\appendix

\section{PKDGRAV3 CosmoGridV1-like vs FLAMINGO L1\_m9 DMO}\label{app: sim comparison}

The two simulations share the D3A cosmology (Table~\ref{tab: D3A_cosmo}) but differ in $N$-body code, initial conditions, box size, neutrino treatment, and particle mass, which is $\sim23$ times larger in the \emph{CosmoGridV1-like} run. Here we compare their matter power spectra and halo mass functions (HMFs), and test the effect of the halo finder.

\subsection{Matter power spectrum}\label{app: matter pk comparison}

\begin{figure}[h]
    \centering
    \includegraphics[width=\textwidth]{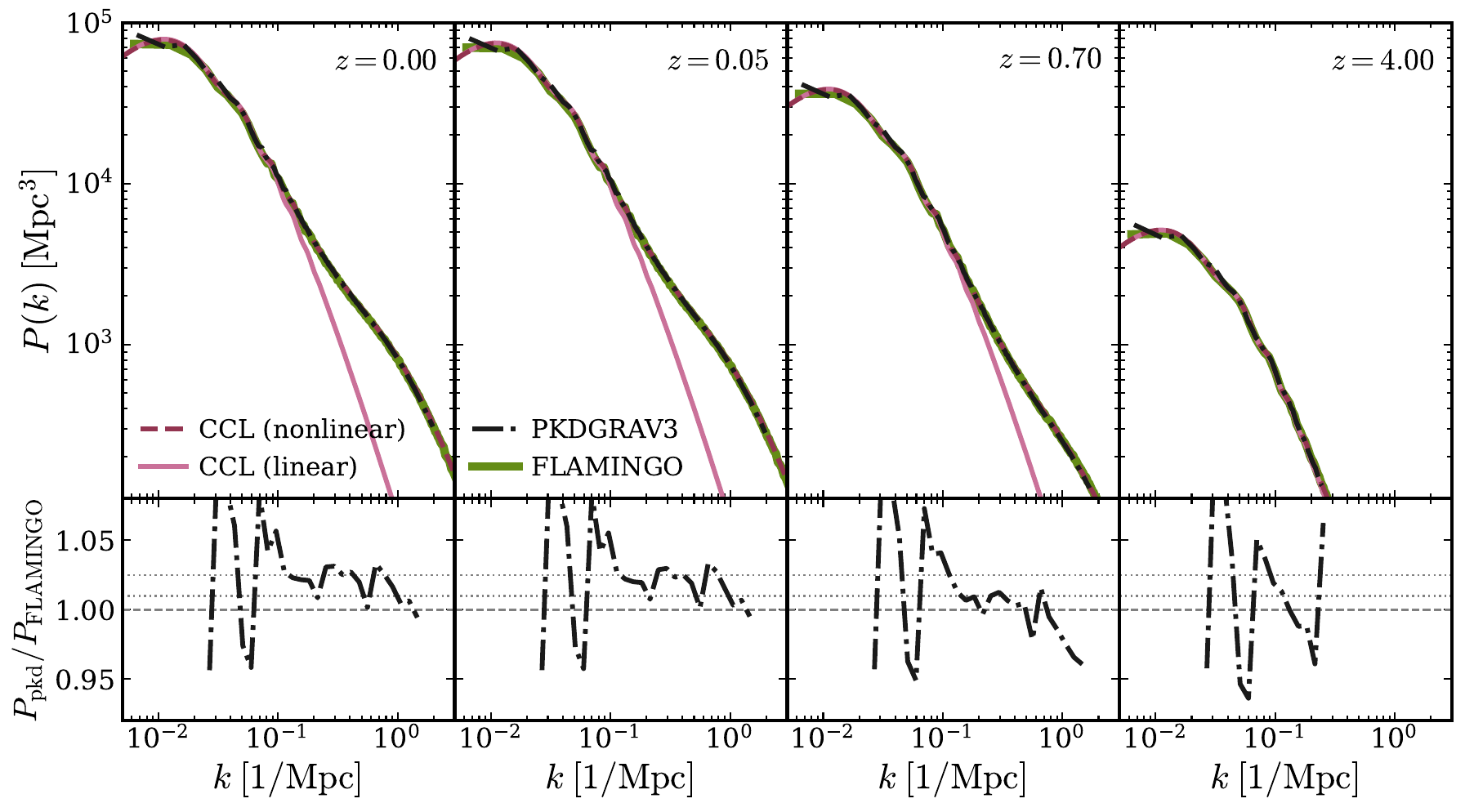}
    \caption{Comparison of the total matter power spectrum between our \textsc{PKDGRAV3} dark-matter-only simulation and the FLAMINGO fiducial dark-matter-only simulation, both run with the same cosmological parameters. \textit{Top row}: measured P(k) from \textsc{PKDGRAV3} (light blue, dash-dotted) and from FLAMINGO (dark blue, dotted), together with the linear (yellow, dashed) and non-linear (orange, solid) matter power spectrum from CCL, computed with CLASS~\citep{Blas_CLASS_2011, Lesgourgues_CLASS_2011} for the transfer function, at redshifts $z=0.00,\,0.05,\,0.70,$ and $4.00$ (left to right). \textit{Bottom row}: ratio of the \textsc{PKDGRAV3} to the FLAMINGO measurements. The two simulations agree within a few percent across the full $k$-range shown: the scatter at $k\lesssim 0.03\,\mathrm{Mpc^{-1}}$ reflects sample variance from the independent (unmatched) phases of the two sets of initial conditions, while the 2-3\% offset seen at intermediate $k$ is consistent with the level of agreement expected between different N-body codes~\citep{Schneider_2016_pkd} combined with differences between the two power-spectrum measurement pipelines and potential small differences in the neutrino treatment. The dashed and dotted horizontal grey lines mark unity and a 2.5\% offset, respectively.}
    \label{fig: pk comparison pkd flamingo}
\end{figure}

Fig.~\ref{fig: pk comparison pkd flamingo} compares the matter power spectra of the two runs. At $z\leq0.05$, \textsc{Pkdgrav3} lies $\sim2$--$3\%$ above FLAMINGO for $0.1\lesssim k\lesssim0.7\,{\rm Mpc}^{-1}$. This offset builds up at late times: it falls to $\sim1\%$ at $z=0.7$ and is consistent with zero within the scatter at $z=4$. At $k\lesssim0.1\,{\rm Mpc}^{-1}$ the ratio scatters by several per cent because the two runs have independent initial phases. Differences of this size are expected between $N$-body codes~\citep{Schneider_2016_pkd}, and we do not attempt to separate them further. They are small compared with the effect of the halo population on $C_\ell^{yy}$.

\subsection{Halo mass function}\label{app: hmf difference in simims}

\begin{figure}
    \centering
    \includegraphics[width=1.\linewidth]{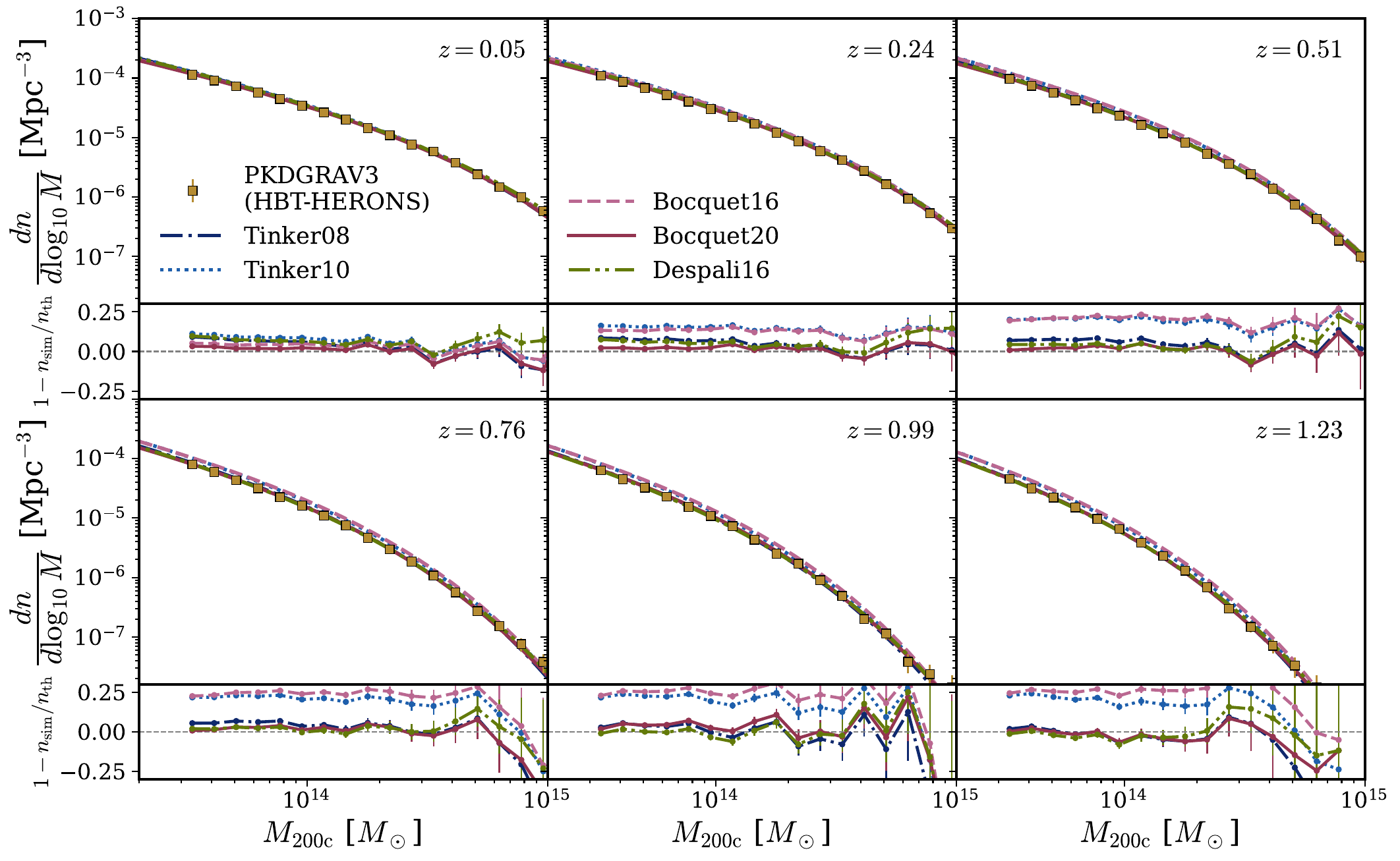}
    \caption{Host-halo mass function of the \emph{CosmoGridV1-like} \textsc{Pkdgrav3} simulation, identified with HBT-HERONS (squares), at six redshifts between $z=0.05$ and $z=1.23$. The lines show the analytic HMFs of Tinker08 (dash-dotted), Tinker10 (dotted), Bocquet16 (dashed), Bocquet20 (solid) and Despali16 (dash-dot-dotted), evaluated in the D3A cosmology. \textit{Lower panels}: fractional difference $1-n_{\rm sim}/n_{\rm th}$ between the measured and each analytic HMF. Positive values mean the simulation has fewer halos than the prediction. Error bars are Poisson. Bocquet20 matches the measurement to within $4\%$ for $3\times10^{13}<M_{200c}/{\rm M_\odot}<3\times10^{14}$ at all redshifts shown.}

    \label{fig: hmf n-body pkdgrav}
\end{figure}

\begin{figure}
    \centering
    \includegraphics[width=1.\linewidth]{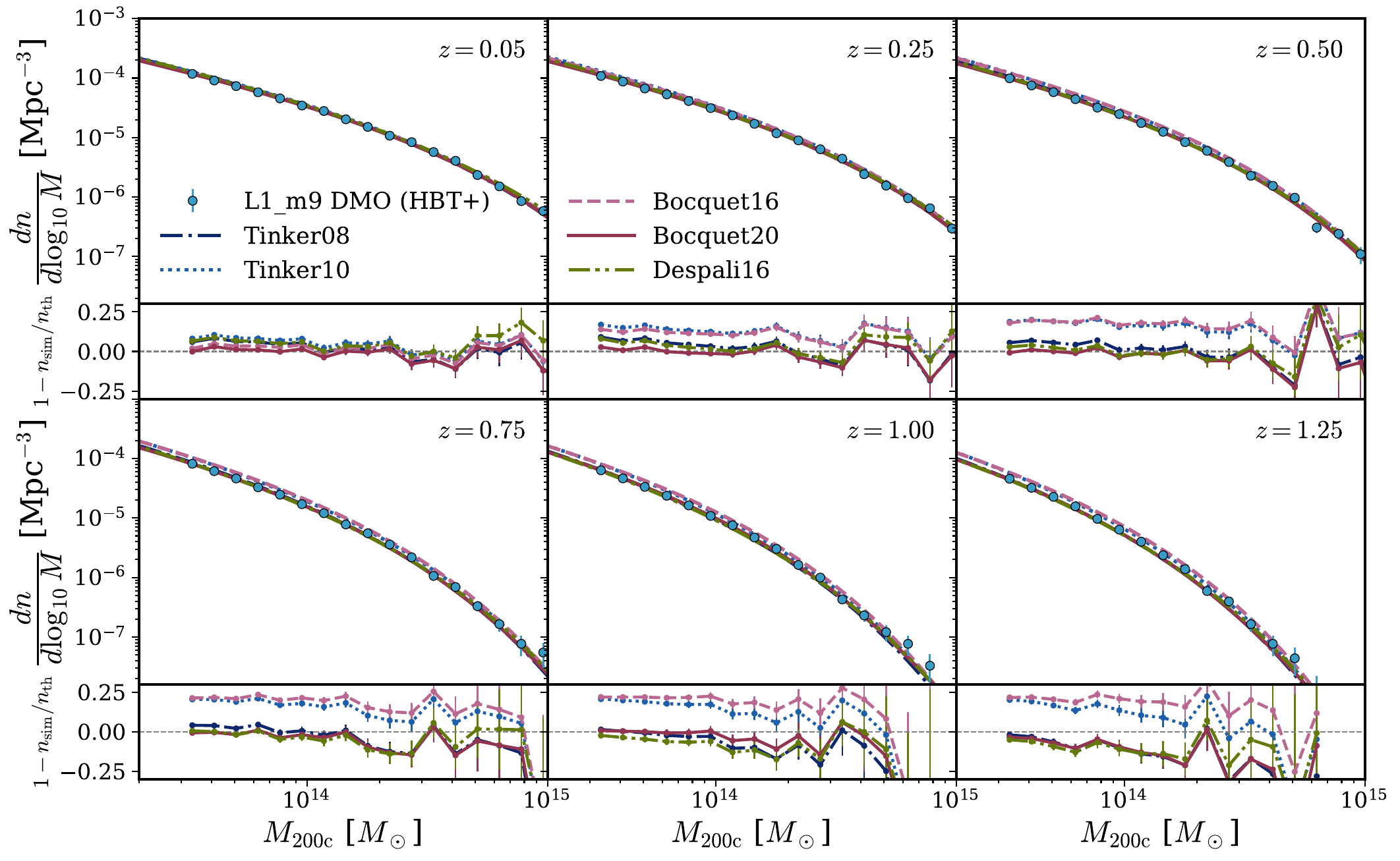}
    \caption{Same as Fig.~\ref{fig: hmf n-body pkdgrav}, but for the FLAMINGO L1\_m9 dark-matter-only run with HBT-HERONS halos (circles), at $z=0.05$--$1.25$. Bocquet20 again gives the best description, to within $2\%$ up to $z=1$. Like the \emph{CosmoGridV1-like} run, FLAMINGO lies $\sim20$--$25\%$ below Tinker10 and Bocquet16 at $z\geq0.5$.}
    \label{fig: hmf dmo flamingo}
\end{figure}

Figs.~\ref{fig: hmf n-body pkdgrav} and~\ref{fig: hmf dmo flamingo} compare the host-halo HMFs of the two runs, both identified with HBT-HERONS~\citep{Forouhar_2025}, with analytic HMFs. Both are described best by Bocquet20~\citep{Bocquet_2020}: for $3\times10^{13}<M_{200c}/{\rm M_\odot}<3\times10^{14}$, the \emph{CosmoGridV1-like} run lies within $4\%$ of it at all redshifts shown, and FLAMINGO within $2\%$ up to $z=1$. Tinker08~\citep{Tinker_2008} and Despali16~\citep{Despali_2016} are within $\sim8\%$, while both runs lie $\sim20$--$25\%$ below Tinker10~\citep{Tinker_2010} and Bocquet16~\citep{Bocquet_2016} at $z\geq0.5$.

At fixed mass, the \emph{CosmoGridV1-like} run contains fewer halos than FLAMINGO. For $10^{14}<M_{200c}/{\rm M_\odot}<3\times10^{14}$, the deficit grows from $\sim3.5\%$ at $z\leq0.25$ to $\sim10\%$ at $z\approx1.25$. Its coarser mass resolution may contribute, but we have not isolated the cause. This deficit sets the high-$\ell$ residual in Fig.~\ref{fig: tsz_pk masscut}.

\subsection{Halo finder: AHF and HBT-HERONS}\label{app: halo finders}

The main text uses HBT-HERONS for both simulations, as this is the finder of the FLAMINGO catalogues. To quantify the impact of this choice, we also identify halos in the \emph{CosmoGridV1-like} simulation with AHF~\citep{AHF}. Because HBT-HERONS requires particle IDs, it was run on a rerun with the same initial conditions, code, and settings.

\begin{figure}
    \centering
    \includegraphics[width=1.\textwidth]{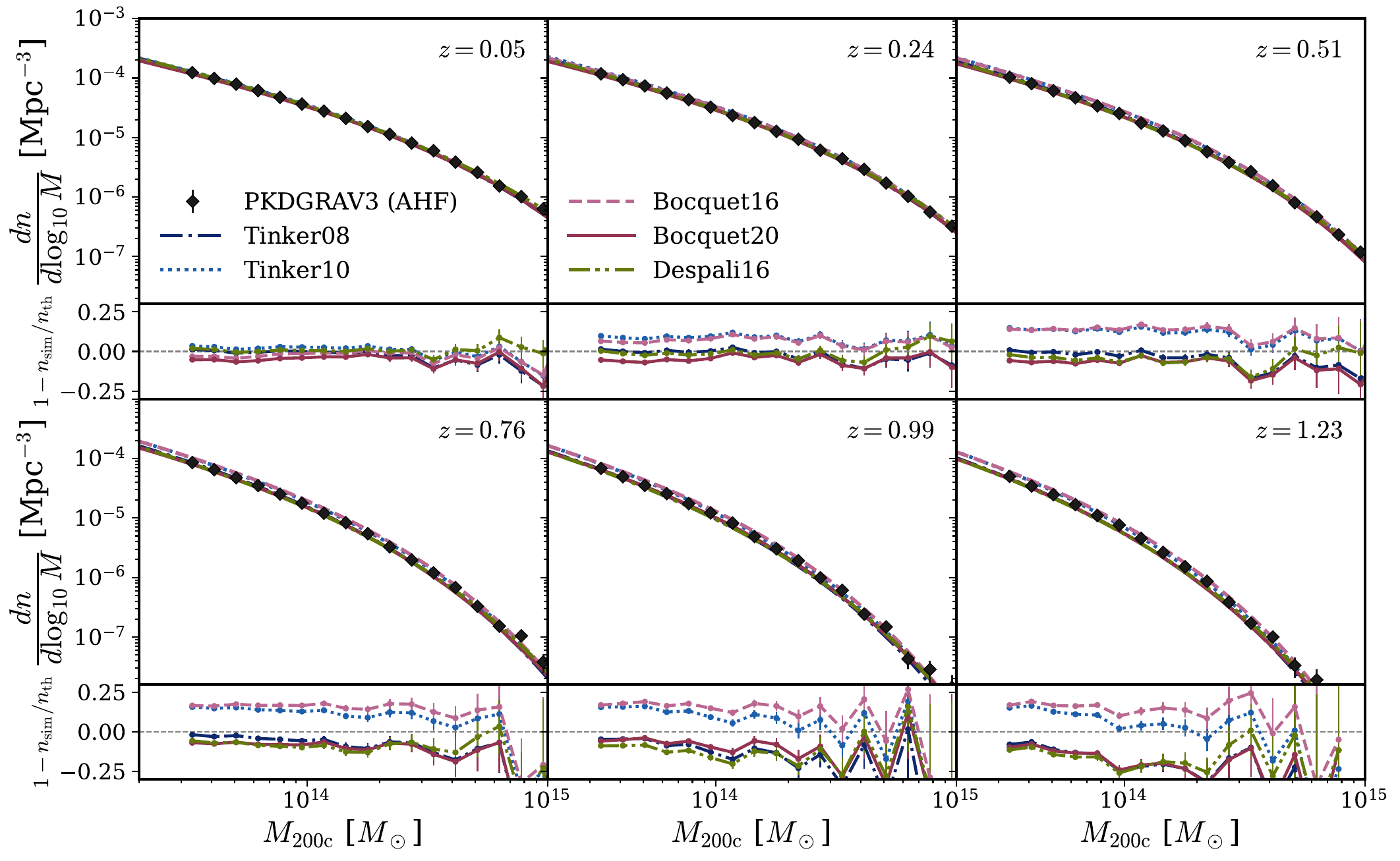}
    \caption{Same as Fig.~\ref{fig: hmf n-body pkdgrav}, but with the halos of the \emph{CosmoGridV1-like} simulation identified with AHF (diamonds) instead of HBT-HERONS. Because AHF assigns slightly higher masses to matched hosts, it finds $5$--$18\%$ more halos at fixed mass than HBT-HERONS, with the excess growing with redshift. As a result, $1-n_{\rm sim}/n_{\rm th}$ moves downwards relative to Fig.~\ref{fig: hmf n-body pkdgrav}, and the AHF HMF lies above Bocquet20 at $z\gtrsim0.5$.}
    \label{fig: ahf hmf pkdgrav3}
\end{figure}

\begin{figure}
    \centering
    \includegraphics[width=\textwidth]{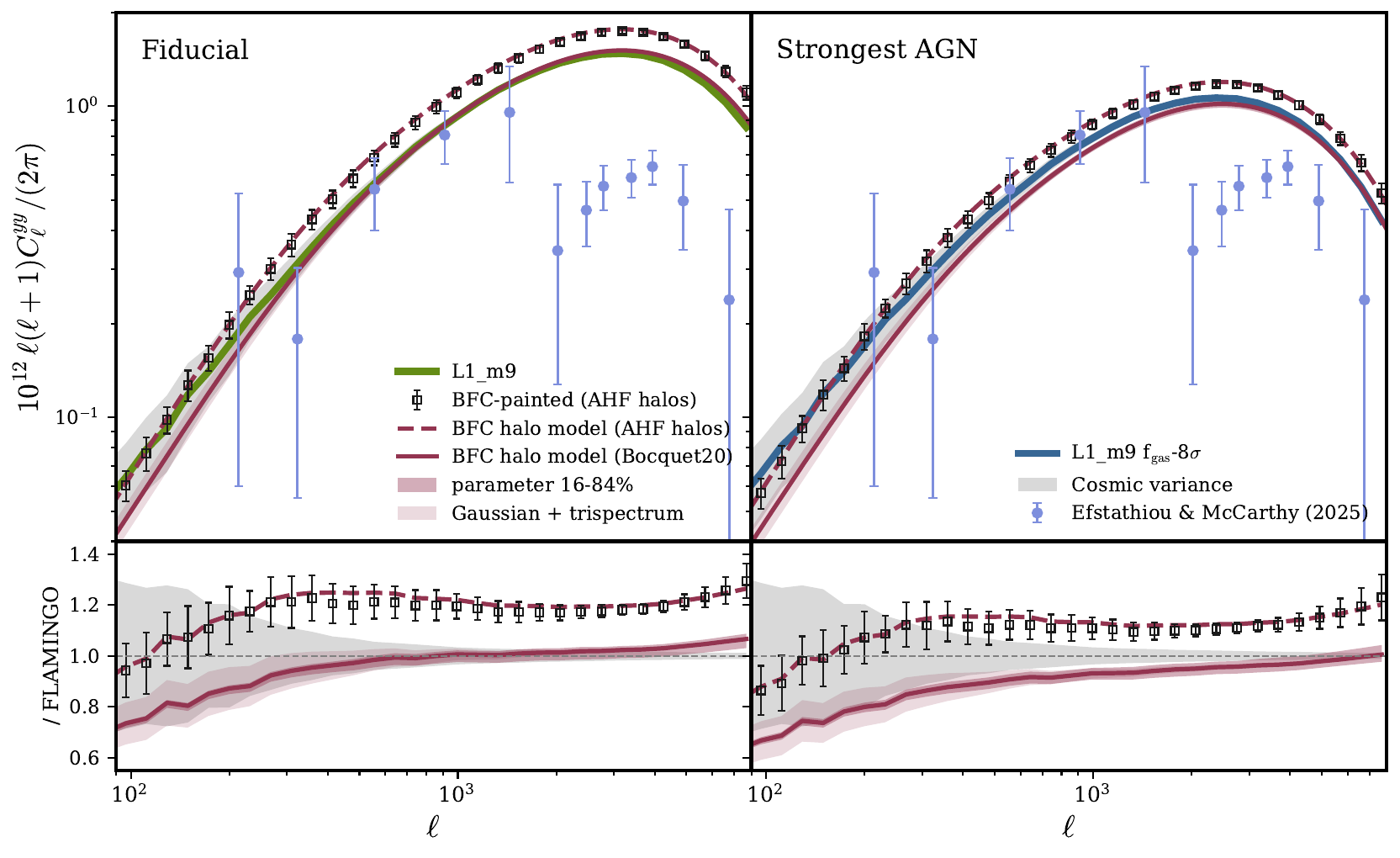}
    \caption{Same as Fig.~\ref{fig: tsz_pk hbt}, but painted onto the AHF halos of the \emph{CosmoGridV1-like} simulation and without the L2p8\_m9 line. The dashed line is the halo model with the one-halo term summed over the AHF catalog and the Bocquet20 two-halo term.}
    \label{fig: tsz_pk ahf appendix}
\end{figure}

At fixed mass, AHF finds more halos than HBT-HERONS (Fig.~\ref{fig: ahf hmf pkdgrav3}): $8$--$11\%$ more for $3\times10^{13}<M_{200c}/{\rm M_\odot}<10^{14}$ and $5$--$18\%$ more for $10^{14}<M_{200c}/{\rm M_\odot}<3\times10^{14}$, increasing with redshift. Painted onto the AHF halos, the same profiles give a $C_\ell^{yy}$ that is $17$--$22\%$ above L1\_m9 and $10$--$16\%$ above $f_{\rm gas}-8\sigma$ for $\ell\gtrsim300$ (Fig.~\ref{fig: tsz_pk ahf appendix}). The HBT-HERONS painting of the same simulation is $\sim10\%$ lower on all scales and in both feedback models. A one-halo sum over the AHF catalog plus a Bocquet20 two-halo term reproduces the painted maps to $\sim2\%$, so this shift comes from the halo catalog rather than from the painting or the pressure model.

\section{Sensitivity of the tSZ power spectrum}\label{app: tsz pk sensitivity}

All spectra in this appendix are BFC halo-model predictions with the Bocquet20 HMF and the L1\_m9 parameters of Table~\ref{tab: parameters}, integrated over $0.05<z<2$ and $10^{10.5}<M_{200c}/{\rm M_\odot}<10^{15.5}$. We vary one quantity at a time.

\subsection{Convergence of the tSZ power spectrum with mass and redshift cuts}\label{app: tsz pk mass and z convergence}
\begin{figure}
    \centering
    \includegraphics[width=0.8\textwidth]{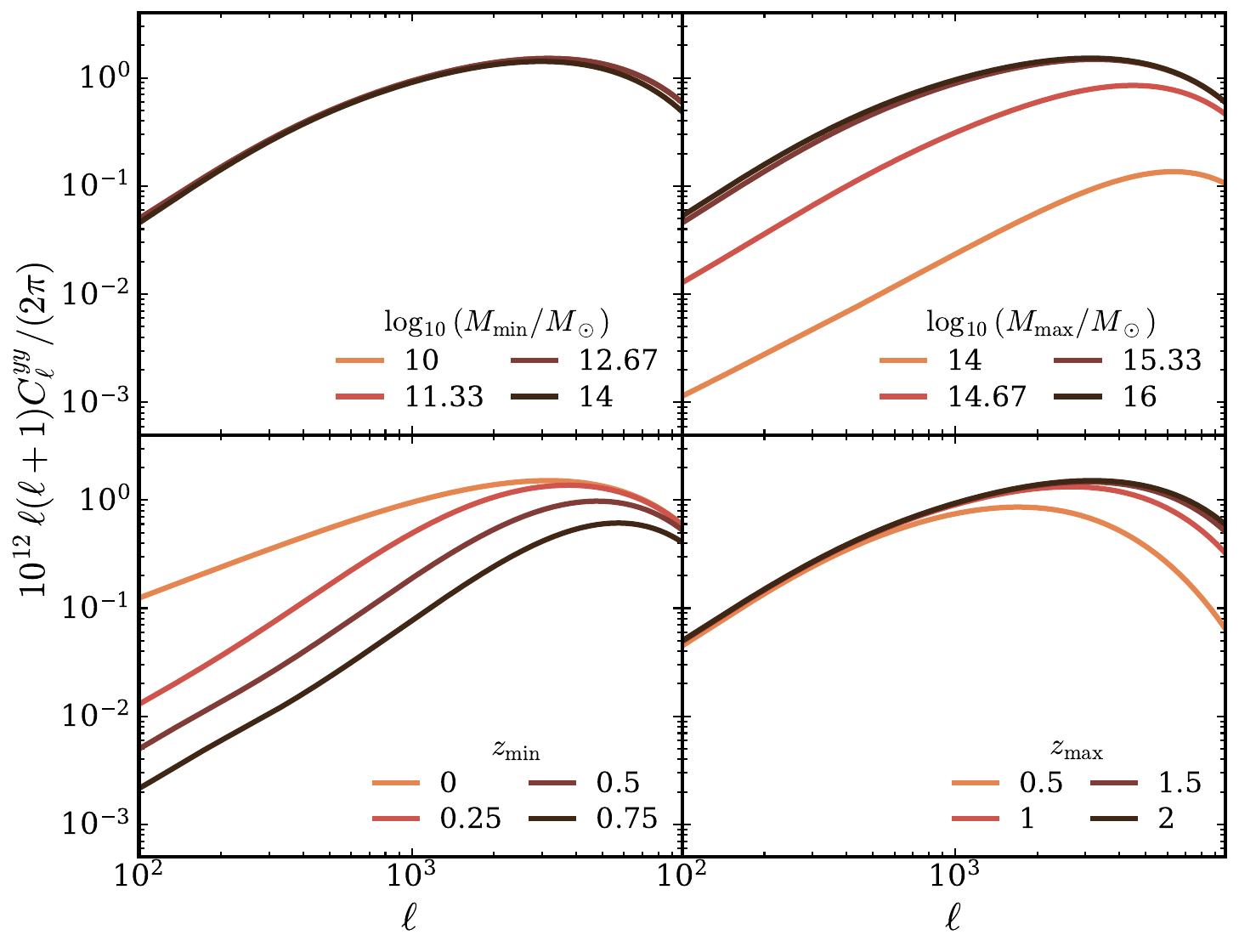}
    \caption{This figure shows how the tSZ power spectrum changes if we change the boundary conditions of the simulation and the theoretical modeling. As shown in previous work, the lower-mass boundary is mostly irrelevant as long as the largest halos are included in the analysis. For the redshift cuts, we see that the lower redshift cut $z_{\rm min}$ has a very big influence on the power spectrum, while the $z_{\rm max}$ becomes important for cuts below $z\approx1.0$.}
    \label{fig: tsz_pk mass and z sensitivity}
\end{figure}

In Fig~\ref{fig: tsz_pk mass and z sensitivity} we show the convergence of the tSZ power spectrum with respect to the lower and upper mass and redshift limits used in the halo-model integration. In each panel, we fix all limits to their fiducial values except the one being varied.

The top row shows the dependence on the halo mass range. Increasing $M_{\rm min}$ has almost no effect on the power spectrum, indicating that low-mass halos contribute negligibly to the tSZ signal. In contrast, the amplitude is highly sensitive to $M_{\rm max}$: increasing the upper mass limit from $10^{14}$ to $10^{16}\, \mathrm{M_\odot}$ increases the power by more than an order of magnitude, reflecting the steep mass dependence of the electron pressure (and hence Compton-$y$) profile; the tSZ signal is dominated by the most massive halos in the sample. We accordingly extend the mass integration to the highest halo masses resolved by the simulation, to avoid artificially truncating this dominant contribution.

The bottom row shows the corresponding dependence on the redshift integration range. The signal is dominated by lower redshifts: raising $z_{\rm min}$ progressively removes the low-$z$ contribution and substantially suppresses the power spectrum by more than an order of magnitude in the range of $z_{\rm min}=0.0-0.75$. On the other hand, the result is largely converged with respect to $z_{\rm max}$ already by $z_{\rm max}\sim 1.5-2.0$, with only a small residual contribution from higher redshifts. Based on this behavior, we adopt $z_{\rm min}=0.05$, excluding the lowest-redshift shell where the signal is most affected by large-scale sample variance, and $z_{\rm max}=2.00$, since the power spectrum is essentially converged by this point and extending further increases computational cost with negligible gain.

\subsection{BFC parameter dependence of the tSZ power spectrum}\label{app: tsz pk bfc params}
\begin{figure}
    \centering
   \includegraphics[width=\textwidth]{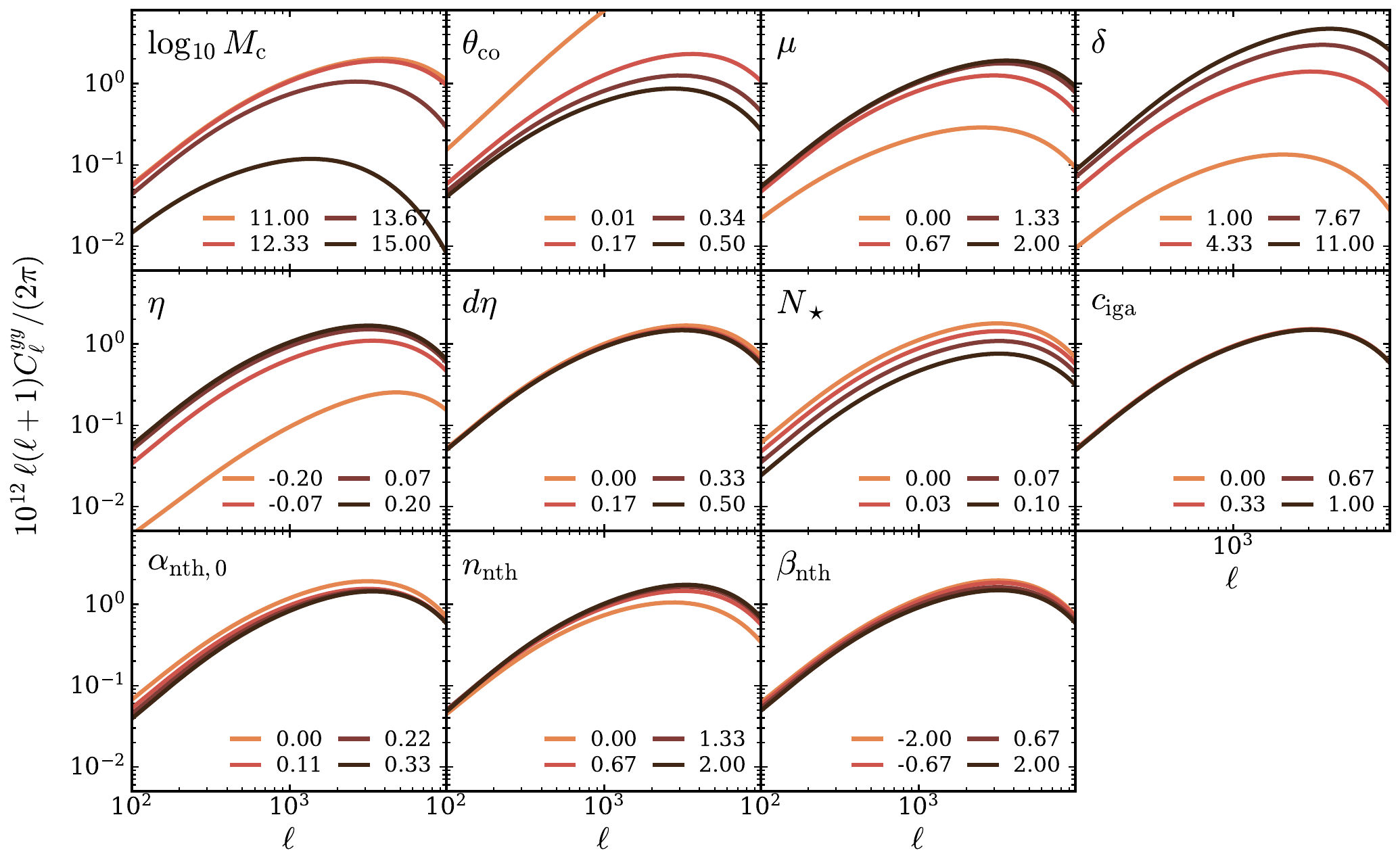}
    \caption{In this figure, we see the influence of each BFC parameter on the tSZ power spectrum. The biggest influence comes from the $\rho_{\rm gas}$ parameters, which are the most flexible part in the calculation of the pressure in the BFC model.}
    \label{fig: tsz_pk param sensitivity}
\end{figure}

In Fig.~\ref{fig: tsz_pk param sensitivity} we show the dependence of the tSZ power spectrum on the BFC parameters. We vary one parameter at a time, keeping the others fixed at the best-fit values from the fiducial FLAMINGO simulation. The lowest value of each parameter is shown in red, the highest in dark blue. 

The top row shows the hot gas parameters $M_{\rm c},\,\theta_{\rm co},\,\mu$, and $\delta$, which together set the shape of the hot gas density profile (Eq.~\eqref{eq: rhoHGA} and~\eqref{eq: rhoHGA beta}). $M_{\rm c}$ sets the characteristic halo mass at which the profile slope $\beta(M)$ transitions from shallow to steep (Eq.~\eqref{eq: rhoHGA beta}); lower $M_{\rm c}$ steepens the profile at fixed halo mass, concentrating more gas, hence more pressure, in the inner regions. $\mu$ controls how fast this transition is. $\theta_{\rm co}$ sets the core radius, and $\delta$ the sharpness of the outer truncation; both therefore redistribute gas between the core and the outskirts. Because the tSZ signal is weighted by pressure ($\propto \rho_{\rm HGA}\times M_{\rm DMB}$), we see the influence of the gas profile directly in the tSZ power spectrum.

The second row shows the stellar parameters $\eta,\,\ d\eta,$ and $N_{\star}$, and the cold gas parameter $c_{\rm iga}$. $\eta$ sets the high-mass slope of the total stellar fraction, $d\eta$ sets the slope of the fraction of the central galaxies, and $N_\star$ the overall normalization; all of them regulate how much of the baryonic budget is locked into the stars rather than hot gas. $c_{\rm iga}$ similarly removes a fraction of the central gas into a cold, non-ionized phase. Although $\eta,\, N_\star$, and $c_{\rm iga}$ act in the same way on the tSZ signal, reducing the amount of gas and thus changing the gas amplitude, we only see an influence from the stellar parameters. This comes from the fact that the amount of cold gas in halos with masses $M_{\rm 200c}>1e13\, \mathrm{M_\odot}/h$ is negligible. 

The bottom row shows the non-thermal pressure parameters $\alpha_{\rm nth,0},\, n_{\rm nth}$, and $\beta_{\rm nth}$. Their effect on the tSZ power spectrum is the smallest of the three groups, but not negligible: unlike a simple amplitude rescaling, they can independently vary the overall normalization ($\alpha_{\rm nth,0}$), its slope of the mass/radial dependence ($n_{\rm nth}$), and its redshift evolution ($\beta_{\rm nth}$), which changes the shape of the tSZ spectrum rather than just its amplitude. This makes them relevant when aiming for percent-level agreement, even though their overall impact is subdominant to the gas and stellar parameters. 

\subsection{Difference in halo mass functions}\label{app: tsz pk hmf dependence}

\begin{figure}
    \centering
    \includegraphics[width=0.8\textwidth]{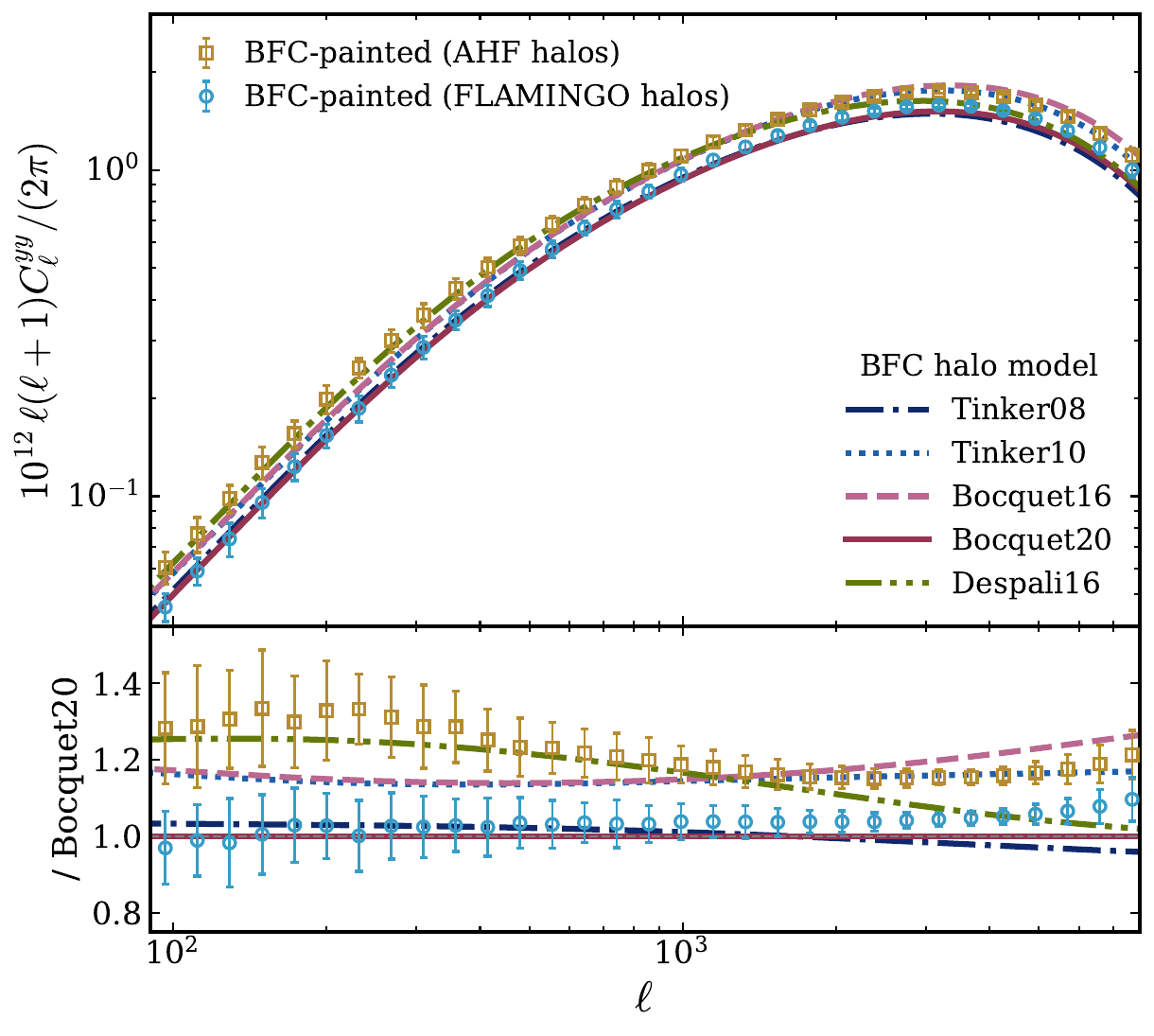}
    \caption{Dependence of the BFC halo-model tSZ power spectrum on the choice of HMF. \textit{Top}: $D_\ell^{yy}$ for $0.05<z<2$ with the L1\_m9 parameters, computed with the Tinker08 (dash-dotted), Tinker10 (dotted), Bocquet16 (dashed), Bocquet20 (solid) and Despali16 (dash-dot-dotted) HMFs. For comparison, the symbols show BFC profiles painted onto the AHF halos of the \emph{CosmoGridV1-like} simulation (squares) and onto the FLAMINGO HBT-HERONS halos (circles). \textit{Bottom}: ratio to the Bocquet20 prediction. The HMF alone changes $C_\ell^{yy}$ by up to $\sim25\%$. The FLAMINGO painting lies $0$--$6\%$ above Bocquet20, and the AHF painting lies $16$--$31\%$ above it.}
    \label{fig: tSZ pk hmf dependence}
\end{figure}

Fig.~\ref{fig: tSZ pk hmf dependence} shows how strongly the predicted $C_\ell^{yy}$ depends on the HMF. Relative to Bocquet20, Tinker08 stays within $3\%$, Tinker10 and Bocquet16 are $14$--$24\%$ higher, and Despali16 falls from $+25\%$ at $\ell<300$ to $+4\%$ at $\ell>3000$. Because $C_\ell^{yy}$ is dominated by the exponential tail of the HMF, small differences in the abundance of massive clusters are strongly amplified. The spread largely reflects the different simulations, halo finders, and mass definitions on which these fits were calibrated, rather than a fundamental uncertainty in the abundance of halos. Some fits, such as Despali16, also require a concentration-dependent conversion to $M_{200c}$. Our simulations show the same effect directly: painting the halos of the same matter field identified with different finders, the FLAMINGO HBT-HERONS halos lie $0$--$6\%$ above the Bocquet20 prediction and the AHF halos $16$--$31\%$ above it. We therefore use Bocquet20 throughout, as it matches the measured HBT-HERONS HMFs (App.~\ref{app: hmf difference in simims}).

For cosmological inference, the HMF therefore has to be calibrated on the same halo definition as the pressure model, with any remaining uncertainty treated as a systematic. This is the approach of current cluster analyses, which use emulators calibrated on large simulation suites \citep[e.g.][]{Bocquet_2020, Castro_2023} and marginalise over residual HMF parameters. Forward-modelling approaches that paint profiles directly onto simulated halos, such as the one presented here, avoid an analytic HMF altogether. They move the systematic to the simulation volume, resolution, and halo finder, which can be tested explicitly, as in Sec.~\ref{subsec: tsz pk results}.

\subsection{Comparison to additional observational data}\label{app: tsz data}

\begin{figure}
    \centering
    \includegraphics[width=0.65\textwidth]{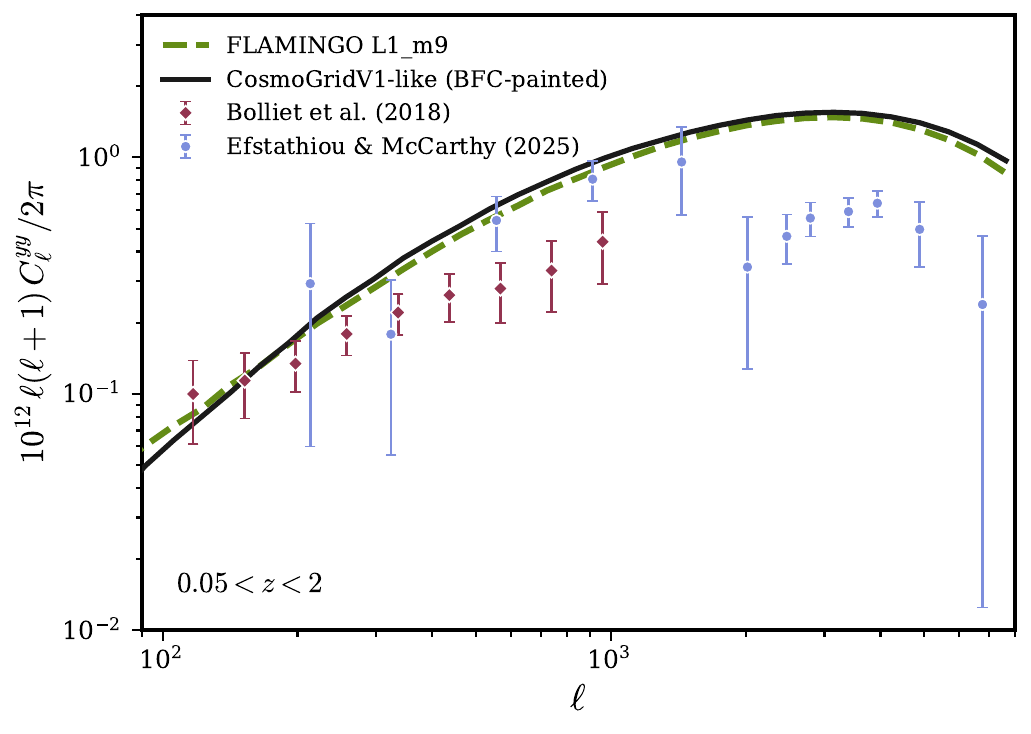}
    \caption{$D_\ell^{yy}$ over $0.05<z<2$ from FLAMINGO L1\_m9 (dashed green) and the BFC-painted \emph{CosmoGridV1-like} map (solid black), compared with the \textit{Planck} $y$-map spectrum of Ref.~\citep{Bolliet_2018} (red diamonds) and the \textit{Planck}+ACT+SPT reconstruction of Ref.~\citep{Efstathiou_McCarthy_25} (blue circles). Error bars are $1\sigma$.}
    \label{fig: tsz pk + bolliet}
\end{figure}

Fig.~\ref{fig: tsz pk + bolliet} compares our spectra with two observational estimates of $C_\ell^{yy}$: the spectrum inferred from the \textit{Planck} Compton-$y$ map~\citep{Bolliet_2018}, and the reconstruction of Ref.~\citep{Efstathiou_McCarthy_25} from \textit{Planck}, ACT, and SPT temperature power spectra, which extends to $\ell\sim7000$. At $\ell\lesssim200$ both simulations agree with the data. At $300\lesssim\ell\lesssim10^3$, the $y$-map spectrum lies a factor of $\sim2$ below the simulations, while the reconstruction of Ref.~\citep{Efstathiou_McCarthy_25} is consistent with them up to $\ell\approx1500$. The two measurements disagree by a similar factor, which points to foreground removal in $y$-map-based spectra as a major systematic~\citep{Efstathiou_McCarthy_25}. At $\ell\gtrsim2000$, the data lie a factor of $\sim2$--$2.5$ below both simulations. This deficit relative to \textit{Planck}-like $\Lambda$CDM simulations is well established~\citep{McCarthy_2014, Efstathiou_McCarthy_25}. It may indicate stronger feedback or residual CIB and kSZ contamination at high $\ell$. Since BFC is calibrated on FLAMINGO, the painted map reproduces the simulation, not the data, and our spectra omit the $z<0.05$ contribution at the lowest multipoles.

\section{Non-thermal pressure in FLAMINGO}
\label{app:pnt}

We measure the non-thermal pressure in the L1\_m9 and $f_{\rm gas}-8\sigma$ runs at $z=0$, $0.2$, and $0.5$. In each of four $0.5$-dex bins of $M_{200c}$ between $10^{13}$ and $10^{15}\,{\rm M_\odot}$, we select 100 random central halos and bin their gas into 30 logarithmic shells out to $10\,R_{200c}$. In each shell, $\mathscr{P}_{\rm th}=n_{\rm gas}k_{\rm B}T$ follows from the mass-weighted temperature and mean gas density, and $\mathscr{P}_{\rm nth}=\tfrac13\rho_{\rm gas}\langle|\mathbf v-\bar{\mathbf v}|^2\rangle$, where $\bar{\mathbf v}$ is the mass-weighted mean gas velocity of the shell~\citep{Lau_2009,Battaglia_2012,Nelson_2014}. We then fit Eq.~\eqref{eq: Pnth} to the median $f_{\rm nth}\equiv\mathscr{P}_{\rm nth}/(\mathscr{P}_{\rm th}+\mathscr{P}_{\rm nth})$, jointly over the three redshifts and over $0.4<r/R_{200c}<1$. Table~\ref{tab:pnt} lists the results, and our fiducial model uses the $10^{14}$--$10^{14.5}\,{\rm M_\odot}$ bin.

\begin{figure}
    \centering
    \includegraphics[width=0.85\linewidth]{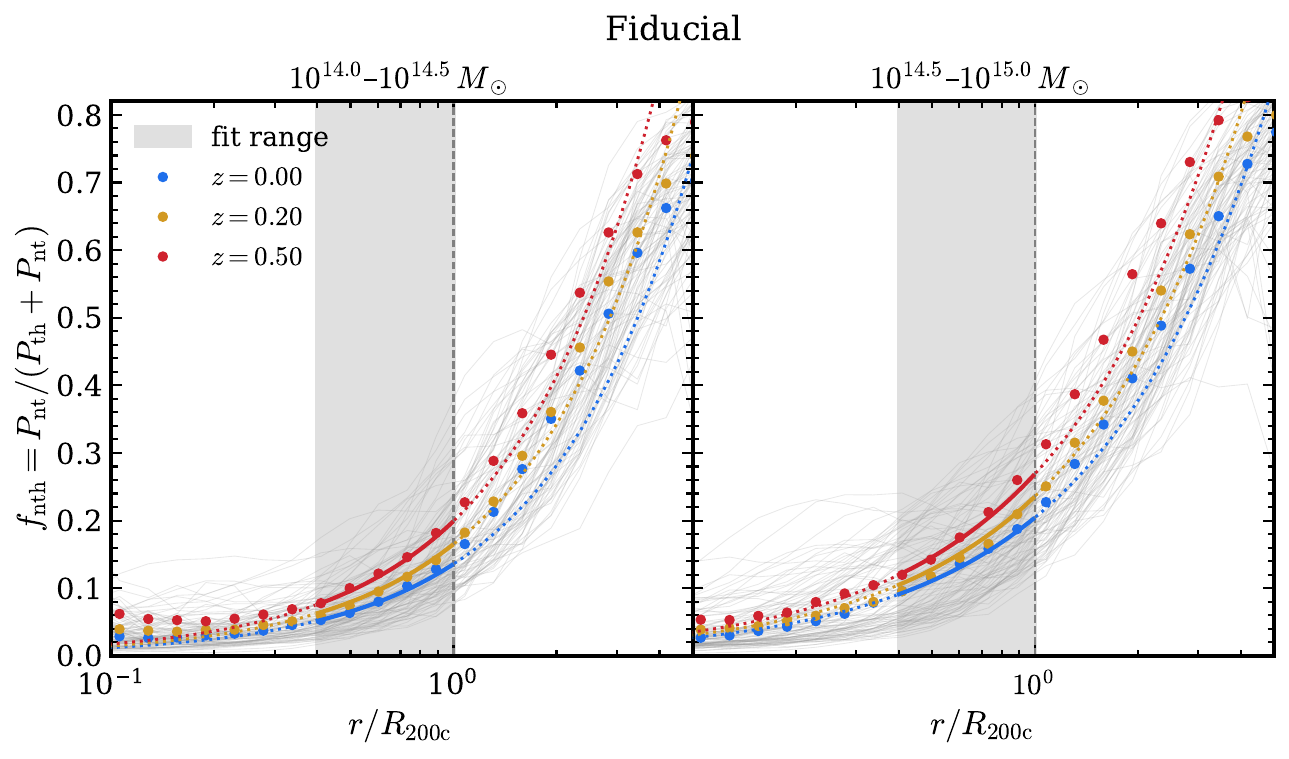}
    \caption{Non-thermal pressure fraction $f_{\rm nth}=\mathscr{P}_{\rm nth}/(\mathscr{P}_{\rm th}+\mathscr{P}_{\rm nth})$ as a function of $r/R_{200c}$ for central halos with $10^{14}<M_{200c}/{\rm M_\odot}<10^{14.5}$ (left) and $10^{14.5}<M_{200c}/{\rm M_\odot}<10^{15}$ (right) in the L1\_m9 run. Thin grey lines show individual halos. Filled circles show the median at $z=0$ (blue), $0.2$ (yellow) and $0.5$ (red). The lines show the joint best fit of Eq.~\eqref{eq: Pnth} (Table~\ref{tab:pnt}): solid inside the fit range $0.4<r/R_{200c}<1$ (shaded) and dotted outside it. Beyond $\sim1.5\,R_{200c}$ the measured fraction rises more steeply than the power law.}
    \label{fig: f_nth}
\end{figure}

The power law describes $f_{\rm nth}$ well inside the fit range ($\chi^2/{\rm dof}\sim1$--2). Beyond $\sim1.5\,R_{200c}$ the measured fraction rises more steeply (Fig.~\ref{fig: f_nth}), and extending the fit to $2$ ($4$) $R_{200c}$ raises $\chi^2/{\rm dof}$ to $\sim4$--10 ($\sim7$--11). The best-fit parameters also vary with halo mass and feedback strength (Table~\ref{tab:pnt}), so a single parameter set is an approximation that is least reliable in the halo outskirts.

\begin{table}
\centering
\caption{Best-fit \citep{Shaw_L_2010} non-thermal pressure parameters
(Eq.~\ref{eq: Pnth}), fitted to the halo-median $f_{\rm nth}(r)$ over
$0.4<r/R_{200\rm c}<1$ and $z=0,0.2,0.5$.}
\label{tab:pnt}
\begin{tabular}{llccc}
\hline
Run & $M_{200\rm c}$ bin & $\alpha_{\rm nth,0}$ & $\beta_{\rm nth}$ & $n_{\rm nth}$ \\
\hline
Fiducial & $10^{13.0}$--$10^{13.5}$ & 0.169 & 0.66 & 0.24 \\
 & $10^{13.5}$--$10^{14.0}$ & 0.111 & 0.97 & 0.65 \\
 & $10^{14.0}$--$10^{14.5}$ & 0.136 & 1.58 & 0.79 \\
 & $10^{14.5}$--$10^{15.0}$ & 0.204 & 1.82 & 0.88 \\
\hline
Strongest AGN & $10^{13.0}$--$10^{13.5}$ & 0.231 & 0.60 & 0.28 \\
 & $10^{13.5}$--$10^{14.0}$ & 0.166 & 0.66 & 0.28 \\
 & $10^{14.0}$--$10^{14.5}$ & 0.107 & 0.94 & 0.54 \\
 & $10^{14.5}$--$10^{15.0}$ & 0.148 & 4.21 & 1.10 \\
\hline
\end{tabular}
\end{table}

\section{Painting/theory Poisson catalog}

\begin{figure}
    \centering
    \includegraphics[width=0.8\textwidth]{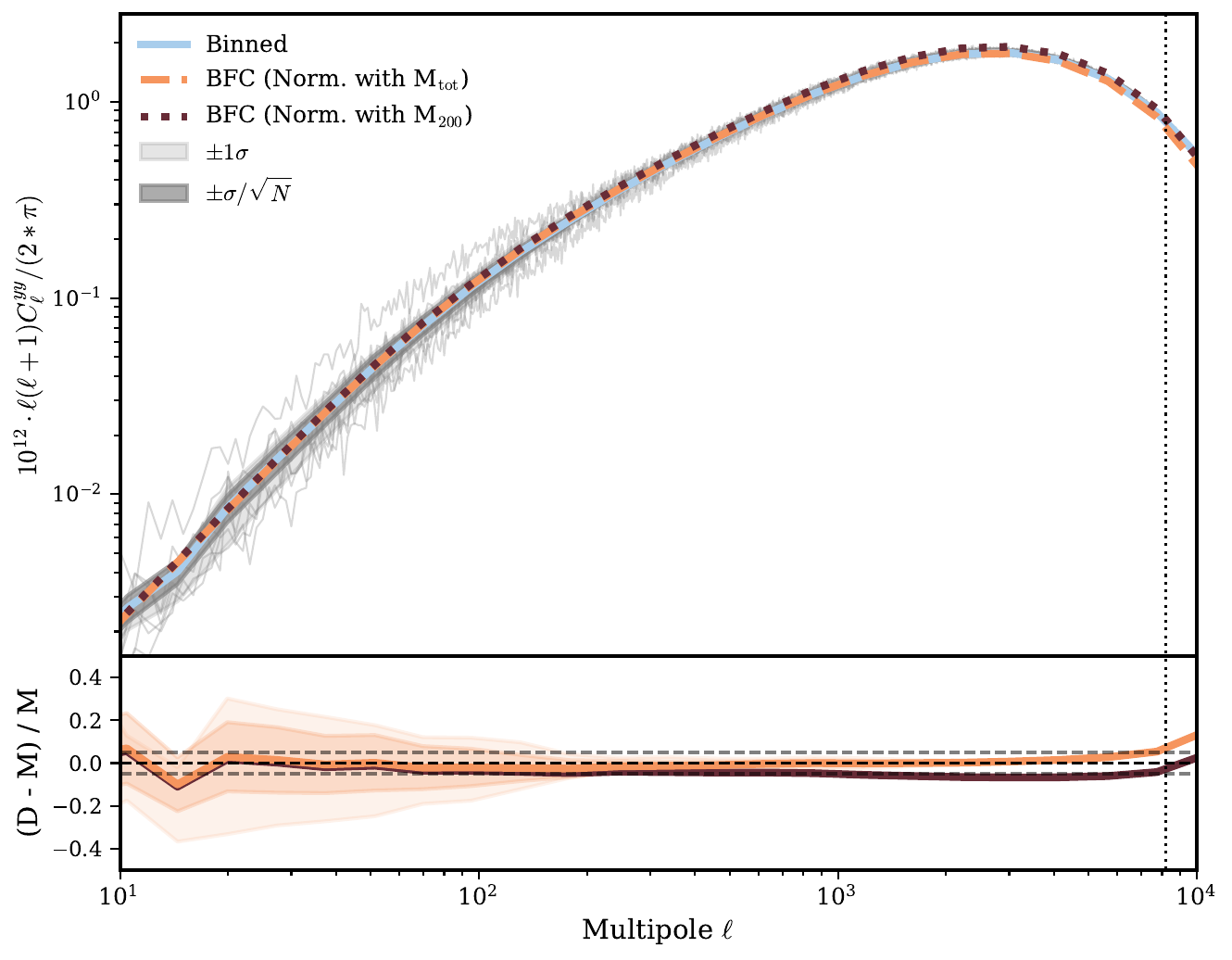}
    \caption{Test of the painting on random halo positions. BFC profiles with the L1\_m9 parameters are painted onto seven Poisson-sampled halo catalogs drawn from the Tinker08 HMF ($M_{200c}$). These maps contain only the one-halo term. \textit{Top}: tSZ power spectra of the individual maps (thin grey lines) and their mean (light blue). The light and dark grey bands show the $\pm1\sigma$ scatter between maps and the uncertainty on the mean, $\pm\sigma/\sqrt{N}$. These are compared with the one-halo BFC prediction for the same HMF, with the gas normalized by the truncated-NFW total mass $M_{\rm tot}$ (orange dashed) or by $M_{200c}$ (dark red dotted). \textit{Bottom}: fractional difference between the mean painted spectrum and each prediction, with the same uncertainty bands. The grey dashed lines mark $\pm5\%$, and the vertical dotted line marks $\ell=2N_{\rm side}$. With the $M_{\rm tot}$ normalization, which our halo-model code uses, the maps match the prediction to within a few per cent up to $\ell\approx8000$. With the $M_{200c}$ normalization, the prediction is $5$--$7\%$ too high.}
\label{fig: poisson painting}
\end{figure}

To test the painting method independently of any simulation, we paint BFC profiles onto seven Poisson-sampled halo catalogs drawn from the Tinker08 HMF. Because the positions are random, the maps contain only the one-halo term and their HMF is known exactly, so any difference from the halo model comes from the painting. Fig.~\ref{fig: poisson painting} shows that the mean of the maps agrees with the prediction to within a few per cent up to $\ell\approx8000$ when the gas is normalized with the truncated-NFW total mass $M_{\rm tot}$, as in our halo-model code. Normalizing with $M_{200}$ instead overpredicts the power by $5$--$7\%$.

The same holds for the simulated halos: a one-halo sum over the painted catalogs plus a Bocquet20 two-halo term reproduces the painted maps to within $\sim3\%$ for $\ell\gtrsim300$. Differences between the painted maps and the analytic halo model in Sec.~\ref{subsec: tsz pk results} therefore reflect the halo catalogs rather than the painting.

\bibliographystyle{JHEP}
\bibliography{biblio.bib}

@ARTICLE{Fluri_J_2022,
       author = {{Fluri}, Janis and {Kacprzak}, Tomasz and {Lucchi}, Aurelien and {Schneider}, Aurel and {Refregier}, Alexandre and {Hofmann}, Thomas},
        title = "{Full w CDM analysis of KiDS-1000 weak lensing maps using deep learning}",
      journal = {\prd},
         year = 2022,
        month = apr,
       volume = {105},
       number = {8},
          eid = {083518},
        pages = {083518},
          doi = {10.1103/PhysRevD.105.083518},
archivePrefix = {arXiv},
       eprint = {2201.07771},
 primaryClass = {astro-ph.CO},
       adsurl = {https://ui.adsabs.harvard.edu/abs/2022PhRvD.105h3518F}
}

@ARTICLE{Schneider2019_code,
       author = {{Schneider}, Aurel and {Teyssier}, Romain and {Stadel}, Joachim and {Chisari}, Nora Elisa and {Le Brun}, Amandine M.~C. and {Amara}, Adam and {Refregier}, Alexandre},
        title = "{Quantifying baryon effects on the matter power spectrum and the weak lensing shear correlation}",
      journal = {\jcap},
         year = 2019,
        month = mar,
       volume = {2019},
       number = {3},
          eid = {020},
        pages = {020},
          doi = {10.1088/1475-7516/2019/03/020},
archivePrefix = {arXiv},
       eprint = {1810.08629},
 primaryClass = {astro-ph.CO},
       adsurl = {https://ui.adsabs.harvard.edu/abs/2019JCAP...03..020S}
}

@ARTICLE{Schneider2015_code,
       author = {{Schneider}, Aurel and {Teyssier}, Romain},
        title = "{A new method to quantify the effects of baryons on the matter power spectrum}",
      journal = {\jcap},
         year = 2015,
        month = dec,
       volume = {2015},
       number = {12},
        pages = {049-049},
          doi = {10.1088/1475-7516/2015/12/049},
archivePrefix = {arXiv},
       eprint = {1510.06034},
 primaryClass = {astro-ph.CO},
       adsurl = {https://ui.adsabs.harvard.edu/abs/2015JCAP...12..049S}
}

@article{Schneider_2016_pkd,
    author = "Schneider, Aurel and Teyssier, Romain and Potter, Doug and Stadel, Joachim and Onions, Julian and Reed, Darren S. and Smith, Robert E. and Springel, Volker and Pearce, Frazer R. and Scoccimarro, Roman",
    title = "{Matter power spectrum and the challenge of percent accuracy}",
    eprint = "1503.05920",
    archivePrefix = "arXiv",
    primaryClass = "astro-ph.CO",
    doi = "10.1088/1475-7516/2016/04/047",
    journal = "JCAP",
    volume = "04",
    pages = "047",
    year = "2016"
}

@ARTICLE{Giri2021_BCemu,
       author = {{Giri}, Sambit K. and {Schneider}, Aurel},
        title = "{Emulation of baryonic effects on the matter power spectrum and constraints from galaxy cluster data}",
      journal = {\jcap},
         year = 2021,
        month = dec,
       volume = {2021},
       number = {12},
          eid = {046},
        pages = {046},
          doi = {10.1088/1475-7516/2021/12/046},
archivePrefix = {arXiv},
       eprint = {2108.08863},
 primaryClass = {astro-ph.CO},
       adsurl = {https://ui.adsabs.harvard.edu/abs/2021JCAP...12..046G}
}

@ARTICLE{Schneider_A_2025,
       author = {{Schneider}, Aurel and {Kova{\v{c}}}, Michael and {Bucko}, Jozef and {Nicola}, Andrina and {Reischke}, Robert and {Giri}, Sambit K. and {Teyssier}, Romain and {Tr{\"o}ster}, Tilman and {Refregier}, Alexandre and {Schaller}, Matthieu and {Schaye}, Joop},
        title = "{Baryonification: an alternative to hydrodynamical simulations for cosmological studies}",
      journal = {\jcap},
         year = 2025,
        month = dec,
       volume = {2025},
       number = {12},
          eid = {043},
        pages = {043},
          doi = {10.1088/1475-7516/2025/12/043},
archivePrefix = {arXiv},
       eprint = {2507.07892},
 primaryClass = {astro-ph.CO},
       adsurl = {https://ui.adsabs.harvard.edu/abs/2025JCAP...12..043S}
}

@ARTICLE{Schneider_2020,
       author = {{Schneider}, Aurel and {Refregier}, Alexandre and {Grandis}, Sebastian and {Eckert}, Dominique and {Stoira}, Nicola and {Kacprzak}, Tomasz and {Knabenhans}, Mischa and {Stadel}, Joachim and {Teyssier}, Romain},
        title = "{Baryonic effects for weak lensing. Part II. Combination with X-ray data and extended cosmologies}",
      journal = {\jcap},
         year = 2020,
        month = apr,
       volume = {2020},
       number = {4},
          eid = {020},
        pages = {020},
          doi = {10.1088/1475-7516/2020/04/020},
archivePrefix = {arXiv},
       eprint = {1911.08494},
 primaryClass = {astro-ph.CO},
       adsurl = {https://ui.adsabs.harvard.edu/abs/2020JCAP...04..020S}
}

@ARTICLE{Schneider2022_WL_xray_kSZ,
       author = {{Schneider}, Aurel and {Giri}, Sambit K. and {Amodeo}, Stefania and {Refregier}, Alexandre},
        title = "{Constraining baryonic feedback and cosmology with weak-lensing, X-ray, and kinematic Sunyaev-Zeldovich observations}",
      journal = {\mnras},
         year = 2022,
        month = aug,
       volume = {514},
       number = {3},
        pages = {3802-3814},
          doi = {10.1093/mnras/stac1493},
archivePrefix = {arXiv},
       eprint = {2110.02228},
 primaryClass = {astro-ph.CO},
       adsurl = {https://ui.adsabs.harvard.edu/abs/2022MNRAS.514.3802S}
}

@ARTICLE{Osato_K_2023,
       author = {{Osato}, Ken and {Nagai}, Daisuke},
        title = "{Baryon pasting algorithm: halo-based and particle-based pasting methods}",
      journal = {\mnras},
         year = 2023,
        month = feb,
       volume = {519},
       number = {2},
        pages = {2069-2082},
          doi = {10.1093/mnras/stac3669},
archivePrefix = {arXiv},
       eprint = {2201.02632},
 primaryClass = {astro-ph.CO},
       adsurl = {https://ui.adsabs.harvard.edu/abs/2023MNRAS.519.2069O}
}

@ARTICLE{Semboloni_2011,
       author = {{Semboloni}, Elisabetta and {Hoekstra}, Henk and {Schaye}, Joop and {van Daalen}, Marcel P. and {McCarthy}, Ian G.},
        title = "{Quantifying the effect of baryon physics on weak lensing tomography}",
      journal = {\mnras},
         year = 2011,
        month = nov,
       volume = {417},
       number = {3},
        pages = {2020-2035},
          doi = {10.1111/j.1365-2966.2011.19385.x},
archivePrefix = {arXiv},
       eprint = {1105.1075},
 primaryClass = {astro-ph.CO},
       adsurl = {https://ui.adsabs.harvard.edu/abs/2011MNRAS.417.2020S}
}

@ARTICLE{Anbajagane_D_2024,
       author = {{Anbajagane}, Dhayaa and {Pandey}, Shivam and {Chang}, Chihway},
        title = "{Map-level baryonification: Efficient modelling of higher-order correlations in the weak lensing and thermal Sunyaev-Zeldovich fields}",
      journal = {The Open Journal of Astrophysics},
         year = 2024,
        month = dec,
       volume = {7},
          eid = {108},
        pages = {108},
          doi = {10.33232/001c.126788},
archivePrefix = {arXiv},
       eprint = {2409.03822},
 primaryClass = {astro-ph.CO},
       adsurl = {https://ui.adsabs.harvard.edu/abs/2024OJAp....7E.108A}
}

@ARTICLE{Zhou_A_2025,
       author = {{Zhou}, Alan Junzhe and {Gatti}, Marco and {Anbajagane}, Dhayaa and {Dodelson}, Scott and {Schaller}, Matthieu and {Schaye}, Joop},
        title = "{Map-level baryonification: unified treatment of weak lensing two-point and higher-order statistics}",
      journal = {\jcap},
         year = 2025,
        month = sep,
       volume = {2025},
       number = {9},
          eid = {073},
        pages = {073},
          doi = {10.1088/1475-7516/2025/09/073},
archivePrefix = {arXiv},
       eprint = {2505.07949},
 primaryClass = {astro-ph.CO},
       adsurl = {https://ui.adsabs.harvard.edu/abs/2025JCAP...09..073Z}
}

@article{Arico_2020_BCM,
    author = "Aric{\`o}, Giovanni and Angulo, Raul E. and Hern{\'a}ndez-Monteagudo, Carlos and Contreras, Sergio and Zennaro, Matteo",
    title = "{Simultaneous modelling of matter power spectrum and bispectrum in the presence of baryons}",
    eprint = "2009.14225",
    archivePrefix = "arXiv",
    primaryClass = "astro-ph.CO",
    doi = "10.1093/mnras/stab699",
    journal = "Mon. Not. Roy. Astron. Soc.",
    volume = "503",
    number = "3",
    pages = "3596--3609",
    year = "2021"
}

@article{Arico_2020_BACCO,
    author = "Aric{\`o}, Giovanni and Angulo, Raul E. and Contreras, Sergio and Ondaro-Mallea, Lurdes and Pellejero-Iba{\~n}ez, Marcos and Zennaro, Matteo",
    title = "{The BACCO simulation project: a baryonification emulator with neural networks}",
    eprint = "2011.15018",
    archivePrefix = "arXiv",
    primaryClass = "astro-ph.CO",
    doi = "10.1093/mnras/stab1911",
    journal = "Mon. Not. Roy. Astron. Soc.",
    volume = "506",
    number = "3",
    pages = "4070--4082",
    year = "2021"
}

@ARTICLE{Arico_G_2024,
       author = {{Aric{\`o}}, Giovanni and {Angulo}, Raul E.},
        title = "{Baryonification extended to thermal Sunyaev Zel'dovich}",
      journal = {\aap},
         year = 2024,
        month = oct,
       volume = {690},
          eid = {A188},
        pages = {A188},
          doi = {10.1051/0004-6361/202451055},
archivePrefix = {arXiv},
       eprint = {2406.01672},
 primaryClass = {astro-ph.CO},
       adsurl = {https://ui.adsabs.harvard.edu/abs/2024A&A...690A.188A}
}

@ARTICLE{Shaw_L_2010,
       author = {{Shaw}, Laurie D. and {Nagai}, Daisuke and {Bhattacharya}, Suman and {Lau}, Erwin T.},
        title = "{Impact of Cluster Physics on the Sunyaev-Zel'dovich Power Spectrum}",
      journal = {\apj},
         year = 2010,
        month = dec,
       volume = {725},
       number = {2},
        pages = {1452-1465},
          doi = {10.1088/0004-637X/725/2/1452},
archivePrefix = {arXiv},
       eprint = {1006.1945},
 primaryClass = {astro-ph.CO},
       adsurl = {https://ui.adsabs.harvard.edu/abs/2010ApJ...725.1452S}
}

@ARTICLE{Schaye_J_2023,
       author = {{Schaye}, Joop and {Kugel}, Roi and {Schaller}, Matthieu and {Helly}, John C. and {Braspenning}, Joey and {Elbers}, Willem and {McCarthy}, Ian G. and {van Daalen}, Marcel P. and {Vandenbroucke}, Bert and {Frenk}, Carlos S. and {Kwan}, Juliana and {Salcido}, Jaime and {Bah{\'e}}, Yannick M. and {Borrow}, Josh and {Chaikin}, Evgenii and {Hahn}, Oliver and {Hu{\v{s}}ko}, Filip and {Jenkins}, Adrian and {Lacey}, Cedric G. and {Nobels}, Folkert S.~J.},
        title = "{The FLAMINGO project: cosmological hydrodynamical simulations for large-scale structure and galaxy cluster surveys}",
      journal = {\mnras},
         year = 2023,
        month = dec,
       volume = {526},
       number = {4},
        pages = {4978-5020},
          doi = {10.1093/mnras/stad2419},
archivePrefix = {arXiv},
       eprint = {2306.04024},
 primaryClass = {astro-ph.CO},
       adsurl = {https://ui.adsabs.harvard.edu/abs/2023MNRAS.526.4978S}
}

@ARTICLE{Chisari_E_2019,
       author = {{Chisari}, Nora Elisa and {Alonso}, David and {Krause}, Elisabeth and {Leonard}, C. Danielle and {Bull}, Philip and {Neveu}, J{\'e}r{\'e}my and {Villarreal}, Antonia Sierra and {Singh}, Sukhdeep and {McClintock}, Thomas and {Ellison}, John and {Du}, Zilong and {Zuntz}, Joe and {Mead}, Alexander and {Joudaki}, Shahab and {Lorenz}, Christiane S. and {Tr{\"o}ster}, Tilman and {Sanchez}, Javier and {Lanusse}, Francois and {Ishak}, Mustapha and {Hlozek}, Ren{\'e}e and {Blazek}, Jonathan and {Campagne}, Jean-Eric and {Almoubayyed}, Husni and {Eifler}, Tim and {Kirby}, Matthew and {Kirkby}, David and {Plaszczynski}, St{\'e}phane and {Slosar}, An{\v{z}}e and {Vrastil}, Michal and {Wagoner}, Erika L. and {LSST Dark Energy Science Collaboration}},
        title = "{Core Cosmology Library: Precision Cosmological Predictions for LSST}",
      journal = {\apjs},
         year = 2019,
        month = may,
       volume = {242},
       number = {1},
          eid = {2},
        pages = {2},
          doi = {10.3847/1538-4365/ab1658},
archivePrefix = {arXiv},
       eprint = {1812.05995},
 primaryClass = {astro-ph.CO},
       adsurl = {https://ui.adsabs.harvard.edu/abs/2019ApJS..242....2C}
}

@ARTICLE{Kacprzak_T_2023,
       author = {{Kacprzak}, Tomasz and {Fluri}, Janis and {Schneider}, Aurel and {Refregier}, Alexandre and {Stadel}, Joachim},
        title = "{CosmoGridV1: a simulated $\Lambda$CDM theory prediction for map-level cosmological inference}",
      journal = {\jcap},
         year = 2023,
        month = feb,
       volume = {2023},
       number = {2},
          eid = {050},
        pages = {050},
          doi = {10.1088/1475-7516/2023/02/050},
archivePrefix = {arXiv},
       eprint = {2209.04662},
 primaryClass = {astro-ph.CO},
       adsurl = {https://ui.adsabs.harvard.edu/abs/2023JCAP...02..050K}
}

@ARTICLE{Potter_D_2017,
       author = {{Potter}, Douglas and {Stadel}, Joachim and {Teyssier}, Romain},
        title = "{PKDGRAV3: beyond trillion particle cosmological simulations for the next era of galaxy surveys}",
      journal = {Computational Astrophysics and Cosmology},
         year = 2017,
        month = may,
       volume = {4},
       number = {1},
          eid = {2},
        pages = {2},
          doi = {10.1186/s40668-017-0021-1},
archivePrefix = {arXiv},
       eprint = {1609.08621},
 primaryClass = {astro-ph.IM},
       adsurl = {https://ui.adsabs.harvard.edu/abs/2017ComAC...4....2P}
}

@ARTICLE{van_Daalen_2011,
       author = {{van Daalen}, Marcel P. and {Schaye}, Joop and {Booth}, C.~M. and {Dalla Vecchia}, Claudio},
        title = "{The effects of galaxy formation on the matter power spectrum: a challenge for precision cosmology}",
      journal = {\mnras},
         year = 2011,
        month = aug,
       volume = {415},
       number = {4},
        pages = {3649-3665},
          doi = {10.1111/j.1365-2966.2011.18981.x},
archivePrefix = {arXiv},
       eprint = {1104.1174},
 primaryClass = {astro-ph.CO},
       adsurl = {https://ui.adsabs.harvard.edu/abs/2011MNRAS.415.3649V}
}

@ARTICLE{Chisari_2019,
       author = {{Chisari}, Nora Elisa and {Mead}, Alexander J. and {Joudaki}, Shahab and {Ferreira}, Pedro G. and {Schneider}, Aurel and {Mohr}, Joseph and {Tr{\"o}ster}, Tilman and {Alonso}, David and {McCarthy}, Ian G. and {Martin-Alvarez}, Sergio and {Devriendt}, Julien and {Slyz}, Adrianne and {van Daalen}, Marcel P.},
        title = "{Modelling baryonic feedback for survey cosmology}",
      journal = {The Open Journal of Astrophysics},
         year = 2019,
        month = jun,
       volume = {2},
       number = {1},
          eid = {4},
        pages = {4},
          doi = {10.21105/astro.1905.06082},
archivePrefix = {arXiv},
       eprint = {1905.06082},
 primaryClass = {astro-ph.CO},
       adsurl = {https://ui.adsabs.harvard.edu/abs/2019OJAp....2E...4C}
}

@ARTICLE{van_Daalen_2020,
       author = {{van Daalen}, Marcel P. and {McCarthy}, Ian G. and {Schaye}, Joop},
        title = "{Exploring the effects of galaxy formation on matter clustering through a library of simulation power spectra}",
      journal = {\mnras},
         year = 2020,
        month = jan,
       volume = {491},
       number = {2},
        pages = {2424-2446},
          doi = {10.1093/mnras/stz3199},
archivePrefix = {arXiv},
       eprint = {1906.00968},
 primaryClass = {astro-ph.CO},
       adsurl = {https://ui.adsabs.harvard.edu/abs/2020MNRAS.491.2424V}
}

@article{Springel_2017,
    author = "Springel, Volker and others",
    title = "{First results from the IllustrisTNG simulations: matter and galaxy clustering}",
    eprint = "1707.03397",
    archivePrefix = "arXiv",
    primaryClass = "astro-ph.GA",
    doi = "10.1093/mnras/stx3304",
    journal = "Mon. Not. Roy. Astron. Soc.",
    volume = "475",
    number = "1",
    pages = "676--698",
    year = "2018"
}

@article{Mccarthy_2016bahamas,
    author = "McCarthy, Ian G. and Schaye, Joop and Bird, Simeon and Le Brun, Amandine M. C.",
    title = "{The BAHAMAS project: Calibrated hydrodynamical simulations for large-scale structure cosmology}",
    eprint = "1603.02702",
    archivePrefix = "arXiv",
    primaryClass = "astro-ph.CO",
    doi = "10.1093/mnras/stw2792",
    journal = "Mon. Not. Roy. Astron. Soc.",
    volume = "465",
    number = "3",
    pages = "2936--2965",
    year = "2017"
}

@article{Dolag2016sz,
    author = "Dolag, Klaus and Komatsu, Eiichiro and Sunyaev, Rashid",
    title = "{SZ effects in the Magneticum Pathfinder Simulation: Comparison with the Planck, SPT, and ACT results}",
    eprint = "1509.05134",
    archivePrefix = "arXiv",
    primaryClass = "astro-ph.CO",
    doi = "10.1093/mnras/stw2035",
    journal = "Mon. Not. Roy. Astron. Soc.",
    volume = "463",
    number = "2",
    pages = "1797--1811",
    year = "2016"
}

@article{leBrun2014owls,
    author = "Le Brun, Amandine M. C. and McCarthy, Ian G. and Schaye, Joop and Ponman, Trevor J.",
    title = "{Towards a realistic population of simulated galaxy groups and clusters}",
    eprint = "1312.5462",
    archivePrefix = "arXiv",
    primaryClass = "astro-ph.CO",
    doi = "10.1093/mnras/stu608",
    journal = "Mon. Not. Roy. Astron. Soc.",
    volume = "441",
    number = "2",
    pages = "1270--1290",
    year = "2014"
}

@ARTICLE{Dubois_HorizonAGN_2014,
       author = {{Dubois}, Y. and {Pichon}, C. and {Welker}, C. and {Le Borgne}, D. and {Devriendt}, J. and {Laigle}, C. and {Codis}, S. and {Pogosyan}, D. and {Arnouts}, S. and {Benabed}, K. and {Bertin}, E. and {Blaizot}, J. and {Bouchet}, F. and {Cardoso}, J. -F. and {Colombi}, S. and {de Lapparent}, V. and {Desjacques}, V. and {Gavazzi}, R. and {Kassin}, S. and {Kimm}, T. and {McCracken}, H. and {Milliard}, B. and {Peirani}, S. and {Prunet}, S. and {Rouberol}, S. and {Silk}, J. and {Slyz}, A. and {Sousbie}, T. and {Teyssier}, R. and {Tresse}, L. and {Treyer}, M. and {Vibert}, D. and {Volonteri}, M.},
        title = "{Dancing in the dark: galactic properties trace spin swings along the cosmic web}",
      journal = {\mnras},
         year = 2014,
        month = oct,
       volume = {444},
       number = {2},
        pages = {1453-1468},
          doi = {10.1093/mnras/stu1227},
archivePrefix = {arXiv},
       eprint = {1402.1165},
 primaryClass = {astro-ph.CO},
       adsurl = {https://ui.adsabs.harvard.edu/abs/2014MNRAS.444.1453D}
}

@ARTICLE{Kugel_2023,
       author = {{Kugel}, Roi and {Schaye}, Joop and {Schaller}, Matthieu and {Helly}, John C. and {Braspenning}, Joey and {Elbers}, Willem and {Frenk}, Carlos S. and {McCarthy}, Ian G. and {Kwan}, Juliana and {Salcido}, Jaime and {van Daalen}, Marcel P. and {Vandenbroucke}, Bert and {Bah{\'e}}, Yannick M. and {Borrow}, Josh and {Chaikin}, Evgenii and {Hu{\v{s}}ko}, Filip and {Jenkins}, Adrian and {Lacey}, Cedric G. and {Nobels}, Folkert S.~J. and {Vernon}, Ian},
        title = "{FLAMINGO: calibrating large cosmological hydrodynamical simulations with machine learning}",
      journal = {\mnras},
         year = 2023,
        month = dec,
       volume = {526},
       number = {4},
        pages = {6103-6127},
          doi = {10.1093/mnras/stad2540},
archivePrefix = {arXiv},
       eprint = {2306.05492},
 primaryClass = {astro-ph.CO},
       adsurl = {https://ui.adsabs.harvard.edu/abs/2023MNRAS.526.6103K}
}

@article{Mccarthy_2018,
    author = "McCarthy, Ian G. and Bird, Simeon and Schaye, Joop and Harnois-Deraps, Joachim and Font, Andreea S. and Van Waerbeke, Ludovic",
    title = "{The BAHAMAS project: the CMB{\textendash}large-scale structure tension and the roles of massive neutrinos and galaxy formation}",
    eprint = "1712.02411",
    archivePrefix = "arXiv",
    primaryClass = "astro-ph.CO",
    doi = "10.1093/mnras/sty377",
    journal = "Mon. Not. Roy. Astron. Soc.",
    volume = "476",
    number = "3",
    pages = "2999--3030",
    year = "2018"
}

@ARTICLE{Bigwood2024,
       author = {{Bigwood}, L. and {Amon}, A. and {Schneider}, A. and {Salcido}, J. and {McCarthy}, I.~G. and {Preston}, C. and {Sanchez}, D. and {Sijacki}, D. and {Schaan}, E. and {Ferraro}, S. and {Battaglia}, N. and {Chen}, A. and {Dodelson}, S. and {Roodman}, A. and {Pieres}, A. and {Fert{\'e}}, A. and {Alarcon}, A. and {Drlica-Wagner}, A. and {Choi}, A. and {Navarro-Alsina}, A. and {Campos}, A. and {Ross}, A.~J. and {Carnero Rosell}, A. and {Yin}, B. and {Yanny}, B. and {S{\'a}nchez}, C. and {Chang}, C. and {Davis}, C. and {Doux}, C. and {Gruen}, D. and {Rykoff}, E.~S. and {Huff}, E.~M. and {Sheldon}, E. and {Tarsitano}, F. and {Andrade-Oliveira}, F. and {Bernstein}, G.~M. and {Giannini}, G. and {Diehl}, H.~T. and {Huang}, H. and {Harrison}, I. and {Sevilla-Noarbe}, I. and {Tutusaus}, I. and {Elvin-Poole}, J. and {McCullough}, J. and {Zuntz}, J. and {Blazek}, J. and {DeRose}, J. and {Cordero}, J. and {Prat}, J. and {Myles}, J. and {Eckert}, K. and {Bechtol}, K. and {Herner}, K. and {Secco}, L.~F. and {Gatti}, M. and {Raveri}, M. and {Kind}, M. Carrasco and {Becker}, M.~R. and {Troxel}, M.~A. and {Jarvis}, M. and {MacCrann}, N. and {Friedrich}, O. and {Alves}, O. and {Leget}, P. -F. and {Chen}, R. and {Rollins}, R.~P. and {Wechsler}, R.~H. and {Gruendl}, R.~A. and {Cawthon}, R. and {Allam}, S. and {Bridle}, S.~L. and {Pandey}, S. and {Everett}, S. and {Shin}, T. and {Hartley}, W.~G. and {Fang}, X. and {Zhang}, Y. and {Aguena}, M. and {Annis}, J. and {Bacon}, D. and {Bertin}, E. and {Bocquet}, S. and {Brooks}, D. and {Carretero}, J. and {Castander}, F.~J. and {da Costa}, L.~N. and {Pereira}, M.~E.~S. and {De Vicente}, J. and {Desai}, S. and {Doel}, P. and {Ferrero}, I. and {Flaugher}, B. and {Frieman}, J. and {Garc{\'\i}a-Bellido}, J. and {Gaztanaga}, E. and {Gutierrez}, G. and {Hinton}, S.~R. and {Hollowood}, D.~L. and {Honscheid}, K. and {Huterer}, D. and {James}, D.~J. and {Kuehn}, K. and {Lahav}, O. and {Lee}, S. and {Marshall}, J.~L. and {Mena-Fern{\'a}ndez}, J. and {Miquel}, R. and {Muir}, J. and {Paterno}, M. and {Plazas Malag{\'o}n}, A.~A. and {Porredon}, A. and {Romer}, A.~K. and {Samuroff}, S. and {Sanchez}, E. and {Sanchez Cid}, D. and {Smith}, M. and {Soares-Santos}, M. and {Suchyta}, E. and {Swanson}, M.~E.~C. and {Tarle}, G. and {To}, C. and {Weaverdyck}, N. and {Weller}, J. and {Wiseman}, P. and {Yamamoto}, M.},
        title = "{Weak lensing combined with the kinetic Sunyaev-Zel'dovich effect: a study of baryonic feedback}",
      journal = {\mnras},
         year = 2024,
        month = oct,
       volume = {534},
       number = {1},
        pages = {655-682},
          doi = {10.1093/mnras/stae2100},
archivePrefix = {arXiv},
       eprint = {2404.06098},
 primaryClass = {astro-ph.CO},
       adsurl = {https://ui.adsabs.harvard.edu/abs/2024MNRAS.534..655B}
}

@ARTICLE{Grandis2023,
       author = {{Grandis}, Sebastian and {Aric{\`o}}, Giovanni and {Schneider}, Aurel and {Linke}, Laila},
        title = "{Determining the baryon impact on the matter power spectrum with galaxy clusters}",
      journal = {\mnras},
         year = 2024,
        month = mar,
       volume = {528},
       number = {3},
        pages = {4379-4392},
          doi = {10.1093/mnras/stae259},
archivePrefix = {arXiv},
       eprint = {2309.02920},
 primaryClass = {astro-ph.CO},
       adsurl = {https://ui.adsabs.harvard.edu/abs/2024MNRAS.528.4379G}
}

@ARTICLE{Kovac_2025,
       author = {{Kova{\v{c}}}, Michael and {Nicola}, Andrina and {Bucko}, Jozef and {Schneider}, Aurel and {Reischke}, Robert and {Giri}, Sambit K. and {Teyssier}, Romain and {Schaller}, Matthieu and {Schaye}, Joop},
        title = "{Baryonification II: constraining feedback with X-ray and kinematic Sunyaev-Zel'dovich observations}",
      journal = {\jcap},
         year = 2025,
        month = nov,
       volume = {2025},
       number = {11},
          eid = {046},
        pages = {046},
          doi = {10.1088/1475-7516/2025/11/046},
archivePrefix = {arXiv},
       eprint = {2507.07991},
 primaryClass = {astro-ph.CO},
       adsurl = {https://ui.adsabs.harvard.edu/abs/2025JCAP...11..046K}
}

@ARTICLE{Navarro_1996,
       author = {{Navarro}, Julio F. and {Frenk}, Carlos S. and {White}, Simon D.~M.},
        title = "{The Structure of Cold Dark Matter Halos}",
      journal = {\apj},
         year = 1996,
        month = may,
       volume = {462},
        pages = {563},
          doi = {10.1086/177173},
archivePrefix = {arXiv},
       eprint = {astro-ph/9508025},
 primaryClass = {astro-ph},
       adsurl = {https://ui.adsabs.harvard.edu/abs/1996ApJ...462..563N}
}

@ARTICLE{Navarro_1997,
       author = {{Navarro}, Julio F. and {Frenk}, Carlos S. and {White}, Simon D.~M.},
        title = "{A Universal Density Profile from Hierarchical Clustering}",
      journal = {\apj},
         year = 1997,
        month = dec,
       volume = {490},
       number = {2},
        pages = {493-508},
          doi = {10.1086/304888},
archivePrefix = {arXiv},
       eprint = {astro-ph/9611107},
 primaryClass = {astro-ph},
       adsurl = {https://ui.adsabs.harvard.edu/abs/1997ApJ...490..493N}
}

@ARTICLE{Baltz_2009,
       author = {{Baltz}, Edward A. and {Marshall}, Phil and {Oguri}, Masamune},
        title = "{Analytic models of plausible gravitational lens potentials}",
      journal = {\jcap},
         year = 2009,
        month = jan,
       volume = {2009},
       number = {1},
          eid = {015},
        pages = {015},
          doi = {10.1088/1475-7516/2009/01/015},
archivePrefix = {arXiv},
       eprint = {0705.0682},
 primaryClass = {astro-ph},
       adsurl = {https://ui.adsabs.harvard.edu/abs/2009JCAP...01..015B}
}

@ARTICLE{Angelinelli_2020,
       author = {{Angelinelli}, M. and {Vazza}, F. and {Giocoli}, C. and {Ettori}, S. and {Jones}, T.~W. and {Brunetti}, G. and {Br{\"u}ggen}, M. and {Eckert}, D.},
        title = "{Turbulent pressure support and hydrostatic mass bias in the intracluster medium}",
      journal = {\mnras},
         year = 2020,
        month = jun,
       volume = {495},
       number = {1},
        pages = {864-885},
          doi = {10.1093/mnras/staa975},
archivePrefix = {arXiv},
       eprint = {1905.04896},
 primaryClass = {astro-ph.CO},
       adsurl = {https://ui.adsabs.harvard.edu/abs/2020MNRAS.495..864A}
}

@ARTICLE{Nelson_2014,
       author = {{Nelson}, Kaylea and {Lau}, Erwin T. and {Nagai}, Daisuke},
        title = "{Hydrodynamic Simulation of Non-thermal Pressure Profiles of Galaxy Clusters}",
      journal = {\apj},
         year = 2014,
        month = sep,
       volume = {792},
       number = {1},
          eid = {25},
        pages = {25},
          doi = {10.1088/0004-637X/792/1/25},
archivePrefix = {arXiv},
       eprint = {1404.4636},
 primaryClass = {astro-ph.CO},
       adsurl = {https://ui.adsabs.harvard.edu/abs/2014ApJ...792...25N}
}

@ARTICLE{Lau_2009,
       author = {{Lau}, Erwin T. and {Kravtsov}, Andrey V. and {Nagai}, Daisuke},
        title = "{Residual Gas Motions in the Intracluster Medium and Bias in Hydrostatic Measurements of Mass Profiles of Clusters}",
      journal = {\apj},
         year = 2009,
        month = nov,
       volume = {705},
       number = {2},
        pages = {1129-1138},
          doi = {10.1088/0004-637X/705/2/1129},
archivePrefix = {arXiv},
       eprint = {0903.4895},
 primaryClass = {astro-ph.CO},
       adsurl = {https://ui.adsabs.harvard.edu/abs/2009ApJ...705.1129L}
}

@ARTICLE{Battaglia_2012,
       author = {{Battaglia}, N. and {Bond}, J.~R. and {Pfrommer}, C. and {Sievers}, J.~L.},
        title = "{On the Cluster Physics of Sunyaev-Zel'dovich and X-Ray Surveys. I. The Influence of Feedback, Non-thermal Pressure, and Cluster Shapes on Y-M Scaling Relations}",
      journal = {\apj},
         year = 2012,
        month = oct,
       volume = {758},
       number = {2},
          eid = {74},
        pages = {74},
          doi = {10.1088/0004-637X/758/2/74},
archivePrefix = {arXiv},
       eprint = {1109.3709},
 primaryClass = {astro-ph.CO},
       adsurl = {https://ui.adsabs.harvard.edu/abs/2012ApJ...758...74B}
}

@ARTICLE{Jeffrey_2024,
       author = {{Jeffrey}, N. and {Whiteway}, L. and {Gatti}, M. and {Williamson}, J. and {Alsing}, J. and {Porredon}, A. and {Prat}, J. and {Doux}, C. and {Jain}, B. and {Chang}, C. and {Cheng}, T.-Y. and {Kacprzak}, T. and {Lemos}, P. and {Alarcon}, A. and {Amon}, A. and {Bechtol}, K. and {Becker}, M.~R. and {Bernstein}, G.~M. and {Campos}, A. and {Carnero Rosell}, A. and {Chen}, R. and {Choi}, A. and {DeRose}, J. and {Drlica-Wagner}, A. and {Eckert}, K. and {Everett}, S. and {Fert{\'e}}, A. and {Gruen}, D. and {Gruendl}, R.~A. and {Herner}, K. and {Jarvis}, M. and {McCullough}, J. and {Myles}, J. and {Navarro-Alsina}, A. and {Pandey}, S. and {Raveri}, M. and {Rollins}, R.~P. and {Rykoff}, E.~S. and {S{\'a}nchez}, C. and {Secco}, L.~F. and {Sevilla-Noarbe}, I. and {Sheldon}, E. and {Shin}, T. and {Troxel}, M.~A. and {Tutusaus}, I. and {Varga}, T.~N. and {Yanny}, B. and {Yin}, B. and {Zuntz}, J. and {Aguena}, M. and {Allam}, S.~S. and {Alves}, O. and {Bacon}, D. and {Bocquet}, S. and {Brooks}, D. and {da Costa}, L.~N. and {Davis}, T.~M. and {De Vicente}, J. and {Desai}, S. and {Diehl}, H.~T. and {Ferrero}, I. and {Frieman}, J. and {Garc{\'\i}a-Bellido}, J. and {Gaztanaga}, E. and {Giannini}, G. and {Gutierrez}, G. and {Hinton}, S.~R. and {Hollowood}, D.~L. and {Honscheid}, K. and {Huterer}, D. and {James}, D.~J. and {Lahav}, O. and {Lee}, S. and {Marshall}, J.~L. and {Mena-Fern{\'a}ndez}, J. and {Miquel}, R. and {Pieres}, A. and {Plazas Malag{\'o}n}, A.~A. and {Roodman}, A. and {Sako}, M. and {Sanchez}, E. and {Sanchez Cid}, D. and {Smith}, M. and {Suchyta}, E. and {Swanson}, M.~E.~C. and {Tarle}, G. and {Tucker}, D.~L. and {Weaverdyck}, N. and {Weller}, J. and {Wiseman}, P. and {Yamamoto}, M.},
        title = "{Dark energy survey year 3 results: likelihood-free, simulation-based wCDM inference with neural compression of weak-lensing map statistics}",
      journal = {\mnras},
         year = 2025,
        month = jan,
       volume = {536},
       number = {2},
        pages = {1303-1322},
          doi = {10.1093/mnras/stae2629},
archivePrefix = {arXiv},
       eprint = {2403.02314},
 primaryClass = {astro-ph.CO},
       adsurl = {https://ui.adsabs.harvard.edu/abs/2025MNRAS.536.1303J}
}

@ARTICLE{Bucko_2026,
       author = {{Bucko}, Jozef and {Nicola}, Andrina and {Schneider}, Aurel and {Kova{\v{c}}}, Michael and {Giri}, Sambit K. and {Reischke}, Robert and {Bulbul}, Esra and {Clerc}, Nicolas and {Liu}, Ang and {Bartalucci}, Iacopo and {Refregier}, Alexandre},
        title = "{Baryonification IV: Constraining baryonic feedback with X-ray gas fractions}",
      journal = {arXiv e-prints},
         year = 2026,
        month = sep,
          eid = {arXiv:2609.09144},
        pages = {arXiv:2609.09144},
archivePrefix = {arXiv},
       eprint = {2609.09144},
 primaryClass = {astro-ph.CO},
       adsurl = {https://ui.adsabs.harvard.edu/abs/2026arXiv260909144B}
}

@ARTICLE{Tinker_2008,
       author = {{Tinker}, Jeremy and {Kravtsov}, Andrey V. and {Klypin}, Anatoly and {Abazajian}, Kevork and {Warren}, Michael and {Yepes}, Gustavo and {Gottl{\"o}ber}, Stefan and {Holz}, Daniel E.},
        title = "{Toward a Halo Mass Function for Precision Cosmology: The Limits of Universality}",
      journal = {\apj},
         year = 2008,
        month = dec,
       volume = {688},
       number = {2},
        pages = {709-728},
          doi = {10.1086/591439},
archivePrefix = {arXiv},
       eprint = {0803.2706},
 primaryClass = {astro-ph},
       adsurl = {https://ui.adsabs.harvard.edu/abs/2008ApJ...688..709T}
}

@ARTICLE{Tinker_2010,
       author = {{Tinker}, Jeremy L. and {Robertson}, Brant E. and {Kravtsov}, Andrey V. and {Klypin}, Anatoly and {Warren}, Michael S. and {Yepes}, Gustavo and {Gottl{\"o}ber}, Stefan},
        title = "{The Large-scale Bias of Dark Matter Halos: Numerical Calibration and Model Tests}",
      journal = {\apj},
         year = 2010,
        month = dec,
       volume = {724},
       number = {2},
        pages = {878-886},
          doi = {10.1088/0004-637X/724/2/878},
archivePrefix = {arXiv},
       eprint = {1001.3162},
 primaryClass = {astro-ph.CO},
       adsurl = {https://ui.adsabs.harvard.edu/abs/2010ApJ...724..878T}
}

@ARTICLE{Bocquet_2016,
       author = {{Bocquet}, Sebastian and {Saro}, Alex and {Dolag}, Klaus and {Mohr}, Joseph J.},
        title = "{Halo mass function: baryon impact, fitting formulae, and implications for cluster cosmology}",
      journal = {\mnras},
         year = 2016,
        month = mar,
       volume = {456},
       number = {3},
        pages = {2361-2373},
          doi = {10.1093/mnras/stv2657},
archivePrefix = {arXiv},
       eprint = {1502.07357},
 primaryClass = {astro-ph.CO},
       adsurl = {https://ui.adsabs.harvard.edu/abs/2016MNRAS.456.2361B}
}

@ARTICLE{Despali_2016,
       author = {{Despali}, Giulia and {Giocoli}, Carlo and {Angulo}, Raul E. and {Tormen}, Giuseppe and {Sheth}, Ravi K. and {Baso}, Giacomo and {Moscardini}, Lauro},
        title = "{The universality of the virial halo mass function and models for non-universality of other halo definitions}",
      journal = {\mnras},
         year = 2016,
        month = mar,
       volume = {456},
       number = {3},
        pages = {2486-2504},
          doi = {10.1093/mnras/stv2842},
archivePrefix = {arXiv},
       eprint = {1507.05627},
 primaryClass = {astro-ph.CO},
       adsurl = {https://ui.adsabs.harvard.edu/abs/2016MNRAS.456.2486D}
}

@ARTICLE{Bocquet_2020,
       author = {{Bocquet}, Sebastian and {Heitmann}, Katrin and {Habib}, Salman and {Lawrence}, Earl and {Uram}, Thomas and {Frontiere}, Nicholas and {Pope}, Adrian and {Finkel}, Hal},
        title = "{The Mira-Titan Universe. III. Emulation of the Halo Mass Function}",
      journal = {\apj},
         year = 2020,
        month = sep,
       volume = {901},
       number = {1},
          eid = {5},
        pages = {5},
          doi = {10.3847/1538-4357/abac5c},
archivePrefix = {arXiv},
       eprint = {2003.12116},
 primaryClass = {astro-ph.CO},
       adsurl = {https://ui.adsabs.harvard.edu/abs/2020ApJ...901....5B}
}

@ARTICLE{AHF,
       author = {{Knollmann}, Steffen R. and {Knebe}, Alexander},
        title = "{AHF: Amiga's Halo Finder}",
      journal = {\apjs},
         year = 2009,
        month = jun,
       volume = {182},
       number = {2},
        pages = {608-624},
          doi = {10.1088/0067-0049/182/2/608},
archivePrefix = {arXiv},
       eprint = {0904.3662},
 primaryClass = {astro-ph.CO},
       adsurl = {https://ui.adsabs.harvard.edu/abs/2009ApJS..182..608K}
}

@ARTICLE{Forouhar_2025,
       author = {{Forouhar Moreno}, Victor J. and {Helly}, John and {McGibbon}, Robert and {Schaye}, Joop and {Schaller}, Matthieu and {Han}, Jiaxin and {Kugel}, Roi and {Bah{\'e}}, Yannick M.},
        title = "{Assessing subhalo finders in cosmological hydrodynamical simulations}",
      journal = {\mnras},
         year = 2025,
        month = oct,
       volume = {543},
       number = {2},
        pages = {1339-1372},
          doi = {10.1093/mnras/staf1478},
archivePrefix = {arXiv},
       eprint = {2502.06932},
 primaryClass = {astro-ph.CO},
       adsurl = {https://ui.adsabs.harvard.edu/abs/2025MNRAS.543.1339F}
}

@ARTICLE{Mead_2021,
       author = {{Mead}, A.~J. and {Verde}, L.},
        title = "{Including beyond-linear halo bias in halo models}",
      journal = {\mnras},
         year = 2021,
        month = may,
       volume = {503},
       number = {2},
        pages = {3095-3111},
          doi = {10.1093/mnras/stab748},
archivePrefix = {arXiv},
       eprint = {2011.08858},
 primaryClass = {astro-ph.CO},
       adsurl = {https://ui.adsabs.harvard.edu/abs/2021MNRAS.503.3095M}
}

@ARTICLE{Asgari_2023,
       author = {{Asgari}, Marika and {Mead}, Alexander J. and {Heymans}, Catherine},
        title = "{The halo model for cosmology: a pedagogical review}",
      journal = {The Open Journal of Astrophysics},
         year = 2023,
        month = nov,
       volume = {6},
          eid = {39},
        pages = {39},
          doi = {10.21105/astro.2303.08752},
archivePrefix = {arXiv},
       eprint = {2303.08752},
 primaryClass = {astro-ph.CO},
       adsurl = {https://ui.adsabs.harvard.edu/abs/2023OJAp....6E..39A}
}

@ARTICLE{Komatsu_2002,
       author = {{Komatsu}, E. and {Seljak}, U.},
        title = "{The Sunyaev-Zel'dovich angular power spectrum as a probe of cosmological parameters}",
      journal = {\mnras},
         year = 2002,
        month = nov,
       volume = {336},
       number = {4},
        pages = {1256-1270},
          doi = {10.1046/j.1365-8711.2002.05889.x},
archivePrefix = {arXiv},
       eprint = {astro-ph/0205468},
 primaryClass = {astro-ph},
       adsurl = {https://ui.adsabs.harvard.edu/abs/2002MNRAS.336.1256K}
}

@ARTICLE{Seljak_2000,
       author = {{Seljak}, Uro{\v{s}}},
        title = "{Analytic model for galaxy and dark matter clustering}",
      journal = {\mnras},
         year = 2000,
        month = oct,
       volume = {318},
       number = {1},
        pages = {203-213},
          doi = {10.1046/j.1365-8711.2000.03715.x},
archivePrefix = {arXiv},
       eprint = {astro-ph/0001493},
 primaryClass = {astro-ph},
       adsurl = {https://ui.adsabs.harvard.edu/abs/2000MNRAS.318..203S}
}

@ARTICLE{Peacock_2000,
       author = {{Peacock}, J.~A. and {Smith}, R.~E.},
        title = "{Halo occupation numbers and galaxy bias}",
      journal = {\mnras},
         year = 2000,
        month = nov,
       volume = {318},
       number = {4},
        pages = {1144-1156},
          doi = {10.1046/j.1365-8711.2000.03779.x},
archivePrefix = {arXiv},
       eprint = {astro-ph/0005010},
 primaryClass = {astro-ph},
       adsurl = {https://ui.adsabs.harvard.edu/abs/2000MNRAS.318.1144P}
}

@ARTICLE{Cooray_2002,
       author = {{Cooray}, Asantha and {Sheth}, Ravi},
        title = "{Halo models of large scale structure}",
      journal = {\physrep},
         year = 2002,
        month = dec,
       volume = {372},
       number = {1},
        pages = {1-129},
          doi = {10.1016/S0370-1573(02)00276-4},
archivePrefix = {arXiv},
       eprint = {astro-ph/0206508},
 primaryClass = {astro-ph},
       adsurl = {https://ui.adsabs.harvard.edu/abs/2002PhR...372....1C}
}

@ARTICLE{Oppenheimer:2018,
       author = {{Oppenheimer}, Benjamin D.},
        title = "{Deviations from hydrostatic equilibrium in the circumgalactic medium: spinning hot haloes and accelerating flows}",
      journal = {\mnras},
         year = 2018,
        month = nov,
       volume = {480},
       number = {3},
        pages = {2963-2975},
          doi = {10.1093/mnras/sty1918},
archivePrefix = {arXiv},
       eprint = {1801.00788},
 primaryClass = {astro-ph.GA},
       adsurl = {https://ui.adsabs.harvard.edu/abs/2018MNRAS.480.2963O}
}

@ARTICLE{Braspenning_2025,
       author = {{Braspenning}, Joey and {Schaye}, Joop and {Schaller}, Matthieu and {Kugel}, Roi and {Kay}, Scott T.},
        title = "{Hydrostatic mass bias for galaxy groups and clusters in the FLAMINGO simulations}",
      journal = {\mnras},
         year = 2025,
        month = feb,
       volume = {536},
       number = {4},
        pages = {3784-3802},
          doi = {10.1093/mnras/stae2798},
archivePrefix = {arXiv},
       eprint = {2409.07849},
 primaryClass = {astro-ph.CO},
       adsurl = {https://ui.adsabs.harvard.edu/abs/2025MNRAS.536.3784B}
}

@book{Dodelson_2003,
    title = {{Modern Cosmology}},
    year = {2003},
    booktitle = {Modern Cosmology},
    author = {Dodelson, Scott},
    doi = {10.1016/B978-0-12-219141-1.X5019-0}
}

@ARTICLE{Thomsen_2025,
       author = {{Thomsen}, A. and {Bucko}, J. and {Kacprzak}, T. and {Ajani}, V. and {Fluri}, J. and {Refregier}, A. and {Anbajagane}, D. and {Castander}, F.~J. and {Fert{\'e}}, A. and {Gatti}, M. and {Jeffrey}, N. and {Alarcon}, A. and {Amon}, A. and {Bechtol}, K. and {Becker}, M.~R. and {Bernstein}, G.~M. and {Campos}, A. and {Carnero Rosell}, A. and {Chang}, C. and {Chen}, R. and {Choi}, A. and {Crocce}, M. and {Davis}, C. and {DeRose}, J. and {Dodelson}, S. and {Doux}, C. and {Eckert}, K. and {Elvin-Poole}, J. and {Everett}, S. and {Fosalba}, P. and {Gruen}, D. and {Harrison}, I. and {Herner}, K. and {Huff}, E.~M. and {Jarvis}, M. and {Kuropatkin}, N. and {Leget}, P.-F. and {MacCrann}, N. and {McCullough}, J. and {Myles}, J. and {Navarro-Alsina}, A. and {Pandey}, S. and {Porredon}, A. and {Prat}, J. and {Raveri}, M. and {Rodriguez-Monroy}, M. and {Rollins}, R.~P. and {Roodman}, A. and {Rykoff}, E.~S. and {S{\'a}nchez}, C. and {Secco}, L.~F. and {Sheldon}, E. and {Shin}, T. and {Troxel}, M.~A. and {Tutusaus}, I. and {Varga}, T.~N. and {Weaverdyck}, N. and {Wechsler}, R.~H. and {Yanny}, B. and {Yin}, B. and {Zhang}, Y. and {Zuntz}, J. and {Aguena}, M. and {Allam}, S. and {Andrade-Oliveira}, F. and {Bacon}, D. and {Blazek}, J. and {Brooks}, D. and {Camilleri}, R. and {Carretero}, J. and {Cawthon}, R. and {da Costa}, L.~N. and {da Silva Pereira}, M.~E. and {Davis}, T.~M. and {De Vicente}, J. and {Desai}, S. and {Doel}, P. and {Garc{\'\i}a-Bellido}, J. and {Gutierrez}, G. and {Hinton}, S.~R. and {Hollowood}, D.~L. and {Honscheid}, K. and {James}, D.~J. and {Kuehn}, K. and {Lahav}, O. and {Lee}, S. and {Marshall}, J.~L. and {Mena-Fern{\'a}ndez}, J. and {Menanteau}, F. and {Miquel}, R. and {Muir}, J. and {Ogando}, R.~L.~C. and {Plazas Malag{\'o}n}, A.~A. and {Sanchez}, E. and {Sanchez Cid}, D. and {Sevilla-Noarbe}, I. and {Smith}, M. and {Suchyta}, E. and {Swanson}, M.~E.~C. and {Thomas}, D. and {To}, C. and {Tucker}, D.~L. and {DES Collaboration}},
        title = "{Dark Energy Survey Year 3 results: Simulation-based wCDM inference from weak lensing and galaxy clustering maps with deep learning: Analysis design}",
      journal = {\prd},
         year = 2026,
        month = apr,
       volume = {113},
       number = {8},
          eid = {083501},
        pages = {083501},
          doi = {10.1103/3sj1-1l9f},
archivePrefix = {arXiv},
       eprint = {2511.04681},
 primaryClass = {astro-ph.CO},
       adsurl = {https://ui.adsabs.harvard.edu/abs/2026PhRvD.113h3501T}
}

@ARTICLE{Efstathiou_McCarthy_25,
       author = {{Efstathiou}, George and {McCarthy}, Fiona},
        title = "{The power spectrum of the thermal Sunyaev-Zeldovich effect}",
      journal = {\mnras},
         year = 2025,
        month = jun,
       volume = {540},
       number = {1},
        pages = {1055-1068},
          doi = {10.1093/mnras/staf709},
archivePrefix = {arXiv},
       eprint = {2502.10232},
 primaryClass = {astro-ph.CO},
       adsurl = {https://ui.adsabs.harvard.edu/abs/2025MNRAS.540.1055E}
}

@ARTICLE{Rotti_Bolliet_Chluba_2021,
       author = {{Rotti}, Aditya and {Bolliet}, Boris and {Chluba}, Jens and {Remazeilles}, Mathieu},
        title = "{Removing the giants and learning from the crowd: A new SZ power spectrum method and revised Compton y-map analysis}",
      journal = {\mnras},
         year = 2021,
        month = jun,
       volume = {503},
       number = {4},
        pages = {5310-5328},
          doi = {10.1093/mnras/stab469},
archivePrefix = {arXiv},
       eprint = {2010.07797},
 primaryClass = {astro-ph.CO},
       adsurl = {https://ui.adsabs.harvard.edu/abs/2021MNRAS.503.5310R}
}

@ARTICLE{Pandey_2024,
       author = {{Pandey}, S. and {Hill}, J.~C. and {Alarcon}, A. and {Alves}, O. and {Amon}, A. and {Anbajagane}, D. and {Andrade-Oliveira}, F. and {Battaglia}, N. and {Baxter}, E. and {Bechtol}, K. and {Becker}, M.~R. and {Bernstein}, G.~M. and {Blazek}, J. and {Bridle}, S.~L. and {Calabrese}, E. and {Camacho}, H. and {Campos}, A. and {Carnero Rosell}, A. and {Carrasco Kind}, M. and {Cawthon}, R. and {Chang}, C. and {Chen}, R. and {Chintalapati}, P. and {Choi}, A. and {Cordero}, J. and {Coulton}, W. and {Crocce}, M. and {Davis}, C. and {DeRose}, J. and {Devlin}, M. and {Diehl}, H.~T. and {Dodelson}, S. and {Doux}, C. and {Drlica-Wagner}, A. and {Eckert}, K. and {Eifler}, T.~F. and {Elvin-Poole}, J. and {Everett}, S. and {Fang}, X. and {Fert{\'e}}, A. and {Fosalba}, P. and {Friedrich}, O. and {Gatti}, M. and {Gaztanaga}, E. and {Giannini}, G. and {Gluscevic}, V. and {Gruen}, D. and {Gruendl}, R.~A. and {Ried Guachalla}, B. and {Harrison}, I. and {Hartley}, W.~G. and {Herner}, K. and {Huang}, H. and {Huff}, E.~M. and {Huterer}, D. and {Jain}, B. and {Jarvis}, M. and {Krause}, E. and {Kuropatkin}, N. and {Kusiak}, A. and {Leget}, P. and {Lemos}, P. and {Liddle}, A.~R. and {Lokken}, M. and {MacCrann}, N. and {McCullough}, J. and {Moodley}, K. and {Muir}, J. and {Myles}, J. and {Navarro-Alsina}, A. and {Omori}, Y. and {Park}, Y. and {Partridge}, B. and {Porredon}, A. and {Prat}, J. and {Raveri}, M. and {Refregier}, A. and {Rollins}, R.~P. and {Roodman}, A. and {Rosenfeld}, R. and {Ross}, A.~J. and {Rykoff}, E.~S. and {Samuroff}, S. and {Sanchez}, J. and {S{\'a}nchez}, C. and {Secco}, L.~F. and {Sevilla-Noarbe}, I. and {Shaikh}, S. and {Sheldon}, E. and {Shin}, T. and {Sif{\'o}n}, Crist{\'o}bal and {To}, C. and {Troja}, A. and {Troxel}, M.~A. and {Tutusaus}, I. and {Varga}, T.~N. and {Weaverdyck}, N. and {Wechsler}, R.~H. and {Wollack}, E.~J. and {Yanny}, B. and {Yin}, B. and {Zhang}, Y. and {Zuntz}, J. and {Allam}, S.~S. and {Bacon}, D. and {Bocquet}, S. and {Brooks}, D. and {Burke}, D.~L. and {Carretero}, J. and {Cawthon}, R. and {Costanzi}, M. and {da Costa}, L.~N. and {da Silva Pereira}, M.~E. and {Davis}, T.~M. and {Desai}, S. and {Frieman}, J. and {Garc{\'\i}a-Bellido}, J. and {Gutierrez}, G. and {Hinton}, S.~R. and {Hollowood}, D.~L. and {Honscheid}, K. and {James}, D.~J. and {Jeffrey}, N. and {Lee}, S. and {Marshall}, J.~L. and {Mena-Fern{\'a}ndez}, J. and {Miquel}, R. and {Mohr}, J.~J. and {Ogando}, R.~L.~C. and {Plazas Malag'on}, A.~A. and {Romer}, A.~K. and {Sanchez}, E. and {Santiago}, B. and {Smith}, M. and {Suchyta}, E. and {Swanson}, M.~E.~C. and {Thomas}, D. and {Vikram}, V. and {Walker}, A.~R. and {Weller}, J. and {Wiseman}, P.},
        title = "{Constraints on cosmology and baryonic feedback with joint analysis of Dark Energy Survey Year 3 lensing data and ACT DR6 thermal Sunyaev-Zel'dovich effect observations}",
      journal = {arXiv e-prints},
         year = 2025,
        month = jun,
          eid = {arXiv:2506.07432},
        pages = {arXiv:2506.07432},
          doi = {10.48550/arXiv.2506.07432},
archivePrefix = {arXiv},
       eprint = {2506.07432},
 primaryClass = {astro-ph.CO},
       adsurl = {https://ui.adsabs.harvard.edu/abs/2025arXiv250607432P}
}

@ARTICLE{Troester_2022,
       author = {{Tr{\"o}ster}, Tilman and {Mead}, Alexander J. and {Heymans}, Catherine and {Yan}, Ziang and {Alonso}, David and {Asgari}, Marika and {Bilicki}, Maciej and {Dvornik}, Andrej and {Hildebrandt}, Hendrik and {Joachimi}, Benjamin and {Kannawadi}, Arun and {Kuijken}, Konrad and {Schneider}, Peter and {Shan}, Huan Yuan and {van Waerbeke}, Ludovic and {Wright}, Angus H.},
        title = "{Joint constraints on cosmology and the impact of baryon feedback: Combining KiDS-1000 lensing with the thermal Sunyaev-Zeldovich effect from Planck and ACT}",
      journal = {\aap},
         year = 2022,
        month = apr,
       volume = {660},
          eid = {A27},
        pages = {A27},
          doi = {10.1051/0004-6361/202142197},
archivePrefix = {arXiv},
       eprint = {2109.04458},
 primaryClass = {astro-ph.CO},
       adsurl = {https://ui.adsabs.harvard.edu/abs/2022A&A...660A..27T}
}

@ARTICLE{Gatti_2022,
       author = {{Gatti}, M. and {Pandey}, S. and {Baxter}, E. and {Hill}, J.~C. and {Moser}, E. and {Raveri}, M. and {Fang}, X. and {DeRose}, J. and {Giannini}, G. and {Doux}, C. and {Huang}, H. and {Battaglia}, N. and {Alarcon}, A. and {Amon}, A. and {Becker}, M. and {Campos}, A. and {Chang}, C. and {Chen}, R. and {Choi}, A. and {Eckert}, K. and {Elvin-Poole}, J. and {Everett}, S. and {Ferte}, A. and {Harrison}, I. and {Maccrann}, N. and {Mccullough}, J. and {Myles}, J. and {Navarro Alsina}, A. and {Prat}, J. and {Rollins}, R.~P. and {Sanchez}, C. and {Shin}, T. and {Troxel}, M. and {Tutusaus}, I. and {Yin}, B. and {Abbott}, T. and {Aguena}, M. and {Allam}, S. and {Andrade-Oliveira}, F. and {Annis}, J. and {Bernstein}, G. and {Bertin}, E. and {Bolliet}, B. and {Bond}, J.~R. and {Brooks}, D. and {Burke}, D.~L. and {Calabrese}, E. and {Carnero Rosell}, A. and {Carrasco Kind}, M. and {Carretero}, J. and {Cawthon}, R. and {Costanzi}, M. and {Crocce}, M. and {da Costa}, L.~N. and {da Silva Pereira}, M.~E. and {De Vicente}, J. and {Desai}, S. and {Diehl}, H.~T. and {Dietrich}, J.~P. and {Doel}, P. and {Dunkley}, J. and {Evrard}, A.~E. and {Ferraro}, S. and {Ferrero}, I. and {Flaugher}, B. and {Fosalba}, P. and {Frieman}, J. and {Garc{\'\i}a-Bellido}, J. and {Gaztanaga}, E. and {Gerdes}, D.~W. and {Giannantonio}, T. and {Gruen}, D. and {Gruendl}, R.~A. and {Gschwend}, J. and {Gutierrez}, G. and {Herner}, K. and {Hincks}, A.~D. and {Hinton}, S.~R. and {Hollowood}, D.~L. and {Honscheid}, K. and {Hughes}, J.~P. and {Huterer}, D. and {Jain}, B. and {James}, D.~J. and {Krause}, E. and {Kuehn}, K. and {Kuropatkin}, N. and {Lahav}, O. and {Lidman}, C. and {Lima}, M. and {Lokken}, M. and {Madhavacheril}, M.~S. and {Maia}, M.~A.~G. and {Marshall}, J.~L. and {Mcmahon}, J.~J. and {Melchior}, P. and {Moodley}, K. and {Mohr}, J.~J. and {Morgan}, R. and {Nati}, F. and {Niemack}, M.~D. and {Page}, L. and {Palmese}, A. and {Paz-Chinch{\'o}n}, F. and {Pieres}, A. and {Plazas Malag{\'o}n}, A.~A. and {Rodriguez-Monroy}, M. and {Romer}, A.~K. and {Sanchez}, E. and {Scarpine}, V. and {Schaan}, E. and {Secco}, L.~F. and {Serrano}, S. and {Sheldon}, E. and {Sherwin}, B.~D. and {Sif{\'o}n}, C. and {Smith}, M. and {Soares-Santos}, M. and {Spergel}, D. and {Suchyta}, E. and {Tarle}, G. and {Thomas}, D. and {To}, C. and {Tucker}, D.~L. and {Varga}, T.~N. and {Weller}, J. and {Wilkinson}, R.~D. and {Wollack}, E.~J. and {Xu}, Z. and {DES} and {ACT Collaboration}},
        title = "{Cross-correlation of Dark Energy Survey Year 3 lensing data with ACT and Planck thermal Sunyaev-Zel'dovich effect observations. I. Measurements, systematics tests, and feedback model constraints}",
      journal = {\prd},
         year = 2022,
        month = jun,
       volume = {105},
       number = {12},
          eid = {123525},
        pages = {123525},
          doi = {10.1103/PhysRevD.105.123525},
archivePrefix = {arXiv},
       eprint = {2108.01600},
 primaryClass = {astro-ph.CO},
       adsurl = {https://ui.adsabs.harvard.edu/abs/2022PhRvD.105l3525G}
}

@ARTICLE{Hojjati_2017,
       author = {{Hojjati}, Alireza and {Tr{\"o}ster}, Tilman and {Harnois-D{\'e}raps}, Joachim and {McCarthy}, Ian G. and {van Waerbeke}, Ludovic and {Choi}, Ami and {Erben}, Thomas and {Heymans}, Catherine and {Hildebrandt}, Hendrik and {Hinshaw}, Gary and {Ma}, Yin-Zhe and {Miller}, Lance and {Viola}, Massimo and {Tanimura}, Hideki},
        title = "{Cross-correlating Planck tSZ with RCSLenS weak lensing: implications for cosmology and AGN feedback}",
      journal = {\mnras},
         year = 2017,
        month = oct,
       volume = {471},
       number = {2},
        pages = {1565-1580},
          doi = {10.1093/mnras/stx1659},
archivePrefix = {arXiv},
       eprint = {1608.07581},
 primaryClass = {astro-ph.CO},
       adsurl = {https://ui.adsabs.harvard.edu/abs/2017MNRAS.471.1565H}
}

@ARTICLE{McCarthy_2024,
       author = {{McCarthy}, Fiona and {Hill}, J. Colin},
        title = "{Cross-correlation of the thermal Sunyaev-Zel'dovich and CMB lensing signals in Planck PR4 data with robust CIB decontamination}",
      journal = {\prd},
         year = 2024,
        month = jan,
       volume = {109},
       number = {2},
          eid = {023529},
        pages = {023529},
          doi = {10.1103/PhysRevD.109.023529},
archivePrefix = {arXiv},
       eprint = {2308.16260},
 primaryClass = {astro-ph.CO},
       adsurl = {https://ui.adsabs.harvard.edu/abs/2024PhRvD.109b3529M}
}

@ARTICLE{Osato_2020,
       author = {{Osato}, Ken and {Shirasaki}, Masato and {Miyatake}, Hironao and {Nagai}, Daisuke and {Yoshida}, Naoki and {Oguri}, Masamune and {Takahashi}, Ryuichi},
        title = "{Cross-correlation of the thermal Sunyaev-Zel'dovich effect and weak gravitational lensing: Planck and Subaru Hyper Suprime-Cam first-year data}",
      journal = {\mnras},
         year = 2020,
        month = mar,
       volume = {492},
       number = {4},
        pages = {4780-4804},
          doi = {10.1093/mnras/staa117},
archivePrefix = {arXiv},
       eprint = {1910.07526},
 primaryClass = {astro-ph.CO},
       adsurl = {https://ui.adsabs.harvard.edu/abs/2020MNRAS.492.4780O}
}

@ARTICLE{Planck_tsz_2016,
       author = {{Planck Collaboration} and {Aghanim}, N. and {Arnaud}, M. and {Ashdown}, M. and {Aumont}, J. and {Baccigalupi}, C. and {Banday}, A.~J. and {Barreiro}, R.~B. and {Bartlett}, J.~G. and {Bartolo}, N. and {Battaner}, E. and {Battye}, R. and {Benabed}, K. and {Beno{\^\i}t}, A. and {Benoit-L{\'e}vy}, A. and {Bernard}, J.-P. and {Bersanelli}, M. and {Bielewicz}, P. and {Bock}, J.~J. and {Bonaldi}, A. and {Bonavera}, L. and {Bond}, J.~R. and {Borrill}, J. and {Bouchet}, F.~R. and {Burigana}, C. and {Butler}, R.~C. and {Calabrese}, E. and {Cardoso}, J.-F. and {Catalano}, A. and {Challinor}, A. and {Chiang}, H.~C. and {Christensen}, P.~R. and {Churazov}, E. and {Clements}, D.~L. and {Colombo}, L.~P.~L. and {Combet}, C. and {Comis}, B. and {Coulais}, A. and {Crill}, B.~P. and {Curto}, A. and {Cuttaia}, F. and {Danese}, L. and {Davies}, R.~D. and {Davis}, R.~J. and {de Bernardis}, P. and {de Rosa}, A. and {de Zotti}, G. and {Delabrouille}, J. and {D{\'e}sert}, F.-X. and {Dickinson}, C. and {Diego}, J.~M. and {Dolag}, K. and {Dole}, H. and {Donzelli}, S. and {Dor{\'e}}, O. and {Douspis}, M. and {Ducout}, A. and {Dupac}, X. and {Efstathiou}, G. and {Elsner}, F. and {En{\ss}lin}, T.~A. and {Eriksen}, H.~K. and {Fergusson}, J. and {Finelli}, F. and {Forni}, O. and {Frailis}, M. and {Fraisse}, A.~A. and {Franceschi}, E. and {Frejsel}, A. and {Galeotta}, S. and {Galli}, S. and {Ganga}, K. and {G{\'e}nova-Santos}, R.~T. and {Giard}, M. and {Gonz{\'a}lez-Nuevo}, J. and {G{\'o}rski}, K.~M. and {Gregorio}, A. and {Gruppuso}, A. and {Gudmundsson}, J.~E. and {Hansen}, F.~K. and {Harrison}, D.~L. and {Henrot-Versill{\'e}}, S. and {Hern{\'a}ndez-Monteagudo}, C. and {Herranz}, D. and {Hildebrandt}, S.~R. and {Hivon}, E. and {Holmes}, W.~A. and {Hornstrup}, A. and {Huffenberger}, K.~M. and {Hurier}, G. and {Jaffe}, A.~H. and {Jones}, W.~C. and {Juvela}, M. and {Keih{\"a}nen}, E. and {Keskitalo}, R. and {Kneissl}, R. and {Knoche}, J. and {Kunz}, M. and {Kurki-Suonio}, H. and {Lacasa}, F. and {Lagache}, G. and {L{\"a}hteenm{\"a}ki}, A. and {Lamarre}, J.-M. and {Lasenby}, A. and {Lattanzi}, M. and {Leonardi}, R. and {Lesgourgues}, J. and {Levrier}, F. and {Liguori}, M. and {Lilje}, P.~B. and {Linden-V{\o}rnle}, M. and {L{\'o}pez-Caniego}, M. and {Mac{\'\i}as-P{\'e}rez}, J.~F. and {Maffei}, B. and {Maggio}, G. and {Maino}, D. and {Mandolesi}, N. and {Mangilli}, A. and {Maris}, M. and {Martin}, P.~G. and {Mart{\'\i}nez-Gonz{\'a}lez}, E. and {Masi}, S. and {Matarrese}, S. and {Melchiorri}, A. and {Melin}, J.-B. and {Migliaccio}, M. and {Miville-Desch{\^e}nes}, M.-A. and {Moneti}, A. and {Montier}, L. and {Morgante}, G. and {Mortlock}, D. and {Munshi}, D. and {Murphy}, J.~A. and {Naselsky}, P. and {Nati}, F. and {Natoli}, P. and {Noviello}, F. and {Novikov}, D. and {Novikov}, I. and {Paci}, F. and {Pagano}, L. and {Pajot}, F. and {Paoletti}, D. and {Pasian}, F. and {Patanchon}, G. and {Perdereau}, O. and {Perotto}, L. and {Pettorino}, V. and {Piacentini}, F. and {Piat}, M. and {Pierpaoli}, E. and {Pietrobon}, D. and {Plaszczynski}, S. and {Pointecouteau}, E. and {Polenta}, G. and {Ponthieu}, N. and {Pratt}, G.~W. and {Prunet}, S. and {Puget}, J.-L. and {Rachen}, J.~P. and {Reinecke}, M. and {Remazeilles}, M. and {Renault}, C. and {Renzi}, A. and {Ristorcelli}, I. and {Rocha}, G. and {Rossetti}, M. and {Roudier}, G. and {Rubi{\~n}o-Mart{\'\i}n}, J.~A. and {Rusholme}, B. and {Sandri}, M. and {Santos}, D. and {Sauv{\'e}}, A. and {Savelainen}, M. and {Savini}, G. and {Scott}, D. and {Spencer}, L.~D. and {Stolyarov}, V. and {Stompor}, R. and {Sunyaev}, R. and {Sutton}, D. and {Suur-Uski}, A.-S. and {Sygnet}, J.-F. and {Tauber}, J.~A. and {Terenzi}, L. and {Toffolatti}, L. and {Tomasi}, M. and {Tramonte}, D. and {Tristram}, M. and {Tucci}, M. and {Tuovinen}, J. and {Valenziano}, L. and {Valiviita}, J. and {Van Tent}, B. and {Vielva}, P. and {Villa}, F. and {Wade}, L.~A. and {Wandelt}, B.~D. and {Wehus}, I.~K. and {Yvon}, D.},
        title = "{Planck 2015 results. XXII. A map of the thermal Sunyaev-Zeldovich effect}",
      journal = {\aap},
         year = 2016,
        month = sep,
       volume = {594},
          eid = {A22},
        pages = {A22},
          doi = {10.1051/0004-6361/201525826},
archivePrefix = {arXiv},
       eprint = {1502.01596},
 primaryClass = {astro-ph.CO},
       adsurl = {https://ui.adsabs.harvard.edu/abs/2016A&A...594A..22P}
}

@ARTICLE{Choi_act_tsz_2020,
       author = {{Choi}, Steve K. and {Hasselfield}, Matthew and {Ho}, Shuay-Pwu Patty and {Koopman}, Brian and {Lungu}, Marius and {Abitbol}, Maximilian H. and {Addison}, Graeme E. and {Ade}, Peter A.~R. and {Aiola}, Simone and {Alonso}, David and {Amiri}, Mandana and {Amodeo}, Stefania and {Angile}, Elio and {Austermann}, Jason E. and {Baildon}, Taylor and {Battaglia}, Nick and {Beall}, James A. and {Bean}, Rachel and {Becker}, Daniel T. and {Bond}, J. Richard and {Bruno}, Sarah Marie and {Calabrese}, Erminia and {Calafut}, Victoria and {Campusano}, Luis E. and {Carrero}, Felipe and {Chesmore}, Grace E. and {Cho}, Hsiao-mei and {Clark}, Susan E. and {Cothard}, Nicholas F. and {Crichton}, Devin and {Crowley}, Kevin T. and {Darwish}, Omar and {Datta}, Rahul and {Denison}, Edward V. and {Devlin}, Mark J. and {Duell}, Cody J. and {Duff}, Shannon M. and {Duivenvoorden}, Adriaan J. and {Dunkley}, Jo and {D{\"u}nner}, Rolando and {Essinger-Hileman}, Thomas and {Fankhanel}, Max and {Ferraro}, Simone and {Fox}, Anna E. and {Fuzia}, Brittany and {Gallardo}, Patricio A. and {Gluscevic}, Vera and {Golec}, Joseph E. and {Grace}, Emily and {Gralla}, Megan and {Guan}, Yilun and {Hall}, Kirsten and {Halpern}, Mark and {Han}, Dongwon and {Hargrave}, Peter and {Henderson}, Shawn and {Hensley}, Brandon and {Hill}, J. Colin and {Hilton}, Gene C. and {Hilton}, Matt and {Hincks}, Adam D. and {Hlo{\v{z}}ek}, Ren{\'e}e and {Hubmayr}, Johannes and {Huffenberger}, Kevin M. and {Hughes}, John P. and {Infante}, Leopoldo and {Irwin}, Kent and {Jackson}, Rebecca and {Klein}, Jeff and {Knowles}, Kenda and {Kosowsky}, Arthur and {Lakey}, Vincent and {Li}, Dale and {Li}, Yaqiong and {Li}, Zack and {Lokken}, Martine and {Louis}, Thibaut and {MacInnis}, Amanda and {Madhavacheril}, Mathew and {Maldonado}, Felipe and {Mallaby-Kay}, Maya and {Marsden}, Danica and {Maurin}, Lo{\"\i}c and {McMahon}, Jeff and {Menanteau}, Felipe and {Moodley}, Kavilan and {Morton}, Tim and {Naess}, Sigurd and {Namikawa}, Toshiya and {Nati}, Federico and {Newburgh}, Laura and {Nibarger}, John P. and {Nicola}, Andrina and {Niemack}, Michael D. and {Nolta}, Michael R. and {Orlowski-Sherer}, John and {Page}, Lyman A. and {Pappas}, Christine G. and {Partridge}, Bruce and {Phakathi}, Phumlani and {Prince}, Heather and {Puddu}, Roberto and {Qu}, Frank J. and {Rivera}, Jesus and {Robertson}, Naomi and {Rojas}, Felipe and {Salatino}, Maria and {Schaan}, Emmanuel and {Schillaci}, Alessandro and {Schmitt}, Benjamin L. and {Sehgal}, Neelima and {Sherwin}, Blake D. and {Sierra}, Carlos and {Sievers}, Jon and {Sifon}, Cristobal and {Sikhosana}, Precious and {Simon}, Sara and {Spergel}, David N. and {Staggs}, Suzanne T. and {Stevens}, Jason and {Storer}, Emilie and {Sunder}, Dhaneshwar D. and {Switzer}, Eric R. and {Thorne}, Ben and {Thornton}, Robert and {Trac}, Hy and {Treu}, Jesse and {Tucker}, Carole and {Vale}, Leila R. and {Van Engelen}, Alexander and {Van Lanen}, Jeff and {Vavagiakis}, Eve M. and {Wagoner}, Kasey and {Wang}, Yuhan and {Ward}, Jonathan T. and {Wollack}, Edward J. and {Xu}, Zhilei and {Zago}, Fernando and {Zhu}, Ningfeng},
        title = "{The Atacama Cosmology Telescope: a measurement of the Cosmic Microwave Background power spectra at 98 and 150 GHz}",
      journal = {\jcap},
         year = 2020,
        month = dec,
       volume = {2020},
       number = {12},
          eid = {045},
        pages = {045},
          doi = {10.1088/1475-7516/2020/12/045},
archivePrefix = {arXiv},
       eprint = {2007.07289},
 primaryClass = {astro-ph.CO},
       adsurl = {https://ui.adsabs.harvard.edu/abs/2020JCAP...12..045C}
}

@ARTICLE{Reichardt_SPT_2021,
       author = {{Reichardt}, C.~L. and {Patil}, S. and {Ade}, P.~A.~R. and {Anderson}, A.~J. and {Austermann}, J.~E. and {Avva}, J.~S. and {Baxter}, E. and {Beall}, J.~A. and {Bender}, A.~N. and {Benson}, B.~A. and {Bianchini}, F. and {Bleem}, L.~E. and {Carlstrom}, J.~E. and {Chang}, C.~L. and {Chaubal}, P. and {Chiang}, H.~C. and {Chou}, T.~L. and {Citron}, R. and {Moran}, C. Corbett and {Crawford}, T.~M. and {Crites}, A.~T. and {de Haan}, T. and {Dobbs}, M.~A. and {Everett}, W. and {Gallicchio}, J. and {George}, E.~M. and {Gilbert}, A. and {Gupta}, N. and {Halverson}, N.~W. and {Harrington}, N. and {Henning}, J.~W. and {Hilton}, G.~C. and {Holder}, G.~P. and {Holzapfel}, W.~L. and {Hrubes}, J.~D. and {Huang}, N. and {Hubmayr}, J. and {Irwin}, K.~D. and {Knox}, L. and {Lee}, A.~T. and {Li}, D. and {Lowitz}, A. and {Luong-Van}, D. and {McMahon}, J.~J. and {Mehl}, J. and {Meyer}, S.~S. and {Millea}, M. and {Mocanu}, L.~M. and {Mohr}, J.~J. and {Montgomery}, J. and {Nadolski}, A. and {Natoli}, T. and {Nibarger}, J.~P. and {Noble}, G. and {Novosad}, V. and {Omori}, Y. and {Padin}, S. and {Pryke}, C. and {Ruhl}, J.~E. and {Saliwanchik}, B.~R. and {Sayre}, J.~T. and {Schaffer}, K.~K. and {Shirokoff}, E. and {Sievers}, C. and {Smecher}, G. and {Spieler}, H.~G. and {Staniszewski}, Z. and {Stark}, A.~A. and {Tucker}, C. and {Vanderlinde}, K. and {Veach}, T. and {Vieira}, J.~D. and {Wang}, G. and {Whitehorn}, N. and {Williamson}, R. and {Wu}, W.~L.~K. and {Yefremenko}, V.},
        title = "{An Improved Measurement of the Secondary Cosmic Microwave Background Anisotropies from the SPT-SZ + SPTpol Surveys}",
      journal = {\apj},
         year = 2021,
        month = feb,
       volume = {908},
       number = {2},
          eid = {199},
        pages = {199},
          doi = {10.3847/1538-4357/abd407},
archivePrefix = {arXiv},
       eprint = {2002.06197},
 primaryClass = {astro-ph.CO},
       adsurl = {https://ui.adsabs.harvard.edu/abs/2021ApJ...908..199R}
}

@article{Harris_numpy_2020,
   author = {Charles R Harris and K Jarrod Millman and Stéfan J van der Walt and Ralf Gommers and Pauli Virtanen and David Cournapeau and Eric Wieser and Julian Taylor and Sebastian Berg and Nathaniel J Smith and Robert Kern and Matti Picus and Stephan Hoyer and Marten H van Kerkwijk and Matthew Brett and Allan Haldane and Jaime Fernández del Río and Mark Wiebe and Pearu Peterson and Pierre Gérard-Marchant and Kevin Sheppard and Tyler Reddy and Warren Weckesser and Hameer Abbasi and Christoph Gohlke and Travis E Oliphant},
   doi = {10.1038/s41586-020-2649-2},
   journal = {Nature},
   pages = {357},
   title = {Array programming with NumPy},
   volume = {585},
   url = {https://doi.org/10.1038/s41586-020-2649-2},
   year = {2020}
}

@ARTICLE{Virtanen_scipy_2020,
       author = {{Virtanen}, Pauli and {Gommers}, Ralf and {Oliphant}, Travis E. and {Haberland}, Matt and {Reddy}, Tyler and {Cournapeau}, David and {Burovski}, Evgeni and {Peterson}, Pearu and {Weckesser}, Warren and {Bright}, Jonathan and {van der Walt}, St{\'e}fan J. and {Brett}, Matthew and {Wilson}, Joshua and {Millman}, K. Jarrod and {Mayorov}, Nikolay and {Nelson}, Andrew R.~J. and {Jones}, Eric and {Kern}, Robert and {Larson}, Eric and {Carey}, C.~J. and {Polat}, {\.I}lhan and {Feng}, Yu and {Moore}, Eric W. and {VanderPlas}, Jake and {Laxalde}, Denis and {Perktold}, Josef and {Cimrman}, Robert and {Henriksen}, Ian and {Quintero}, E.~A. and {Harris}, Charles R. and {Archibald}, Anne M. and {Ribeiro}, Ant{\^o}nio H. and {Pedregosa}, Fabian and {van Mulbregt}, Paul and {SciPy 1. 0 Contributors}},
        title = "{SciPy 1.0: fundamental algorithms for scientific computing in Python}",
      journal = {Nature Methods},
         year = 2020,
        month = feb,
       volume = {17},
        pages = {261-272},
          doi = {10.1038/s41592-019-0686-2},
archivePrefix = {arXiv},
       eprint = {1907.10121},
 primaryClass = {cs.MS},
       adsurl = {https://ui.adsabs.harvard.edu/abs/2020NatMe..17..261V}
}

@ARTICLE{Hunter_matplotlib_2007,
  author={Hunter, John D.},
  journal={Computing in Science \& Engineering}, 
  title={Matplotlib: A 2D Graphics Environment}, 
  year={2007},
  volume={9},
  number={3},
  pages={90-95},
  doi={10.1109/MCSE.2007.55}}

@ARTICLE{Gorski_Healpix_2005,
       author = {{G{\'o}rski}, K.~M. and {Hivon}, E. and {Banday}, A.~J. and {Wandelt}, B.~D. and {Hansen}, F.~K. and {Reinecke}, M. and {Bartelmann}, M.},
        title = "{HEALPix: A Framework for High-Resolution Discretization and Fast Analysis of Data Distributed on the Sphere}",
      journal = {\apj},
         year = 2005,
        month = apr,
       volume = {622},
       number = {2},
        pages = {759-771},
          doi = {10.1086/427976},
archivePrefix = {arXiv},
       eprint = {astro-ph/0409513},
 primaryClass = {astro-ph},
       adsurl = {https://ui.adsabs.harvard.edu/abs/2005ApJ...622..759G}
}

@ARTICLE{Bolliet_2018,
       author = {{Bolliet}, Boris and {Comis}, Barbara and {Komatsu}, Eiichiro and {Mac{\'\i}as-P{\'e}rez}, Juan Francisco},
        title = "{Dark energy constraints from the thermal Sunyaev-Zeldovich power spectrum}",
      journal = {\mnras},
         year = 2018,
        month = jul,
       volume = {477},
       number = {4},
        pages = {4957-4967},
          doi = {10.1093/mnras/sty823},
archivePrefix = {arXiv},
       eprint = {1712.00788},
 primaryClass = {astro-ph.CO},
       adsurl = {https://ui.adsabs.harvard.edu/abs/2018MNRAS.477.4957B}
}

@ARTICLE{Blas_CLASS_2011,
       author = {{Blas}, Diego and {Lesgourgues}, Julien and {Tram}, Thomas},
        title = "{The Cosmic Linear Anisotropy Solving System (CLASS). Part II: Approximation schemes}",
      journal = {\jcap},
         year = 2011,
        month = jul,
       volume = {2011},
       number = {7},
          eid = {034},
        pages = {034},
          doi = {10.1088/1475-7516/2011/07/034},
archivePrefix = {arXiv},
       eprint = {1104.2933},
 primaryClass = {astro-ph.CO},
       adsurl = {https://ui.adsabs.harvard.edu/abs/2011JCAP...07..034B}
}

@ARTICLE{Lesgourgues_CLASS_2011,
       author = {{Lesgourgues}, Julien},
        title = "{The Cosmic Linear Anisotropy Solving System (CLASS) I: Overview}",
      journal = {arXiv e-prints},
         year = 2011,
        month = apr,
          eid = {arXiv:1104.2932},
        pages = {arXiv:1104.2932},
          doi = {10.48550/arXiv.1104.2932},
archivePrefix = {arXiv},
       eprint = {1104.2932},
 primaryClass = {astro-ph.IM},
       adsurl = {https://ui.adsabs.harvard.edu/abs/2011arXiv1104.2932L}
}

@ARTICLE{SZ_effect_1970,
       author = {{Sunyaev}, R.~A. and {Zeldovich}, Ya. B.},
        title = "{Small-Scale Fluctuations of Relic Radiation}",
      journal = {\apss},
         year = 1970,
        month = apr,
       volume = {7},
       number = {1},
        pages = {3-19},
          doi = {10.1007/BF00653471},
       adsurl = {https://ui.adsabs.harvard.edu/abs/1970Ap&SS...7....3S}
}

@ARTICLE{Delabrouille_2009,
       author = {{Delabrouille}, J. and {Cardoso}, J.-F. and {Le Jeune}, M. and {Betoule}, M. and {Fay}, G. and {Guilloux}, F.},
        title = "{A full sky, low foreground, high resolution CMB map from WMAP}",
      journal = {\aap},
         year = 2009,
        month = jan,
       volume = {493},
       number = {3},
        pages = {835-857},
          doi = {10.1051/0004-6361:200810514},
archivePrefix = {arXiv},
       eprint = {0807.0773},
 primaryClass = {astro-ph},
       adsurl = {https://ui.adsabs.harvard.edu/abs/2009A&A...493..835D}
}

@ARTICLE{Helly_2026,
       author = {{Helly}, John C. and {McGibbon}, Robert J. and {Schaye}, Joop and {Schaller}, Matthieu and {McDonald}, William and {Braspenning}, Joey and {Broxterman}, Jeger C. and {Costello}, Emily E. and {Elbers}, Willem and {Forouhar Moreno}, Victor J. and {Frenk}, Carlos S. and {Jenkins}, Adrian and {Kugel}, Roi and {McCarthy}, Ian G. and {Salcido}, Jaime and {van Daalen}, Marcel P. and {Vandenbroucke}, Bert and {Yang}, Tianyi},
        title = "{The FLAMINGO simulations data release}",
      journal = {Astronomy and Computing},
         year = 2026,
        month = oct,
       volume = {57},
          eid = {101159},
        pages = {101159},
          doi = {10.1016/j.ascom.2026.101159},
archivePrefix = {arXiv},
       eprint = {2604.24324},
 primaryClass = {astro-ph.CO},
       adsurl = {https://ui.adsabs.harvard.edu/abs/2026A&C....5701159H}
}

@ARTICLE{McGibbon_2025,
       author = {{McGibbon}, Robert and {Helly}, John and {Schaye}, Joop and {Schaller}, Matthieu and {Vandenbroucke}, Bert},
        title = "{SOAP: A Python Package for Calculating the Properties of Galaxies and Halos Formed in Cosmological Simulations}",
      journal = {The Journal of Open Source Software},
         year = 2025,
        month = jul,
       volume = {10},
       number = {111},
          eid = {8252},
        pages = {8252},
          doi = {10.21105/joss.08252},
archivePrefix = {arXiv},
       eprint = {2507.22669},
 primaryClass = {astro-ph.IM},
       adsurl = {https://ui.adsabs.harvard.edu/abs/2025JOSS...10.8252M}
}

@ARTICLE{Tram_2019,
       author = {{Tram}, Thomas and {Brandbyge}, Jacob and {Dakin}, Jeppe and {Hannestad}, Steen},
        title = "{Fully relativistic treatment of light neutrinos in N-body simulations}",
      journal = {\jcap},
         year = 2019,
        month = mar,
       volume = {2019},
       number = {3},
          eid = {022},
        pages = {022},
          doi = {10.1088/1475-7516/2019/03/022},
archivePrefix = {arXiv},
       eprint = {1811.00904},
 primaryClass = {astro-ph.CO},
       adsurl = {https://ui.adsabs.harvard.edu/abs/2019JCAP...03..022T}
}

@ARTICLE{McCarthy_2025,
       author = {{McCarthy}, Ian G. and {Amon}, Alexandra and {Schaye}, Joop and {Schaan}, Emmanuel and {Angulo}, Raul E. and {Salcido}, Jaime and {Schaller}, Matthieu and {Bigwood}, Leah and {Elbers}, Willem and {Kugel}, Roi and others},
        title = "{FLAMINGO: combining kinetic SZ effect and galaxy-galaxy lensing measurements to gauge the impact of feedback on large-scale structure}",
      journal = {\mnras},
         year = 2025,
        month = jun,
       volume = {540},
       number = {1},
        pages = {143-163},
          doi = {10.1093/mnras/staf731},
archivePrefix = {arXiv},
       eprint = {2410.19905},
 primaryClass = {astro-ph.CO},
       adsurl = {https://ui.adsabs.harvard.edu/abs/2025MNRAS.540..143M}
}

@ARTICLE{McCarthy_2023,
       author = {{McCarthy}, Ian G. and {Salcido}, Jaime and {Schaye}, Joop and {Kwan}, Juliana and {Elbers}, Willem and {Kugel}, Roi and {Schaller}, Matthieu and {Helly}, John C. and {Braspenning}, Joey and {Frenk}, Carlos S. and {van Daalen}, Marcel P. and {Vandenbroucke}, Bert and {Conley}, Jonah T. and {Font}, Andreea S. and {Upadhye}, Amol},
        title = "{The FLAMINGO project: revisiting the S$_8$ tension and the role of baryonic physics}",
      journal = {\mnras},
         year = 2023,
       volume = {526},
        pages = {5494-5519},
          doi = {10.1093/mnras/stad3107},
archivePrefix = {arXiv},
       eprint = {2309.07959},
 primaryClass = {astro-ph.CO}
}

@ARTICLE{McCarthy_2014,
       author = {{McCarthy}, I.~G. and {Le Brun}, A.~M.~C. and {Schaye}, J. and {Holder}, G.~P.},
        title = "{The thermal Sunyaev-Zel'dovich effect power spectrum in light of Planck}",
      journal = {\mnras},
         year = 2014,
        month = jun,
       volume = {440},
       number = {4},
        pages = {3645-3657},
          doi = {10.1093/mnras/stu543},
archivePrefix = {arXiv},
       eprint = {1312.5341},
 primaryClass = {astro-ph.CO},
       adsurl = {https://ui.adsabs.harvard.edu/abs/2014MNRAS.440.3645M}
}

@ARTICLE{Siegel_2026,
       author = {{Siegel}, Jared and {Bigwood}, Leah and {Amon}, Alexandra and {McCullough}, Jamie and {Yamamoto}, Masaya and {McCarthy}, Ian G. and {Schaller}, Matthieu and {Schaye}, Joop and {Schneider}, Aurel},
        title = "{The suppression of the matter power spectrum: strong feedback from X-ray gas mass fractions, kSZ effect profiles, and galaxy-galaxy lensing}",
      journal = {\mnras},
         year = 2026,
        month = jul,
       volume = {549},
       number = {4},
          eid = {stag993},
        pages = {stag993},
          doi = {10.1093/mnras/stag993},
archivePrefix = {arXiv},
       eprint = {2512.02954},
 primaryClass = {astro-ph.CO},
       adsurl = {https://ui.adsabs.harvard.edu/abs/2026MNRAS.549ag993S}
}

@ARTICLE{Lucie-Smith_2025,
       author = {{Lucie-Smith}, Luisa and {Peiris}, Hiranya V. and {Pontzen}, Andrew and {Halder}, Anik and {Schaye}, Joop and {Schaller}, Matthieu and {Helly}, John and {McGibbon}, Robert J. and {Elbers}, Willem},
        title = "{Cosmological feedback from a halo assembly perspective}",
      journal = {\prd},
         year = 2025,
        month = sep,
       volume = {112},
       number = {6},
          eid = {063541},
        pages = {063541},
          doi = {10.1103/vh8n-9cr2},
archivePrefix = {arXiv},
       eprint = {2505.18258},
 primaryClass = {astro-ph.CO},
       adsurl = {https://ui.adsabs.harvard.edu/abs/2025PhRvD.112f3541L}
}

@ARTICLE{Castro_2023,
       author = {{Euclid Collaboration} and {Castro}, T. and {Fumagalli}, A. and {Angulo}, R.~E. and {Bocquet}, S. and {Borgani}, S. and {Carbone}, C. and {Dakin}, J. and {Dolag}, K. and {Giocoli}, C. and others},
        title = "{Euclid preparation. XXIV. Calibration of the halo mass function in {\ensuremath{\Lambda}}({\ensuremath{\nu}})CDM cosmologies}",
      journal = {\aap},
         year = 2023,
        month = mar,
       volume = {671},
          eid = {A100},
        pages = {A100},
          doi = {10.1051/0004-6361/202244674},
archivePrefix = {arXiv},
       eprint = {2208.02174},
 primaryClass = {astro-ph.CO},
       adsurl = {https://ui.adsabs.harvard.edu/abs/2023A&A...671A.100E}
}

@ARTICLE{Schaller_2025,
       author = {{Schaller}, Matthieu and {Schaye}, Joop and {Kugel}, Roi and {Broxterman}, Jeger C. and {van Daalen}, Marcel P.},
        title = "{The FLAMINGO project: baryon effects on the matter power spectrum}",
      journal = {\mnras},
         year = 2025,
        month = may,
       volume = {539},
       number = {2},
        pages = {1337-1351},
          doi = {10.1093/mnras/staf569},
archivePrefix = {arXiv},
       eprint = {2410.17109},
 primaryClass = {astro-ph.CO},
       adsurl = {https://ui.adsabs.harvard.edu/abs/2025MNRAS.539.1337S}
}

@ARTICLE{Schaller_2024,
       author = {{Schaller}, Matthieu and {Borrow}, Josh and {Draper}, Peter W. and {Ivkovic}, Mladen and {McAlpine}, Stuart and {Vandenbroucke}, Bert and {Bah{\'e}}, Yannick and {Chaikin}, Evgenii and {Chalk}, Aidan B.~G. and {Chan}, Tsang Keung and others},
        title = "{SWIFT: A modern highly-parallel gravity and smoothed particle hydrodynamics solver for astrophysical and cosmological applications}",
      journal = {\mnras},
         year = 2024,
        month = may,
       volume = {530},
       number = {2},
        pages = {2378-2419},
          doi = {10.1093/mnras/stae922},
archivePrefix = {arXiv},
       eprint = {2305.13380},
 primaryClass = {astro-ph.IM},
       adsurl = {https://ui.adsabs.harvard.edu/abs/2024MNRAS.530.2378S}
}

@ARTICLE{Davis_1985,
       author = {{Davis}, M. and {Efstathiou}, G. and {Frenk}, C.~S. and {White}, S.~D.~M.},
        title = "{The evolution of large-scale structure in a universe dominated by cold dark matter}",
      journal = {\apj},
         year = 1985,
        month = may,
       volume = {292},
        pages = {371-394},
          doi = {10.1086/163168},
       adsurl = {https://ui.adsabs.harvard.edu/abs/1985ApJ...292..371D}
}

\end{document}